\documentclass[twocolumn]{aastex63}
\usepackage{amsmath}
\usepackage[caption=false]{subfig}

\newcommand{\ST}{\textsc{StarTrack}}
\newcommand{\msun}{M_\sun}
\newcommand{\rsun}{R_\sun}
\newcommand{\lsun}{L_\sun}

\received{January 3, 2022}
\revised{March 16, 2022}
\accepted{March 25, 2022; Published May 4, 2022}
\shorttitle{Synthetic population of binary Cepheids: metallicity and initial parameter distribution}
\shortauthors{Karczmarek et al.}

\begin{document}

\title{Synthetic population of binary Cepheids. I. The effect of metallicity and initial parameter distribution on characteristics of Cepheids' companions}

\author[0000-0002-0136-0046]{Paulina Karczmarek}
\affiliation{Departamento de Astronom{\'i}a, Universidad de Concepci{\'o}n, Casilla 160-C, Concepci{\'o}n, Chile}

\author{Rados{\l}aw Smolec}
\affiliation{Nicolaus Copernicus Astronomical Center, Bartycka 18, 00-716 Warsaw, Poland}

\author{Gergely Hajdu}
\affiliation{Nicolaus Copernicus Astronomical Center, Bartycka 18, 00-716 Warsaw, Poland}

\author{Grzegorz Pietrzy{\'n}ski}
\affiliation{Nicolaus Copernicus Astronomical Center, Bartycka 18, 00-716 Warsaw, Poland}

\author{Wolfgang Gieren}
\affiliation{Departamento de Astronom{\'i}a, Universidad de Concepci{\'o}n, Casilla 160-C, Concepci{\'o}n, Chile}

\author{Weronika Narloch}
\affiliation{Departamento de Astronom{\'i}a, Universidad de Concepci{\'o}n, Casilla 160-C, Concepci{\'o}n, Chile}

\author{Grzegorz Wiktorowicz}
\affiliation{National Astronomical Observatories, Chinese Academy of Sciences, Beijing 100101, China}
\affiliation{School of Astronomy and Space Science, University of the Chinese Academy of Sciences, Beijing 100012, China}

\author{Krzysztof Belczynski}
\affiliation{Nicolaus Copernicus Astronomical Center, Bartycka 18, 00-716 Warsaw, Poland}


\correspondingauthor{Paulina Karczmarek}
\email{pkarczmarek@astro-udec.cl}



\begin{abstract}
The majority of classical Cepheids are binary stars, yet the contribution of companions' light to the total brightness of the system has been assumed negligible and lacked a thorough, quantitative evaluation. We present an extensive study of synthetic populations of binary Cepheids, which aims to characterize Cepheids' companions (e.g. masses, evolutionary, and spectral types), quantify their contribution to the brightness and color of Cepheid binaries, and assess the relevance of input parameters on the results. We introduce a collection of synthetic populations, which vary in metal content, initial parameter distribution, location of the instability strip edges, and star formation history. Our synthetic populations are free from the selection bias, while the percentage of Cepheid binaries is controlled by the binarity parameter. We successfully reproduce recent theoretical and empirical results: the percentage of binary Cepheids with main-sequence (MS) companions, the contrast-mass ratio relation for binary Cepheids with MS companions, the appearance of binary Cepheids with giant, evolved companions as outlier data points above the period-luminosity relation. Moreover, we present the first estimation of the percentage of binary Cepheids in the Large Magellanic Cloud and announce the quantification of the effect of binarity on the slope and zero-point of multiband period-luminosity relations, which will be reported in the next paper of this series.
\end{abstract}




\section{Introduction}
\label{sec:intro}

Classical Cepheids (hereafter referred to as Cepheids) are among the most famous and widely used cosmic distance calibrators; their high luminosity and characteristic light curves make them easily recognizable, while the period-luminosity relation (PLR) that they follow is considered universal and of superior accuracy to that offered by other types of radial pulsators. Still, Cepheids' PLR suffers from several systematic uncertainties, related to, e.g. binarity, metallicity, number of crossing of the instability strip, reddening, that hinder achieving a subpercent precision in distance determination.

Binary Cepheids, in particular, can be a potent source of systematic errors since they constitute $60-80$\% (and likely even more) of all Galactic Cepheids. This high Cepheid-binary fraction is supported by both theoretical \citep{neilson15,mor17} and empirical studies \citep[e.g. ][]{szabados03,kervella19a}. Both approaches have their limitations; theoretical results are strongly dependent on the input parameters, while empirical ones suffer from the selection bias. Indeed, an ultraviolet (UV) survey of binary Cepheids with hot main-sequence (MS) companions in the Milky Way (MW) was limited to stars with $V< 8$\,mag \citep{evans92_uv}, leaving fainter binaries and binaries with cooler companions undetected. In the Large Magellanic Cloud (LMC), 25 spectroscopic binaries with Cepheids have been reported so far \citep[][and references therein]{pilecki21,szabados12}, five of them being eclipsing binaries \citep{pilecki18}. In the Small Magellanic Cloud (SMC), only nine Cepheid binaries have been reported so far, among which two are spectroscopic binaries, another two show eclipsing variations, and the remaining five are firm candidates for Cepheid--Cepheid binaries \citep[][and references therein]{szabados12}. Such scarcity of binary Cepheids in the LMC and SMC relative to the MW indicates a strong selection bias in the Magellanic Clouds, which favors Cepheids with giant (and possibly pulsating) companions and highly inclined orbits.

While resolved binary Cepheids are remarkable tools to determine geometrical distances and companions' dynamical masses with an astonishing 1\% accuracy \citep{pietrzynski10,gallenne18}, unresolved binary Cepheids can bias the measurements in a number of undesired ways. For example, the presence of a companion can affect the radial-velocity curve of a Cepheid, impeding an accurate radius determination \citep{gieren98}. Spectroscopic analysis can yield inaccurate stellar parameters and abundances if single-star models are fitted to the combined spectrum of unresolved binaries \citep{el-badry18}. Astrometric solutions and parallaxes provided by the Gaia space mission are yet to be corrected for the variability of binary and pulsating stars; until this happens, Gaia parallaxes for binary Cepheids should be inferred from resolved companion stars \citep{kervella19a}. 

Furthermore, contribution of a companion's light to the total brightness of the system causes Cepheids with unresolved companions to seem brighter than their single counterparts; this effect is largest in the near-infrared domain if the companion is a red giant \citep[RG][]{pilecki18}, and in the UV domain if the companion is a hot MS star \citep{evans92_uv}. As a result, Cepheids with unresolved companions can alter the slope and the zero-point of the PLR, which has been predicted and described in a qualitative way \citep[e.g.][]{szabados12_PL} but still lacks quantification. 
Binary Cepheids are expected to have different color indices, especially if the binary components have very dissimilar effective temperatures, which leads to an incorrect estimation of the reddening values toward the system. Luminous companions diminish the observed pulsation amplitudes of Cepheids \citep{pilecki21}, and may be in part responsible for the scatter in the pulsation period-amplitude relations \citep{klagyivik09}. Observed luminosities and amplitudes of unresolved binary Cepheids are unreliable points of reference for theoretical models of stellar pulsations and evolution, which operate within a framework of single stars. Moreover, \citet{evans05} reported that at least 44\% of all binary Cepheids are in fact triple systems and \citet{dinnbier22} suggested that around half of all Cepheids form in triple and quadruple systems, instead of binaries. This information is crucial for mass determination of (assumed) binary components, as the presence of a third component leads to inaccurate estimations.

In order to address some of the aforementioned issues, we employ a theoretical approach called binary population synthesis. This method relies on approximate formulas that govern the evolution of single and binary stars, which makes it fast and efficient. Results of population synthesis are independent of reddening, blending, and selection bias, and therefore provide invaluable insight in the characteristics of binary Cepheids. For example, \citet{mor17} estimated that 68\% of all Cepheids should have companions, and \citet{neilson15} reported that 35\% of MW Cepheids should be detectable as spectroscopic binaries. \citet{anderson18} assessed the impact of the photometric bias from MW Cepheids in wide binaries ($a > 400$\,au) and open clusters on the value of the Hubble constant. While \citet{anderson16_leavittlaw} did not resort to the population synthesis to assess the excess light of a Cepheid binary with a MS star as a function of a Cepheid's pulsation period ($\log P-\Delta M$), they recognized that population synthesis is required in order to characterize this relation more thoroughly.

In this paper, we present the most extensive study of synthetic populations of binary Cepheids up to date. We take full advantage of the population synthesis method that treats metallicity and binarity percentage as free parameters, which can be set to any arbitrary value or a grid of values. We create synthetic populations of three metallicities $Z=$\,0.004, 0.008, 0.02, reflecting the metal content of classical Cepheids in the SMC, LMC, and MW, respectively, and with binarity percentages of 0\%, 25\%, 50\%, 75\%, and 100\%. Such an approach gives us full control over the binarity and metallicity, and allows us to study the impact of these parameters on the observed characteristics of resolved and unresolved binary Cepheids.

Moreover, for every combination of metallicity and binarity, we test four sets of initial conditions and their effect on the outcome. We examine the entire collection of synthetic populations of binary Cepheids for similarities and differences between the variants, and compare our theoretical predictions with features observed in binary Cepheids in the MW and LMC. Following papers in this series will focus on the detailed analysis of PLRs and the quantification of the expected shift of their zero-points due to binarity (Paper II), and the quantification of the shift in Cepheids' color indices, caused by companions' dissimilar effective temperatures, which affects the reddening toward binary Cepheids, and presents an opportunity to detect companions on the color-color diagram (Paper III).

\section{Synthetic populations}
\label{sec:syntpop}

The \ST\ population synthesis code \citep{belczynski02,belczynski08} is based on the revised formulae from \citet{hurley00,hurley02}, fitted to detailed single-star models with convective core overshooting across the entire Hertzsprung-Russell diagram (HRD), created by \citet{pols98}. A number of enhancements implemented to the \ST\ code account for wind accretion through the Bondi-Hoyle mechanism, atmospheric Roche lobe overflow \citep{ritter88} and wind Roche lobe overflow \citep{mohamed12,abate13}.

We used \ST\ to generate populations of 200\,000 binaries on the zero-age main sequence (ZAMS) of three metallicities $Z = 0.004, 0.008, 0.02$, and helium abundances $Y = 0.248, 0.256, 0.280$, reflecting the environmental properties of young stellar populations in the SMC, LMC, and MW, respectively. This metallicity comes from adopting the equation $\mathrm{[Fe/H]} = \log(Z) - \log(Z_\odot)$, assuming solar metallicity $Z_\sun = 0.02$ and $\mbox{[Fe/H]}_\mathrm{SMC} = -0.74 $ dex, $\mbox{[Fe/H]}_\mathrm{LMC} = -0.35$ dex \citep{lemasle17}. Throughout our study, we refer to the primary component (A) as the more massive star on the ZAMS while the secondary (B) is the less massive one. The maximum evolutionary age considered for each binary is 14~Gyr. In order to create a synthetic population that consists purely of binaries with Cepheids, we created a filtering algorithm to test if either component in a (synthetic) binary has stellar parameters characteristic of Cepheids, described below. Only binaries with at least one component that passed the test were included in the final sample.

\begin{figure}
\centering
\includegraphics[width=\columnwidth]{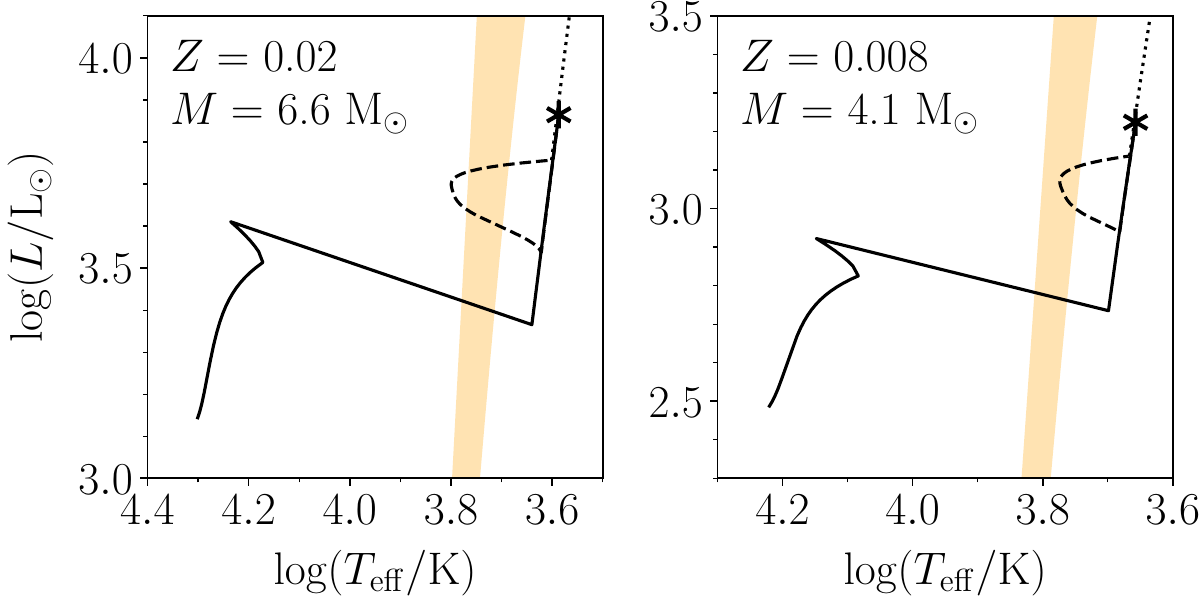}
\caption{Examples of evolutionary tracks of primary components in the $\log T-\log L$ plane. Companions were omitted for clarity. Solid, dashed, and dotted lines indicate evolutionary stages before, during, and after core helium burning, respectively. Asterisk indicates the moment of helium ignition in the core. The instability strip loci and shapes are adopted from \citet{anderson16}.
\label{fig:hr_tracks}}
\end{figure}

First, we selected binaries with a component that crosses the instability strip (IS). Figure \ref{fig:hr_tracks} shows two examples of stars that evolve through a Cepheid stage while inside the IS. The first crossing (IS1) happens when a star traverses IS in the Hertzsprung Gap, the second crossing (IS2) happens at the stage of core helium burning (blue loop) when a star traverses the IS toward the blue edge, and the third crossing (IS3) happens when it makes a blue loop toward the red edge. If a star makes its blue loop and \emph{turns around} while still inside the IS, the turning point divides its evolutionary stage into IS2 and IS3.

We excluded all systems that experienced substantial mass transfer (MT), i.e. the mass lost or gained due to the MT constituted more than 10\% of the star's initial mass\footnote{We chose a value of 10\% to match criteria set by \citet{neilson15}, so that their and our results can be compared.}. More than 10\% of mass lost/gained before the IS crossing would disrupt the physical structure of binary stars, raising a question of whether a Cepheid variable would retain its pulsation properties after the MT, and if so, whether it be justified to label it a Cepheid once its internal structure and evolutionary status has changed \citep{karczmarek17}. However, the mass loss due to stellar winds is common among Cepheids and ranges from $10^{-10}$ to $10^{-6}\,\msun/\mathrm{yr}$ \citep{deasy88,matthews16}; we included every star that experienced mass loss due to stellar winds below this upper limit. Another possible event is a supernova explosion, which however may disrupt the orbit of a binary. Since we focus on Cepheids in stable and uneventful binaries, we did not follow the evolution of disrupted binary components, meaning that all such stars were excluded from our sample.

Every binary selected so far hosts a star that crosses the IS at some point during its evolution, but it can only be regarded as a Cepheid if its birth time stamp equals its age in a \emph{Cepheid stage}. Therefore, the last step was to assign birth time stamps to our synthetic binaries in order to determine their age and therefore the evolutionary status of the components. This procedure is described in detail in Section \ref{sec:SFH}.

Constructing a synthetic population of binary Cepheids is a complex task with an ambiguous outcome, because the output depends strongly on the input parameters. The results can be considered reliable if they agree with the observations or if they do not change much upon uncertainties introduced by the input physics relevant for the formation of binary Cepheids. We selected three areas that we consider are of most relevance to the reliability of the results: (i) distribution of initial parameters; (ii) shape and location of the IS; (iii) star formation history (SFH) that impacts Cepheids' birth time stamps. In the rest of this section, we detail the alterations introduced in these three areas, and we describe the process of augmentation of our synthetic data with pulsation periods, amplitudes, and multiband magnitudes, which allow for a comparison with observed binary Cepheids, and provide a frame of reference for future discoveries.

\subsection{Initial distributions}

A synthetic population is generated by \ST\ based on four initial parameter distributions: mass of the primary, mass ratio (secondary to primary), orbital separation (semi-major axis), and eccentricity. By default these parameters are independent and for every binary are drawn from the following distributions:
\begin{itemize}
    \item broken power-law initial mass function \citep{kroupa03} with the value of a slope $-2.35$ for the mass of the primary $M_\mathrm{A}$ from 2.5 to 12.0\,$\msun$\footnote{
    The upper limit was chosen based on evolutionary models of massive Cepheids, which above $11-12\,\msun$ fail to present a blue loop or their blue loop is erratic. The lower limit was chosen based on evolutionary models of low-mass stars, which below $2.5-3.0\,\msun$ fail to cross the IS as post main-sequence objects.},
    \item flat distribution of mass ratio of secondary to primary $q = M_\mathrm{B}/M_\mathrm{A}$ \citep{kobulnicky07} in a range from $q_\mathrm{min}$ to 1, where $q_\mathrm{min}$ is the mass ratio that results in a secondary with $M_\mathrm{B}=0.08\,\msun$,
    \item flat distribution of the logarithm of semi-major axis of binary orbit \citep{abt83} in a range from  $a_\mathrm{min}$ to $-10^5\,\rsun$, %
    where $a_\mathrm{min}$ is twice the sum of component's radii at periastron,
    \item thermal eccentricity distribution $f(e) = 2e$ \citep{heggie75} in range from 0 to 0.99.
\end{itemize}

The above set of initial parameter distributions is called ``set A''. Alternative distributions were published by \citet{duquennoy91}, and recently by \citet{moe17}. \citet{duquennoy91} reported normal distribution of mass ratios $\mathcal{N}(4.8,\,2.3^{2})\,$, log-normal distribution of orbital periods $\log \mathcal{N}(0.23,\,0.42^{2})\,$, and a mixture of normal $\mathcal{N}(0.27,\,0.13^{2})\,$ (for $10\,\mathrm{d}<P_\mathrm{orb}<1000\,\mathrm{d}$), thermal (for $P_\mathrm{orb} \geqslant 1000$\,d), and uniform (for $P_\mathrm{orb} \leqslant 10\,\mathrm{d}$) distributions of eccentricities, with $M_\mathrm{A}$, $P_\mathrm{orb}$, and $q$ independent, and $e$ dependent on $P_\mathrm{orb}$. They did not state the distribution of $M_\mathrm{A}$, which we decided to keep as in set A. Their initial distributions are used in our work as ``set B''. \citet{moe17} reported $e$ and $q$ distributions that follow power laws with different exponent values, depending on $P_\mathrm{orb}$, and $M_\mathrm{A}$, while keeping $M_\mathrm{A}$ and $P_\mathrm{orb}$ independent. They did not describe distributions for $P_\mathrm{orb}$ and $M_\mathrm{A}$. Thus, for the distribution of $M_\mathrm{A}$, we used the one from set A, while for $P_\mathrm{orb}$ we used two already presented variants: log-uniform (as in set A) and log-normal (as in set B), creating two more sets, C and D, respectively. Table \ref{tab:STmodels} summarizes the four sets of initial distributions, and Figure \ref{apdx:initialdistr} in Appendix \ref{apdx:initialdistr} supplements Table \ref{tab:STmodels} with triangle plots of initial parameters for all four sets.

Binary populations created from the four sets differ especially in the distributions of orbital periods and masses of the companions. In Section \ref{sec:results}, we show how the choice of a set impacts the distributions of orbital periods and effective temperatures of Cepheids' companions.




\begin{deluxetable*}{lllll}
\tablecaption{Models of Initial Parameters Used in \ST.\label{tab:STmodels}
}
\tablehead{
\colhead{Parameter} & \colhead{Set A} & \colhead{Set B} & \colhead{Set C} & \colhead{Set D}
}
\startdata
$M_1$ ($\msun$) & power law (1) & power law (1) & power law (1) & power law (1)\\
$q=M_2/M_1$ & uniform (2) & log-normal (5) & power law (6) & power law (6)\\
$a$ ($\rsun$) & log-uniform (3) & - & log-uniform (3) & - \\
$P$ (d) & - & log-normal (5) & - & log-normal (5)\\
$e$ & thermal (4) & log-normal + thermal (5) & power law (6) & power law (6) \\
\hline
Remarks & all indep. & $M_1$, $P$, $q$ indep., & $M_1,\,P$ indep., & $M_1,\,P$ indep., \\
&& $e$ dep. on $P$ & $q$, $e$ dep. on $M_1,\,P$ & $q$, $e$ dep. on $M_1,\,P$
\enddata
\tablerefs{(1) \citet{kroupa03}; (2) \citet{kobulnicky07}; (3) \citet{abt83}; (4) \citet{heggie75}; (5) \citet{duquennoy91}; (6) \citet{moe17}.}
\tablecomments{Orbital periods and semi-major axes are interchangeable. This means that when one of them was drawn from a distribution, the other one was calculated.}
\end{deluxetable*}

\subsection{Shape and location of the instability strip}
\label{sec:shapeinstabilitystrip}

\begin{figure*}[t]
\centering
\includegraphics[width=0.95\textwidth]{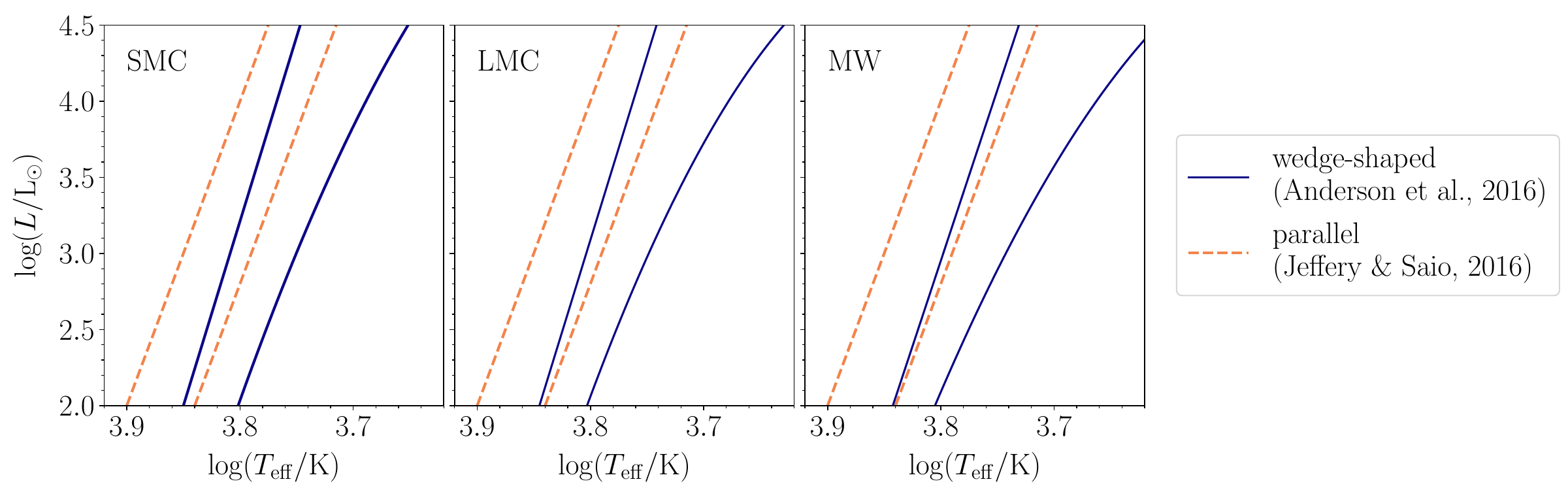}
\caption{Loci and shapes of instability strips for different metallicities. Note that the parallel instability strip has the same shape and location, independently of the metallicity. 
\label{fig:is_z}}
\end{figure*}

All Cepheid binaries were extracted from synthetic populations if they met a basic criterion: an evolved component (either primary or secondary, or both) had an effective temperature and luminosity that placed it inside the IS. We adopted two variants of the IS, shown in Figure \ref{fig:is_z}: (i) simplistic, metallicity-independent parallel IS from \citet[][their Figure 1]{jeffery16}; (ii) metallicity-dependent and wedge-shaped IS from \citet{anderson16}. Although we did not set explicit upper and lower luminosity limits, the upper and lower mass limits imposed limits on luminosity via the mass--luminosity relation (see Section \ref{subsec:MLR}). As a result, the luminosities of the Cepheid components are $2 < \log(L/\lsun) < 4.5$.

The shape and location of the parallel IS is historically driven. The pioneering numerical investigations of the Cepheid IS explored only the blue edge, because of the insufficient computational power and knowledge about the convective processes inside Cepheid variables that impact the location of the red edge \citep[e.g.][]{baker62}. As a result, the red edge was estimated by assuming an ad hoc efficiency of convective transport or by shifting the blue edge redward by a fixed temperature, and thus was parallel to the blue edge. The wedge-like shape of the IS was first reported by \citet{fernie90} based on empirical data of about 100 classical Cepheids, and later supported by a number of numerical studies, which included the effect of convection \citep[e.g.][]{bono00pulsmodels,anderson16}.

In this work, the choice of the IS plays an important role in determining whether a star of a given effective temperature is classified as a Cepheid. For instance, a MW star having $T_\mathrm{eff}=3.8$\,K and $\log(L/\lsun)=3.5$ will be found inside the parallel IS but outside the Anderson IS. The location of Cepheids on the HRD also impacts their pulsation periods and brightnesses, and therefore affects the PLRs; this effect is detailed in Section \ref{sec:sanitycheck}.


\subsection{Star Formation History}
\label{sec:SFH}

The SFH is crucial to determine how many stars are observable as Cepheids \emph{now}, given that they were born at a specific time in the past and have evolved to the point of crossing the IS. In this sense, the distribution of Cepheids' birth time stamps is a function of lookback time, with the current time (now) being 0. For instance, if a star that becomes a Cepheid at the age of 200\,Myr is assigned a birth time stamp of 100\,Myr, that means that at the time of observations (now) the star is only 100\,Myr old and has not become a Cepheid yet. Consequently, such a star is observed as a nonpulsator. On the other hand, if the birth time stamp assigned to the same star was 200\,Myr, this star would be observed as a pulsator, because it would cross the IS at exactly 200\,Myr old.  

Within the frame of population synthesis, we assigned to every binary a birth time stamp drawn from one of two variants of SFH: (i) uniform; (ii) based on Cepheids' ages. The uniform SFH means that all birth time stamps are equally probable for all metallicity environments\footnote{%
Although popular, the uniform SFH has been recently challenged by e.g. \citet{olejak20} who presented that synthetic SFH in the MW depends not only on metallicity but also on Galactic component (disk, bulge, and halo). Their SFH at lookback time $0-5$\,Gyr, i.e. the age span of classical Cepheids, remains uniform.%
}. The SFH based on Cepheids' ages was calculated from a period-age relation of \citet{bono05} using the pulsation periods of fundamental-mode Cepheids observed in the MW \citep{skowron19} and the Magellanic Clouds \citep{soszynski15ccep}. As a result of this calculation, three age distributions for the SMC, LMC, and MW were created, with age bins of width of 10\,Myr. From these distributions, we drew time stamps of birth with the probabilities related to the heights of the bins. If the time stamp of birth (and therefore age) of a system was between the time of entering and exiting the IS for either binary component, then such a system was marked as a Cepheid binary and added to the final sample. Cepheid's location in the IS (closer to the edge or closer to the middle) and all related stellar parameters (i.a. $T_\mathrm{eff}$, $L$, $R$, $P_\mathrm{pul}$) were linearly interpolated on the basis of its age relative to its time of IS entrance and exit. For instance, if a star's age is close to the time of IS entrance/exit, its location on the HRD is closer to the IS edge, and if a star's age is closer to the mean of IS entrance and exit times, it resides in the middle of the IS. This selection of stars based on their birth time stamps was repeated until the final sample reached 10,000 systems. 

Our samples are much larger than the real samples of Cepheids; the Magellanic Clouds have approximately 5000 Cepheids each with completeness of virtually 100\% \citep{soszynski15ccep}, while 3352 MW Cepheids reported so far constitute a somewhat incomplete sample \citep[completeness of about 88\% down to a magnitude $G=18$,][]{pietrukowicz21}. By keeping our synthetic samples this large, we allow binary Cepheids with more exotic and less probable companions to occur, and by keeping them equal in size, we assure that statistical errors affecting the sample size remain the same.

Our attempt to create a realistic mixture of Cepheids of various ages from the period-age relation was only partially successful, meaning that our synthetic Cepheids in the SMC, LMC, and MW are no older than 200, 170, and 100\,Myr, respectively, while the Cepheids' ages calculated from the period-age relation \citep{bono05} are as old as 350, 260, and 200\,Myr for the SMC, LMC and MW, respectively. One of the reasons might be that we used the period-age relations of \citet{bono05} who assumed nonrotating progenitors on the MS. On the contrary, rotating progenitors can experience enhanced internal mixing and therefore spend, on the MS, even twice as long as their nonrotating counterparts before they evolve into Cepheids \citep{anderson16}. Consequently, Cepheids evolved from rotating progenitors are older. For the consistency sake, we calculated ages for our populations, which consist of nonrotating stars, using formulas of \citet{bono05}. The other reason might be the fact that the theoretical models of stellar evolution fail to render extensive blue loops for low-mass (and therefore older) stars, and as a result, the sample consists of young Cepheids (either massive IS2+3 crossers or IS1 crossers). We provide more comment on this caveat in Section \ref{sec:proportions}. Nevertheless, we proceed with our study bearing in mind that our results are relevant only for young and massive Cepheids and should be interpreted with caution in the context of older and low-mass ones.

The end result of all the above computations is 16 variants of synthetic populations of binary Cepheids for each of the three metallicity environments (SMC, LMC, MW), which were created as permutations of four variants of initial parameters distributions, two variants of the IS, and two variants of the SFH ($4 \times 2 \times 2 = 16$).

\subsection{Magnitudes, amplitudes, and pulsation periods}
\label{subsec:magper}

In order to calculate multiband photometry, we used the online YBC database\footnote{\url{http://stev.oapd.inaf.it/YBC/}, accessed January 22, 2022} of stellar bolometric corrections \citep{chen19}. We chose ATLAS9 model atmospheres of \citet{castelli03} and derived UBVRIJHK magnitudes in the Bessell \& Brett photometric system \citep{bessell90,bessell98}. We also created a reddening-free quasi-magnitude Wesenheit index, following the formula from \citet{udalski99}:
\begin{equation}
\label{eq:wesenheit}
    W_{VI} = I - 1.55(V-I)\,.
\end{equation}

$V$-band maximum peak-to-peak amplitudes were estimated based on the $\log P$-$A$ plot of \citet[][their Figure 1]{klagyivik09}. We traced the envelope of highest $V$-band amplitudes as a function of $\log P$, including the dip at $\log(P/\mathrm{d}) \approx 1$. Next, we used theoretical predictions of \citet{bhardwaj17} for amplitudes in $U$, $B$, $V$, $I$, $J$ and $K$ bands in order to calculate ratios of amplitudes in these bands relative to the $V$-band averaged amplitude. The ratio $A_H/A_V$ was interpolated linearly between $A_J/A_V$ and $A_K/A_V$. The results of these calculations are as follows:
$A_U/A_V = 1.94$, 
$A_B/A_V = 1.44$, 
$A_I/A_V = 0.65$, 
$A_J/A_V = 0.42$, 
$A_H/A_V = 0.36$, 
$A_K/A_V = 0.30$.
As the last step, we multiplied the envelope of highest $V$-band amplitudes as a function of $\log P$ by the above constants. The amplitudes in the Wesenheit index were created from $V$, $I$ magnitudes and amplitudes, following Eq. \ref{eq:wesenheit}. Our estimated maximum amplitudes do not depend on metallicity.

Because \ST\ was not tailored to check whether a star is dynamically unstable and prone to pulsations, we assumed that all stars found in the IS pulsate as fundamental-mode Cepheids and calculated the pulsation periods using two external and independent sets of formulas. The first one was taken from \citet{bono00pulsmodels}:
\begin{eqnarray}
\label{eq:logP02_bono}
\log P &=& 9.874 - 3.108 \log T - 0.767 \log M + 0.942 \log L \nonumber\\
&&\mbox{for } Z = 0.02\\
\label{eq:logP008_bono}
\log P &=& 10.557 - 3.279 \log T -0.795 \log M + 0.931 \log L \nonumber\\ 
&&\mbox{for } Z = 0.008\\
\label{eq:logP004_bono}
\log P &=& 10.971 - 3.387 \log T - 0.813 \log M + 0.929 \log L \nonumber\\ 
&&\mbox{for } Z = 0.004
\end{eqnarray}

The other set of formulas for fundamental periods was created for the purpose of this study using Warsaw Pulsational Code \citep{smolec08} with the homogeneous envelope and convective parameters $\alpha=1.5$, $\alpha_m=0.5$, $\alpha_s=1.0$, $\alpha_c=1.0$, $\alpha_d=1.0$, $\alpha_p=0.0$, $\alpha_t=0.0$, $\gamma_r=1$. For each of the three metallicities [Fe/H] = 0.0,\,--0.5,\,--1.0\,dex, we constructed a grid of stellar models of various effective temperatures and luminosities, with mass-luminosity (ML) relations fitted to the synthetic data presented in Figure \ref{fig:is_ml_lit} and detailed in Section \ref{subsec:MLR}. This fit yielded two ML relations (for IS1 and IS2+3 Cepheids) for each of the three metallicities:

\begin{eqnarray}
    \mathrm{for~[Fe/H]}&=&-1.0\,\mathrm{dex} \nonumber\\
    \label{eq:ML_SMC_IS1}
    \log L &=& 0.605 +3.640 \log M \mbox{~~~for IS1} \\
    \label{eq:ML_SMC_IS23}
    \log L &=& 0.887 +3.729 \log M \mbox{~~~for IS2+3} \\[5pt]
    \mathrm{for~[Fe/H]}&=&-0.5\,\mathrm{dex} \nonumber\\
    \label{eq:ML_LMC_IS1}
    \log L &=& 0.512 +3.666 \log M \mbox{~~~for IS1} \\
    \label{eq:ML_LMC_IS23}
    \log L &=& 0.703 +3.855 \log M \mbox{~~~for IS2+3} \\[5pt]
    \mathrm{for~[Fe/H]}&=&0.0\,\mathrm{dex} \nonumber\\
    \label{eq:ML_MW_IS1}
    \log L &=& 0.418 +3.691 \log M \mbox{~~~for IS1} \\
    \label{eq:ML_MW_IS23}
    \log L &=& 0.533 +3.882 \log M \mbox{~~~for IS2+3}
\end{eqnarray}
From the grids, we extracted models on the blue and red edge of the IS, and using the information about their fundamental-mode pulsation periods, we created the period-luminosity-temperature relations, separately for the IS1 Cepheids:

\begin{eqnarray}
    \label{eq:logP02_IS1_smolec}
    \log P &=& 11.485 -3.450 \log T + 0.665 \log L \nonumber \\
    &&\mbox{for } \mathrm{[Fe/H]}=0.0\,\mathrm{dex} \\
    \label{eq:logP008_IS1_smolec}
    \log P &=& 11.833 -3.537 \log T + 0.660 \log L \nonumber \\
    &&\mbox{for } \mathrm{[Fe/H]}=-0.5\,\mathrm{dex} \\
    \label{eq:logP004_IS1_smolec}
    \log P &=& 12.188 -3.627 \log T + 0.658 \log L\nonumber\\
    &&\mbox{for } \mathrm{[Fe/H]}=-1.0\,\mathrm{dex}
\end{eqnarray}
The following relations are for the IS2+3 Cepheids:

\begin{eqnarray}
    \label{eq:logP02_IS23_smolec}
    \log P &=& 11.292 -3.405 \log T + 0.688 \log L \nonumber \\
    &&\mbox{for } \mathrm{[Fe/H]}=0.0\,\mathrm{dex} \\
    \label{eq:logP008_IS23_smolec}
    \log P &=& 11.604 -3.482 \log T + 0.687 \log L \nonumber \\
    &&\mbox{for } \mathrm{[Fe/H]}=-0.5\,\mathrm{dex} \\
    \label{eq:logP004_IS23_smolec}
    \log P &=& 12.616 -3.741 \log T + 0.679 \log L\nonumber\\
    &&\mbox{for } \mathrm{[Fe/H]}=-1.0\,\mathrm{dex}
\end{eqnarray}

Bono's and our prescriptions for fundamental periods yield similar results, which agree best for short-period Cepheids and diverge for long-period ones ($P_\mathrm{pul} \geqslant 40$\,d) with Bono's periods being systematically longer by $4-6$\,d. Such long-period Cepheids are however rare and their contribution to the sample is minuscule. 

\section{Sanity check}
\label{sec:sanitycheck}

We performed three sanity checks in order to evaluate the agreement of selected parameters of our synthetic populations with the observed and literature data. These tests were executed on single Cepheids only, so that the reliability of our sample is endorsed before we introduce the next level of complexity to the analysis, i.e. Cepheids' companions.

\subsection{Mass-luminosity relation}
\label{subsec:MLR}

\begin{figure*}
\centering
\includegraphics[width=0.9\textwidth]{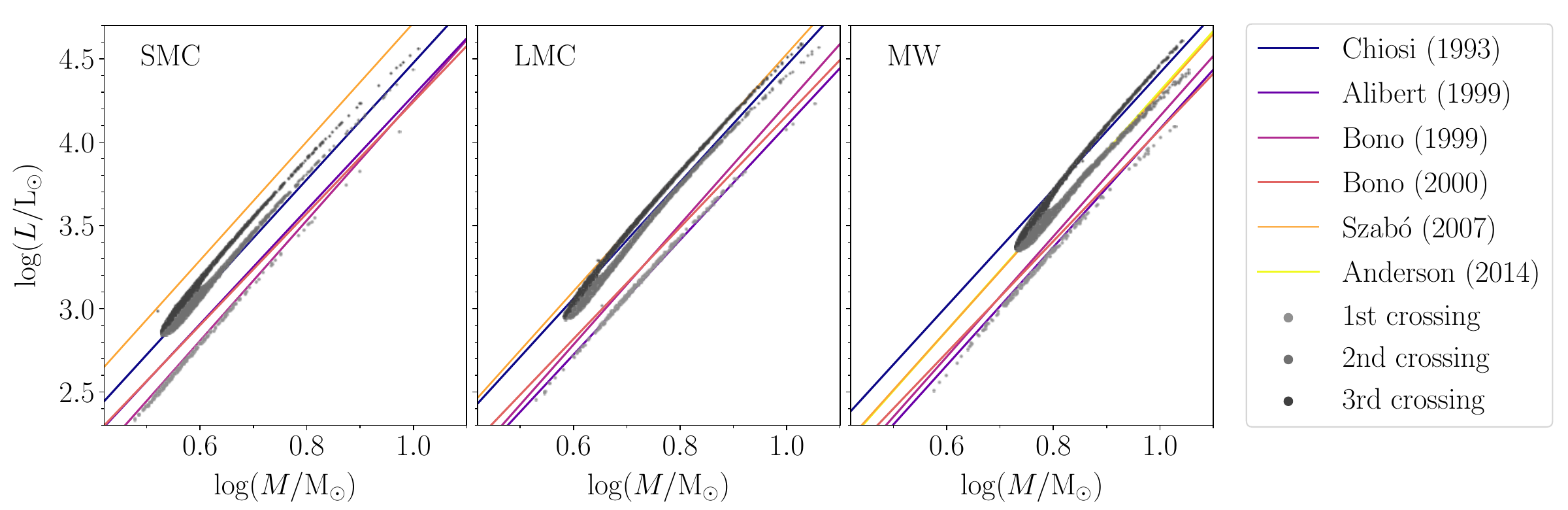}
\caption{Literature mass-luminosity relations, with synthetic populations of single Cepheids on their 1st, 2nd and 3rd instability crossing overplotted. All literature relations refer to IS2+3 Cepheids but differ in the treatment of mixing mechanisms (rotation, overshooting).
\label{fig:is_ml_lit}}
\end{figure*}

Mass-luminosity (ML) relation for Cepheids, presented in a form $\log L = \alpha \log M + \beta$, depends on i.a. the metallicity, helium content, rotation, and/or overshooting on the MS, as well as the mass-loss rate \citep{chiosi93,alibert99,bono99,bono00metalcontent,szabo07,anderson14}. In general, high rotation rate, large overshooting, and low metallicity cause the parameter $\beta$ to increase, making such Cepheids more luminous by as much as $\log L = 0.25$. In Figure \ref{fig:is_ml_lit} the ML relation for our exemplary synthetic population (set D, Anderson prescription for the IS, the SFH based on Cepheids' ages) is compared with ML relations from the literature, showing fair agreement. Noticeably, synthetic Cepheids in IS1 and IS2+3 obey different ML relations, in a sense that IS2+3 Cepheids are systematically brighter. These different ML relations have been taken into account while calculating pulsation periods using Warsaw Pulsational Code, as described in Section \ref{subsec:magper}, and have resulted in slightly different PLRs for IS1 and IS2+3 Cepheids.


\subsection{Proportions of Cepheids in the first, second, and third crossing}
\label{sec:proportions}

\begin{figure}
    \centering
    \includegraphics[width=\columnwidth]{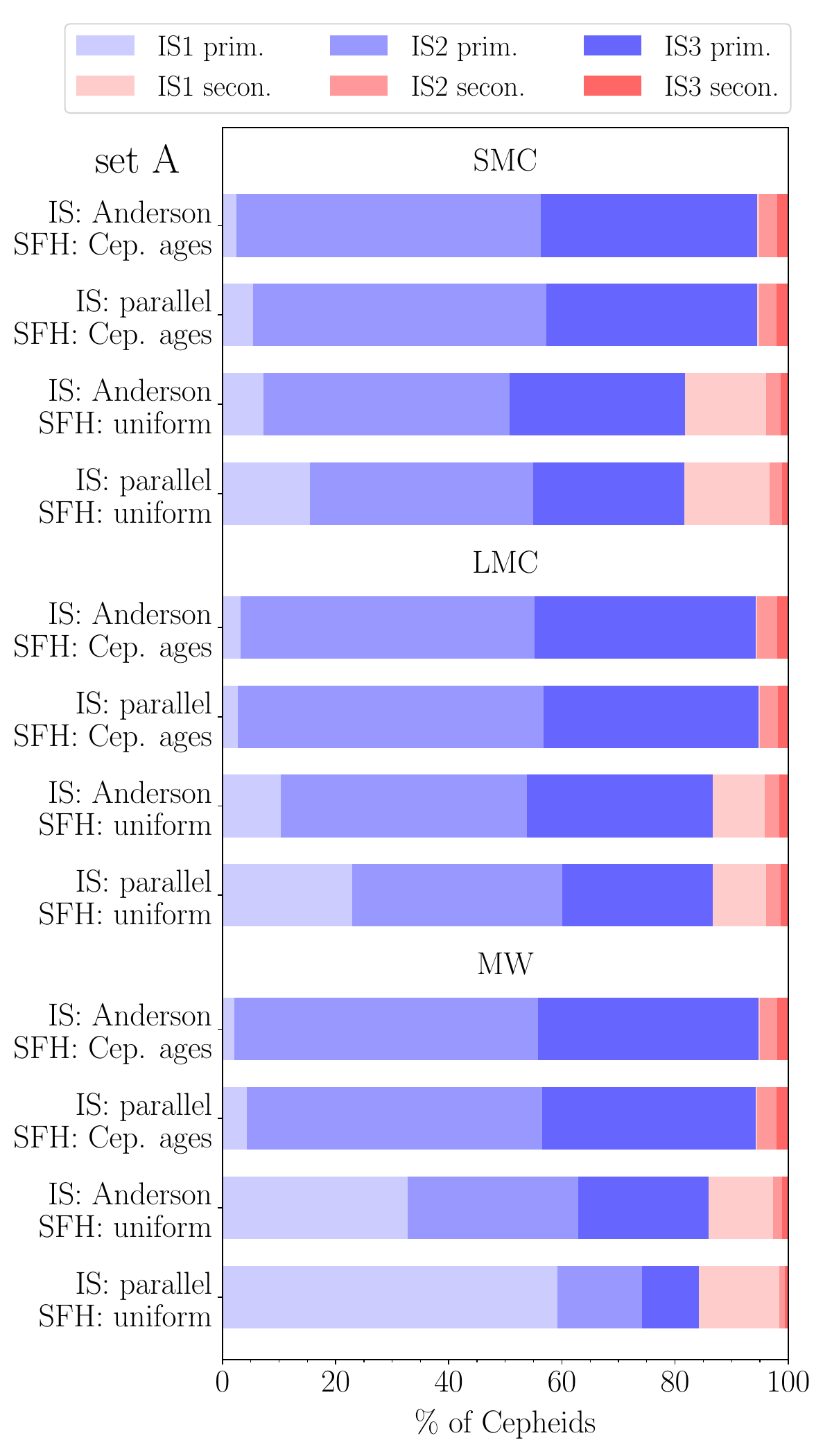}
    \caption{Proportions of Cepheids (as primaries in blue and secondaries in red) on their first, second, and third crossing through the instability strip. Four different variants represent all combinations of two SFHs (uniform and based on Cepheids' ages) and two IS shapes (parallel and Anderson). The distribution of initial parameters has no impact on proportions of Cepheids.
    \label{fig:CepProportions}}
\end{figure}

Since different SFHs favor different time stamps of birth, the choice of the SFH impacts the number of Cepheids on different IS crossings. The shape of the IS also impacts the number of Cepheids on different IS crossings, because it affects the time that stars spend inside the IS, but this effect is much smaller. Finally, the metallicity correlates with the sizes of blue loops and, consequently, the proportions of Cepheids on different IS crossings. The combination of all three factors results in different percentages of IS1, IS2 and IS3 Cepheids, shown in Figure \ref{fig:CepProportions}. Not only primaries (blue areas in the plot) but also secondaries (red areas) can become Cepheids, although this scenario is much rarer. In such cases, the companions to secondary Cepheids are more-evolved asymptotic giant branch (AGB) stars or compact objects: white dwarfs (WD) or neutron stars (NS), and they are described in more detail in Section \ref{subsec:compan}.

Cepheids on their first crossing, i.e. traversing the Hertzsprung Gap, are short-lived and therefore expected to be extremely scarce in the observational data. In our synthetic populations, the percentage of the IS1 Cepheids can vary from negligible to prevalent, depending on the choice of SFH, IS, and metallicity. In some variants (e.g. a MW population generated from parallel IS and uniform SFH), IS1 Cepheids---despite being short-lived---are much more abundant than IS2+3 Cepheids, and therefore could be observed more frequently. 

However, this phenomenon arises from the fact that the theoretical models of stellar evolution for metal-rich stars of masses below $\sim 3.5\,\msun$ fail to render extensive blue loops. Consequently, such stars cross the IS only once in the Hertzsprung Gap, which actually creates a deficiency of IS2+3 crossers. This issue has been approached by many authors \citep[e.g.][and references therein]{xu04}, but remains unresolved despite the evidence for the existence of short-period IS2+3 Cepheids \citep{turner06,rodgiguez-segovia21}.

In the majority of variants, the percentage of IS2 Cepheids is $40-60$\% (except for the MW Cepheids generated from the uniform SFH, where IS1 Cepheids dominate the samples, and thus the IS2 Cepheids are only $15-30$\%). In general, variants with the parallel IS produce fewer IS2 Cepheids than variants with Anderson IS, meaning that slightly more IS2 Cepheids constitute the samples if the red edge of the IS is shifted toward lower temperatures; this observation can again be explained by the aforementioned \emph{blue loop} issue of theoretical models.

First-, second-, and third-crossers can be distinguished based on their rates of period changes (negative for IS2, positive for IS1 and IS3, and larger for IS1 than for IS3), because their pulsation periods decrease as they cross the IS toward the blue edge, and increase as they cross the IS toward the red edge. \citet{turner06} measured rates of period change for 200 MW Cepheids and reported that only 33\% were negative (belonged to IS2 Cepheids). \citet{poleski08} estimated that only 15\% from 655 analyzed LMC Cepheids show consistent period change, and from those $\sim 57$\% have negative period changes. Theoretical estimations of the percentage of IS2 Cepheids with the metallicity of $Z=0.02$ yielded only $10-15$\% \citep{neilson12,miller20}, and increased to $40-45$\% only after a significant initial rotation was introduced \citep{miller20}. Recently the most comprehensive study of pulsation period change of LMC Cepheids \citep{rodgiguez-segovia21} shows that among 1303 objects 43\% are IS2 crossers, 53\% are IS3 crossers, and the remaining 4\% are objects with inconclusive period changes, and two candidates for IS1 crossers. Our results agree with \citet{poleski08} and \citet{rodgiguez-segovia21} on the percentage of IS2 crossers, but we find IS3 crossers underrepresented, and IS1 crossers overrepresented with respect to the results of \citet{rodgiguez-segovia21}.

\subsection{Multiband Period--Luminosity relations}

\begin{figure}
\centering
\includegraphics[width=0.99\columnwidth]{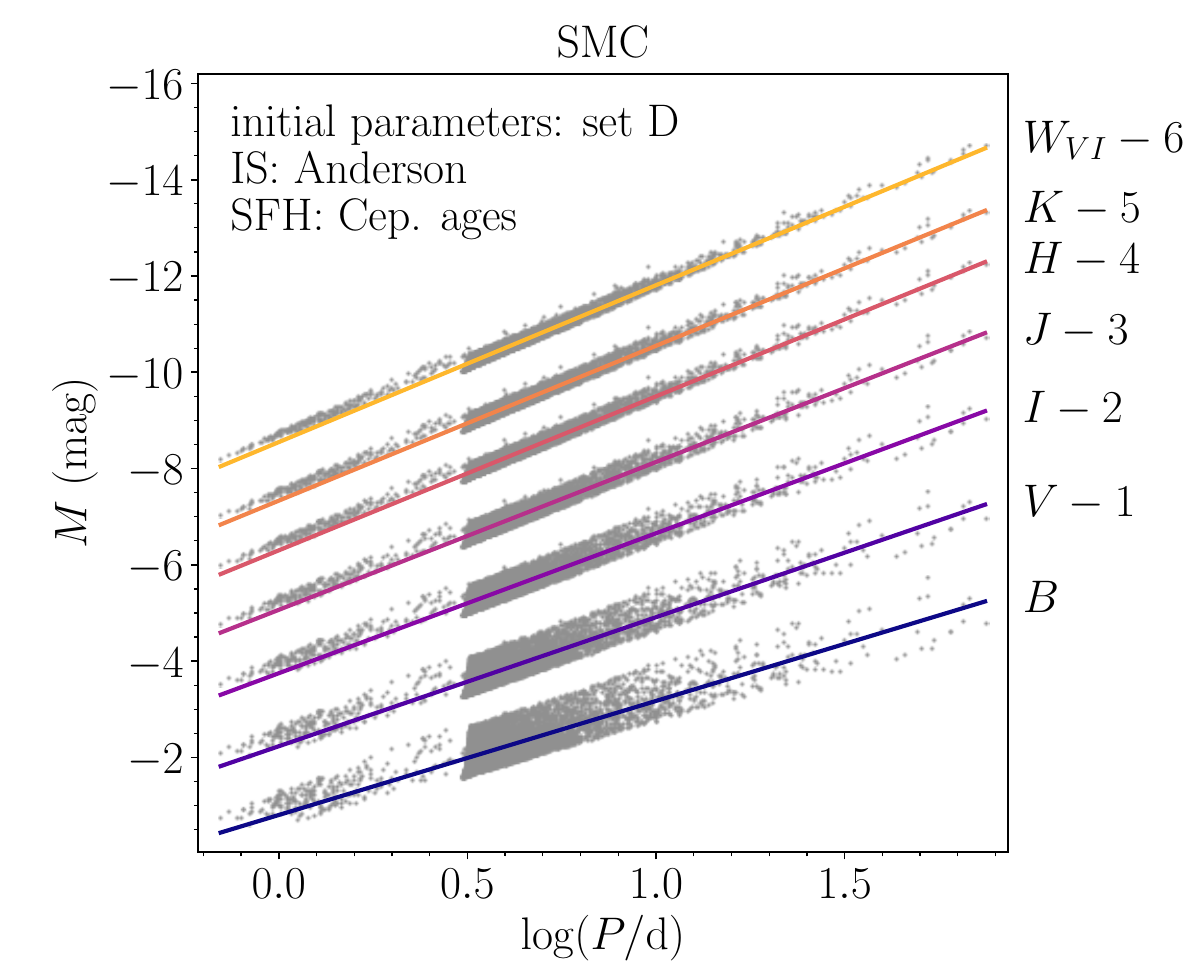}\\
\includegraphics[width=0.99\columnwidth]{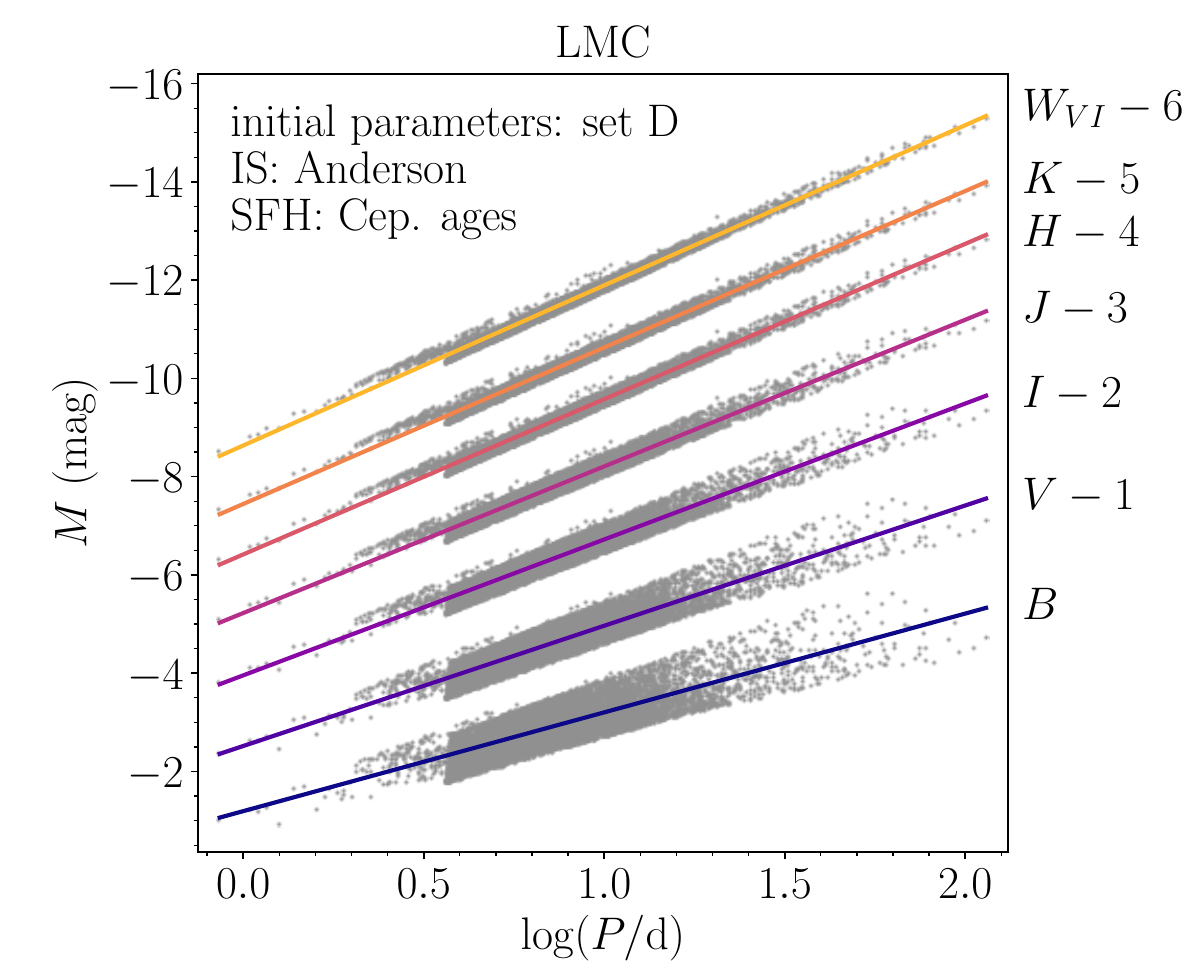}\\
\includegraphics[width=0.99\columnwidth]{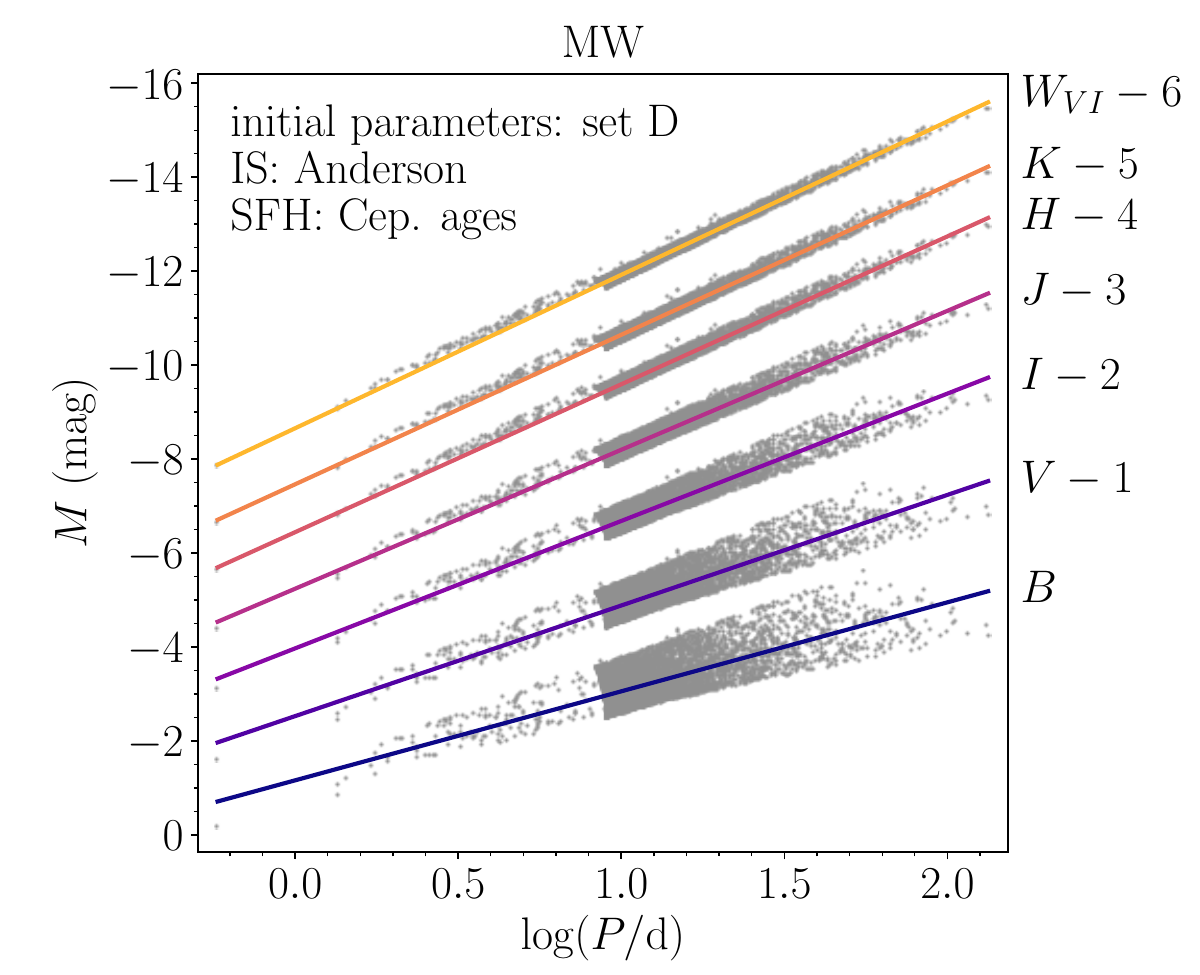}
\caption{Example of period-luminosity relations for the SMC, LMC, and MW. All passbands except $B$ are shifted for clarity by as many magnitudes as indicated on the right side of the images. Solid lines represent linear least-squares fits.
\label{fig:plr_examples}}
\end{figure}

Having calculated pulsation periods and magnitudes, we created multiband PLRs for all variants of our synthetic populations. Figure \ref{fig:plr_examples} illustrates the results for the variant with initial parameters from set D, Anderson IS, SFH based on Cepheids' ages, and our prescription for the pulsation periods. For a clearer comparison between the three metallicities, the absolute magnitudes are provided instead of observed ones.

The scatter of PLRs reflects the fact that Cepheids populate the entire width of the instability strip. \citet{madore12} determined theoretical scatters of multiband PLRs for LMC Cepheids (in mag): 0.36 ($B$), 0.27 ($V$), 0.18 ($I$), 0.14 ($J$), 0.12 ($H$), 0.11 ($K$). They agree with the observed scatters recently reported by \citet{breuval21}: 0.23 ($V$), 0.15 ($I$), 0.12 ($J$), 0.11 ($H$), 0.10 ($K$), 0.08 ($W_{VI}$). The scatter of PLRs of our synthetic populations, calculated as $1\sigma$ standard deviation from the linear least-squares fit, agrees well with both theoretical and observed values. However, it tends to be smaller in variants of synthetic populations with the parallel IS%
: 0.23 ($B$), 0.18 ($V$), 0.13 ($I$), 0.10 ($J$), 0.07 ($H$), 0.07 ($K$), 0.05 ($W_{VI}$). It tends to be larger in variants of synthetic populations with Anderson's wedge-like IS: 0.29 ($B$), 0.21 ($V$), 0.16 ($I$), 0.12 ($J$), 0.09 ($H$), 0.08 ($K$), 0.07 ($W_{VI}$). We also notice that, in variants with parallel IS, the scatter remains constant for all values of $\log (P/\mathrm{d})$, but it grows with larger $\log (P/\mathrm{d})$ in variants with Anderson IS; this effect is especially visible in the $B$ band, where the scatter is the largest in general. The varying scatter as a function of $\log (P/\mathrm{d})$ reflects the wedge-like shape of the Anderson IS. Empirical PLRs for short wavelengths \citep[e.g.][]{musella97,bhardwaj16} do not show larger scatter for larger $\log (P/\mathrm{d})$, which either supports the parallel variant of the IS over Anderson's wedge-like variant or suggests that not enough Cepheids have been observed to populate the PLR on the long-period end.

Synthetic Cepheids cluster at $\log(P/\mathrm{d}) \approx 0.5,\,0.6,\,1.0$ for the SMC, LMC, and MW, respectively, which is the minimum pulsation period for IS2+3 Cepheids, while all stars with shorter periods are IS1 Cepheids. Observational data support the existence of IS2+3 Cepheids with shorter periods \citep{turner06,poleski08}, but theoretical models fail to render extensive blue loops for such stars \citep[see ][and discussion in Section \ref{sec:proportions} of this paper]{xu04}.

IS1 Cepheids show a slightly steeper slope for their PLR than IS2+3 Cepheids, which was expected since we used different period formulas for IS1 (Eqs. \ref{eq:logP02_IS1_smolec}--\ref{eq:logP004_IS1_smolec}) and IS2+3 Cepheids (Eqs. \ref{eq:logP02_IS23_smolec}--\ref{eq:logP004_IS23_smolec}). This difference in slopes is best visible in the NIR passbands and $W_{VI}$, and could be potentially used to distinguish between IS1 and IS2+3 Cepheids and reveal the number of IS1 Cepheids in the population, provided a large sample with accurate magnitudes.

\begin{deluxetable*}{lccccccc}
\tablecaption{Selected-literature Slope Coefficients of the Period–Luminosity Relation, $M = \alpha \log(P) + \beta$, for Different Passbands and Metallicities.
\label{tab:plr_alpha}}
\tablewidth{0pt}
\tablehead{
\colhead{Reference} & \colhead{$B$} & \colhead{$V$} & \colhead{$I$} & \colhead{$J$} & \colhead{$H$} & \colhead{$K$} & \colhead{$W_{VI}$}
}
\startdata
\multicolumn{8}{c}{SMC}\\
\hline
\citet{udalski99} & $-2.207$ & $-2.572$ & $-2.857$ & \nodata & \nodata & \nodata & $-3.303$ \\
\citet{groenewegen00} & \nodata & \nodata & \nodata & $-3.037$ & $-3.160$ & $-3.212$ & $-3.328$ \\
\citet{storm04} & \nodata & $-2.590$ & $-2.865$ & \nodata& \nodata & \nodata & $-3.283$ \\
\citet{sandage09} & $-2.222$ & $-2.588$ & $-2.862$ & \nodata & \nodata & \nodata & \nodata \\
\citet{ripepi17}\tablenotemark{a} & \nodata & \nodata & \nodata & $-3.070$ & \nodata & $-3.513$ & \nodata \\
\citet{wielgorski17} & \nodata & $-2.644$ & $-2.947$ & $-3.087$ & $-3.184$ & $-3.206$ & $-3.330$ \\
\citet{gieren18} & \nodata & $-2.705$ & $-2.934$ & $-2.856$ & \nodata & $-3.179$ & $-3.287$ \\
\citet{breuval21} & \nodata & $-2.594$ & $-2.871$ & $-2.956$ & \nodata & $-3.163$ & $-3.334$ \\
\hline
\multicolumn{8}{c}{LMC}\\
\hline
\citet{madore91} & $-2.53$ & $-2.88$ & $-3.14$ & $-3.31$ & $-3.37$ & $-3.42$ & \nodata \\
\citet{gieren98} & \nodata & \nodata & \nodata & $-3.129$ & $-3.249$ & $-3.267$  & \nodata \\
\citet{udalski99} & \nodata & $-2.760$ & $-2.962$ & \nodata& \nodata & \nodata & $-3.277$ \\
\citet{groenewegen00} & \nodata & \nodata & \nodata & $-3.144$ & $-3.236$ & $-3.246$ & $-3.337$ \\
\citet{sandage04} & $-2.340$ & $-2.702$ & $-2.949$ & \nodata & \nodata & \nodata & \nodata \\
\citet{persson04} & \nodata & \nodata & \nodata & $-3.153$ & $-3.234$ & $-3.281$ & \nodata \\
\citet{fiorentino07} & $-2.44$ & $-2.78$ & $-2.98$ & $-3.15$ & \nodata & $-3.26$ & $-3.29$ \\
\citet{macri15} & \nodata & \nodata & \nodata & $-3.156$ & $-3.187$ & $-3.247$ & \nodata \\
\citet{wielgorski17} & \nodata & $-2.779$ & $-2.977$ & $-3.118$ & $-3.224$ & $-3.247$ & $-3.332$ \\
\citet{gieren18} & \nodata & $-2.775$ & $-3.021$ & \nodata & $-3.220$ & $-3.282$ & $-3.411$ \\
\citet{breuval21} & \nodata & $-2.704$ & $-2.916$ & $-3.127$ & $-3.160$ & $-3.217$ & $-3.281$ \\
\citet{ripepi22} & \nodata & \nodata & \nodata & $-3.084$ & \nodata & $-3.230$ & \nodata \\
\hline
\multicolumn{8}{c}{MW}\\
\hline
\citet{caldwell91} & \nodata & $-2.81$ & \nodata & \nodata & \nodata & \nodata & \nodata \\
\citet{gieren93} & \nodata & $-2.986$ & \nodata & \nodata & \nodata & \nodata & \nodata \\
\citet{laney94} & \nodata & $-2.874$ & \nodata & $-3.306$ & $-3.421$ & $-3.443$ & \nodata \\
\citet{gieren98} & \nodata & $-2.77$ & $-3.04$ & \nodata & \nodata & \nodata & \nodata \\
\citet{freedman01} & \nodata & $-2.760$ & $-2.962$ & \nodata & \nodata & \nodata & $-3.26$ \\
\citet{tammann03} & $-2.757$ & $-3.141$ & $-3.408$ & \nodata & \nodata & \nodata & \nodata \\
\citet{sandage04} & $-2.692$ & $-3.087$ & $-3.348$ & \nodata & \nodata & \nodata & \nodata \\
\citet{storm04} & $-2.74$ & $-3.08$ & $-3.30$ & $-3.53$ & $-3.63$ & $-3.67$ & $-3.63$ \\
\citet{benedict07} & \nodata & $-2.43$ & $-2.81$ & \nodata & \nodata & $-3.32$ & $-3.34$ \\
\citet{storm11} & $-2.13$ & $-2.67$ & $-2.81$ & $-3.18$ & $-3.30$ & $-3.33$ & $-3.26$ \\
\citet{gieren18} & \nodata & $-2.615$ & $-2.664$ & $-3.114$ & \nodata & $-3.258$ & $-3.084$ \\
\citet{breuval21} & \nodata & $-2.443$ & $-2.780$ & $-3.050$ & $-3.160$ & $-3.207$ & $-3.289$ \\
\enddata
\tablenotetext{a}{Slope values for a subset of fundamental-mode Cepheids with $\log(P/\mathrm{d}) \geq 0.47$}
\end{deluxetable*}

\begin{figure*}
    \includegraphics[width=\textwidth]{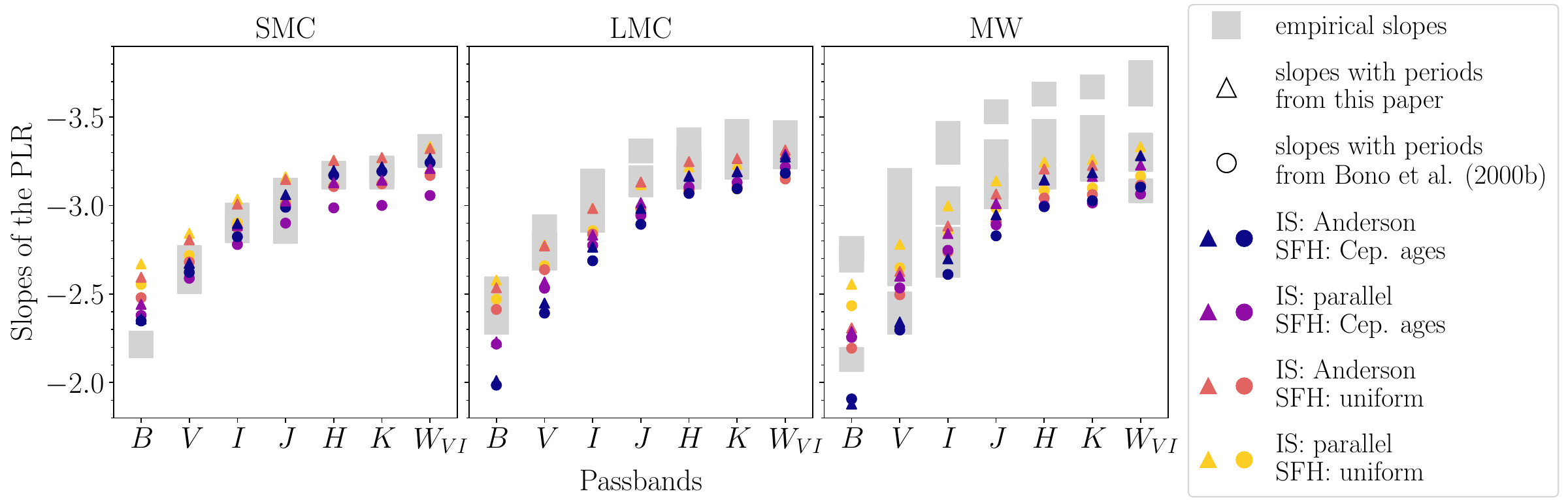}
    \caption{Comparison of empirical and synthetic slopes for single Cepheids of all IS crossings (IS1+2+3). Empirical slopes are presented in Table \ref{tab:plr_alpha}.}
    \label{fig:slopes}
\end{figure*}

Figure \ref{fig:slopes} presents the comparison between a selection of Cepheid PLR slope values taken from the literature (listed in Table \ref{tab:plr_alpha}) and our synthetic Cepheid populations, for the $B$, $V$, $I$, $J$, $H$, and $K$ bands, and the $W_{VI}$ Wesenheit index. The slopes of the synthetic Cepheid populations were calculated by the linear least-squares fit to joint samples of IS1+2+3 Cepheids, as seen in Figure \ref{fig:plr_examples}. We used two different prescriptions for calculating pulsation periods: Bono's (Eqs. \ref{eq:logP02_bono}-\ref{eq:logP004_bono}) and ours (Eqs. \ref{eq:logP02_IS1_smolec}-\ref{eq:logP004_IS23_smolec}), which result in two different sets of slopes, marked in Figure \ref{fig:slopes} as circles and triangles, respectively. Four variants of synthetic populations are marked with different colors. These are all combinations of two IS prescriptions (Anderson's and parallel) and two SFH formulas (uniform and based on the ages of Cepheids). We found that the initial parameters (sets A--D) have no impact on the slope for single Cepheids, and are therefore omitted.

Our slopes (triangles) tend to be slightly steeper than Bono's (circles), and appear to fit the values of empirical slopes better, especially for longer wavelengths. However, the large scatter of the slopes for both the observed and synthetic populations prevents us from excluding or approving any variant of the synthetic population.


Synthetic slopes for a given passband have different values in the three metallicity environments, but the differences get smaller with the longer wavelengths, similarly to the empirical slopes of PLRs reported by \cite{breuval21}. This result partially validates the assumption that the slopes are metallicity-independent and, in practice, can be therefore fixed for the distance determinations to Cepheids in farther galaxies \citep[e.g.][]{wielgorski17,gieren18}, as long as near-infrared PLRs are used.

The above sanity checks have shown a satisfactory agreement between our synthetic populations and the literature/empirical data of classical Cepheids, meaning that our samples resemble the observed populations reasonably well. In the next section, we introduce the companions and perform a statistical analysis of their properties.

\section{Results} 
\label{sec:results}

We present statistical properties of binary Cepheids: orbital periods, eccentricities, mass ratios, evolutionary stages, effective temperatures, and spectral types of the companions. The results have a qualitative character, and their purpose is to anticipate the characteristics of Cepheids' companions, which might help designing future observations to detect them. The results show similarities and differences between 12 variants of populations: four combinations of IS and SFH prescriptions for three metallicities (SMC, LMC, MW), similar to the results presented in Figure \ref{fig:CepProportions}. Every population contains a small fraction of binary systems with two Cepheids; their properties are discussed in Section \ref{subsec:2bincep}.

\subsection{Characteristics of binaries}
\begin{figure}
    \centering
    \includegraphics[width=1.0\columnwidth]{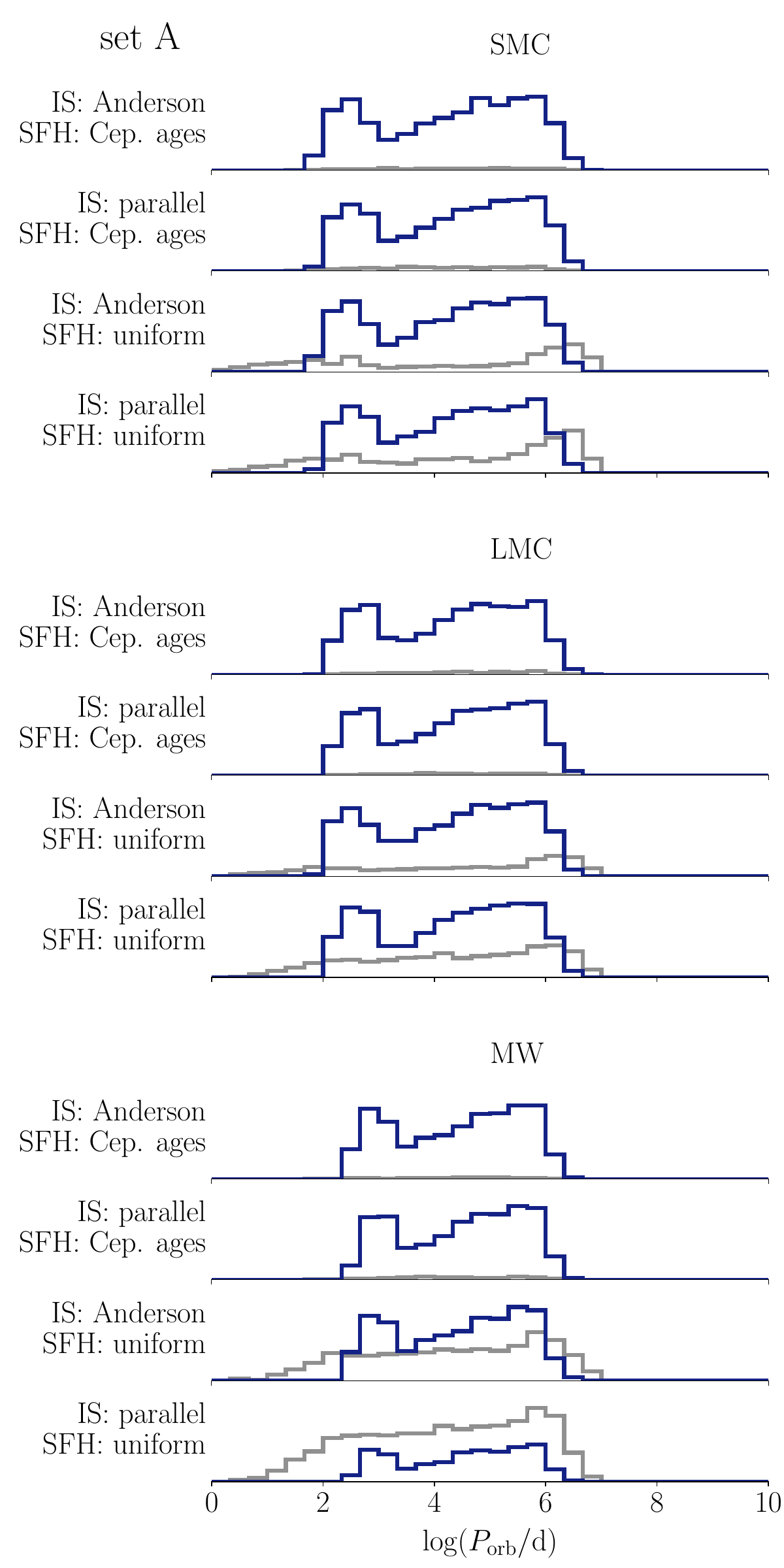}
    \caption{Distributions of orbital periods in 12 variants of synthetic populations, for set A of the initial parameters. IS2+3 Cepheids (navy blue) and IS1 Cepheids (gray) are presented separately. The values on all $y$-axes were scaled linearly from 0 to 1, and then omitted for a clearer comparison of the shapes of the distributions. Images for sets B, C, and D can be found in Appendix \ref{apdx:characteristics}.}
    \label{fig:porb}
\end{figure}

\begin{figure}
    \centering
    \includegraphics[width=1.0\columnwidth]{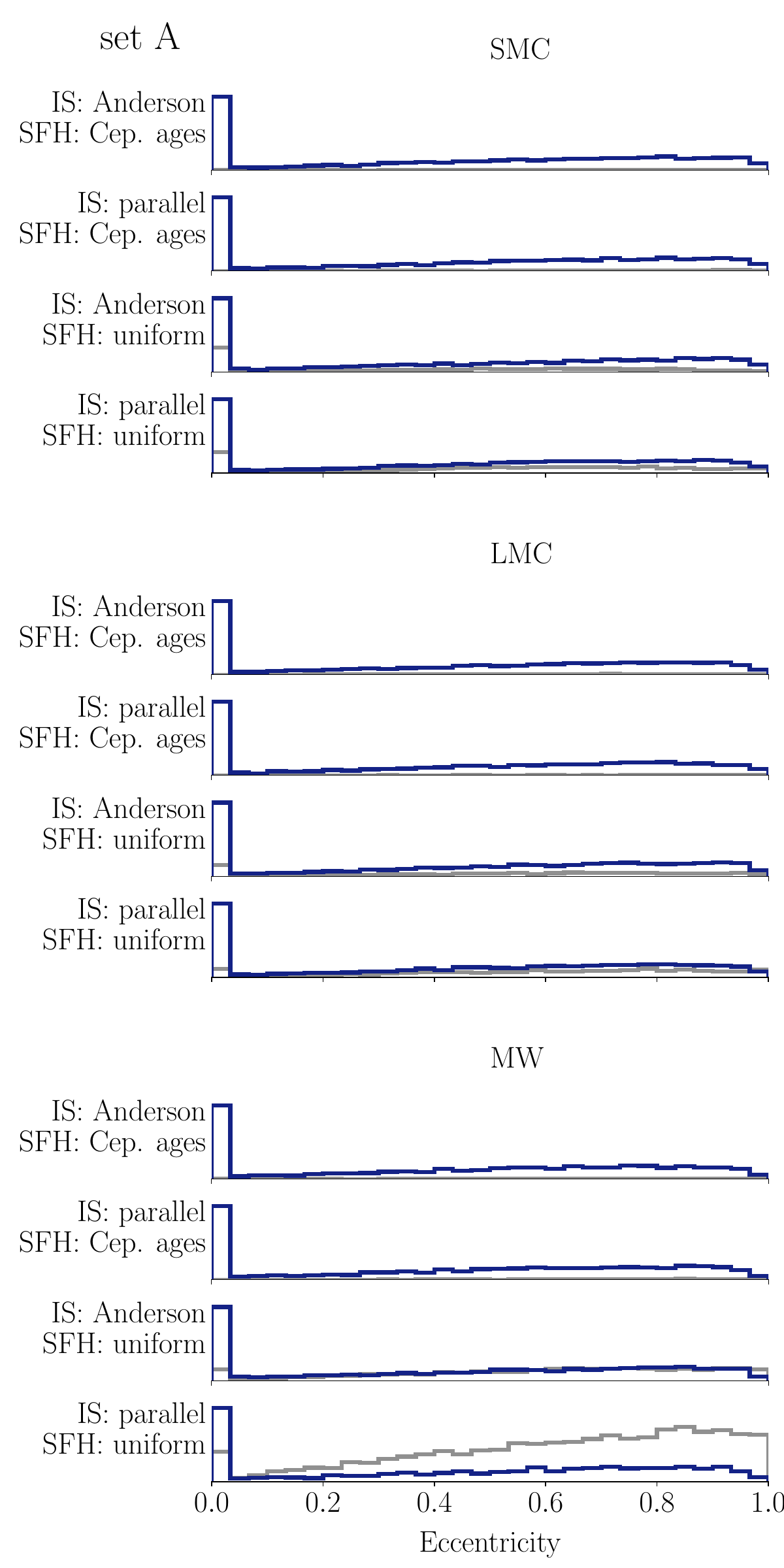}
    \caption{Eccentricities in 12 variants of synthetic populations, for set A of the initial parameters. IS2+3 Cepheids (navy blue) and IS1 Cepheids (gray) are presented separately. The values on all $y$-axes were scaled linearly from 0 to 1, and then omitted for a clearer comparison of the shapes of the distributions. Images for sets B, C, and D can be found in Appendix \ref{apdx:characteristics}.}
    \label{fig:ecc}
\end{figure}

\begin{figure}
    \centering
    \includegraphics[width=1.0\columnwidth]{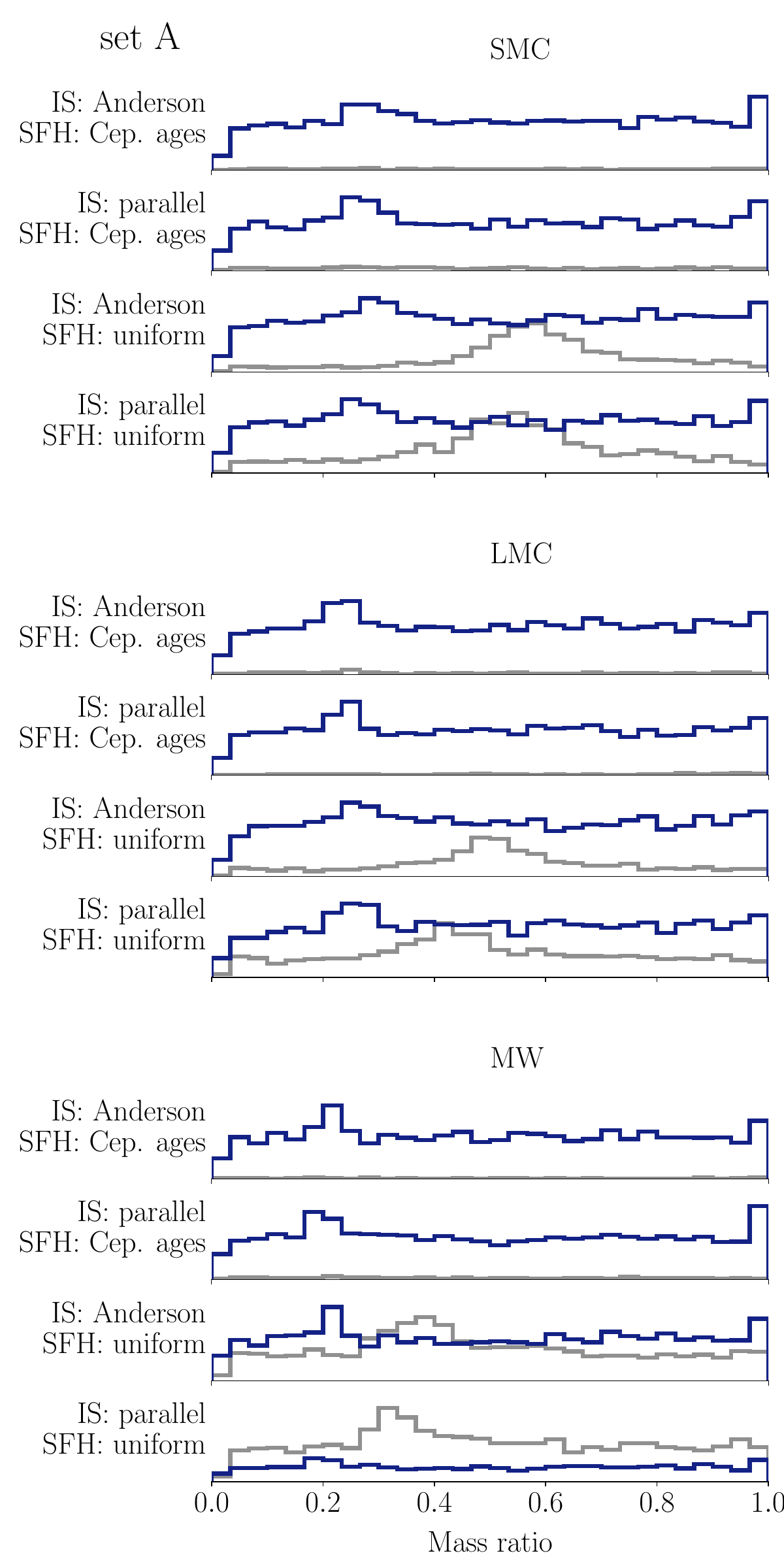}
    \caption{Mass ratios in 12 variants of synthetic populations, for set A of the initial parameters. IS2+3 Cepheids (navy blue) and IS1 Cepheids (gray) are presented separately. The values on all $y$-axes were scaled linearly from 0 to 1, and then omitted for a clearer comparison of the shapes of the distributions. Images for sets B, C, and D can be found in Appendix \ref{apdx:characteristics}.}
    \label{fig:q}
\end{figure}

The distribution of three binary characteristics, orbital period $\log(P/\mathrm{d})$, eccentricity $e$, and mass ratio $q=M_\mathrm{B}/M_\mathrm{A}$, are presented in Figures \ref{fig:porb}, \ref{fig:ecc}, and \ref{fig:q}, respectively. These figures show the results for the set A of the initial parameters, while the results for the sets B--D are shown in Figures \ref{apdxfig:porb+q}, \ref{apdxfig:ecc+teff}, and \ref{apdxfig:evol+spec} of Appendix \ref{apdx:characteristics}.

Distributions of $\log P$ are very similar to their original initial distributions, which means that the orbital periods did not change significantly during the binary evolution. An exception is a peak at $\log (P/\mathrm{d}) \approx 3$ for IS2+3 Cepheids (navy color), which hints that some of binary Cepheids shortened their orbital periods prior to their blue loop. In contrast, IS1 Cepheids (abundant in variants with uniform SFH and color coded as gray), tend to spread uniformly across the entire available range of orbital periods. Such clustering of IS2+3 Cepheids at $\log (P/\mathrm{d}) \approx 3$ was caused by tidal interactions on the red giant branch (RGB), which led to the shrinkage and circularization of their orbits\footnote{%
For particularly large eccentricities ($e > 0.8$), binary components tend to rendezvous at a very close proximity at the periastron, which amplifies the tidal forces. For binary stars with large convective envelopes (e.g. RGB stars), tidal forces in the \ST\ code are amplified by a factor of 50, which was calibrated based on the orbital separations and eccentricities of binaries in the Hyades open cluster \citetext{(Belczynski 2022, private communicatio}. As a result, a considerable fraction of IS2+3 Cepheids, i.e. after the RGB evolutionary phase, have their orbits shrunk and circularized.}.
Indeed, distributions of eccentricities of binary Cepheids peak at $e=0$ even though the initial distributions (sets A, B, D) did not favor this value. Apart from the peak at $e=0$, the eccentricity distributions remain similar to their initial values.

Comparison with a population synthesis study of binary Cepheids in the MW, carried out by \citet{neilson15}, shows a satisfactory agreement for distributions of orbital periods but considerable discordance for distributions of eccentricities. Indeed, \citet{neilson15} did not amplify the tidal forces, causing the values of eccentricities and orbital periods to remain virtually unchanged throughout binary evolution. Eccentricities of observed MW binary Cepheids \citep{evans05} favor the synthetic population of \citet{neilson15}, suggesting that the amplification factor of the tidal forces in our population synthesis is too large for the population of Galactic binary Cepheids.

Analogously to the orbital periods and eccentricities, the mass ratios of binary IS2+3 Cepheids preserve similar distributions as initially injected to the \ST\ code. For sets A and B, an excess of systems with a mass ratio of $0.2-0.3$ can be observed, being larger and more clustered for MW binary Cepheids, and less visible for the SMC. Within one metallicity environment, the four combinations of IS and SFH prescriptions do not show noticeable differences for IS2+3 Cepheids. On the other hand, IS1 Cepheids show distributions of mass ratios that tend to cluster around $0.5-0.6$ for the SMC, $0.4-0.5$ for the LMC, and $0.3-0.4$ for the MW, regardless of the set of initial parameters. We caution the reader that the characteristics of binary IS1 Cepheids have not been studied in detail yet, and therefore our results need further investigation with more precise tools, like the Modules for Experiments in Stellar Astrophysics \citep{paxton19}.

The key message from the above analysis of physical and orbital parameters on binaries is that the output of the binary population synthesis method is strongly affected by the input values and distributions. Therefore, one needs to be cautious not to trust the result based on only one set of input parameters, but instead investigate different sets and their outputs, in order to reliably assess the impact of initial parameters on the results.

\subsection{Characteristics of companions}
\label{subsec:compan}
\begin{figure}
    \centering
    \includegraphics[width=1.0\columnwidth]{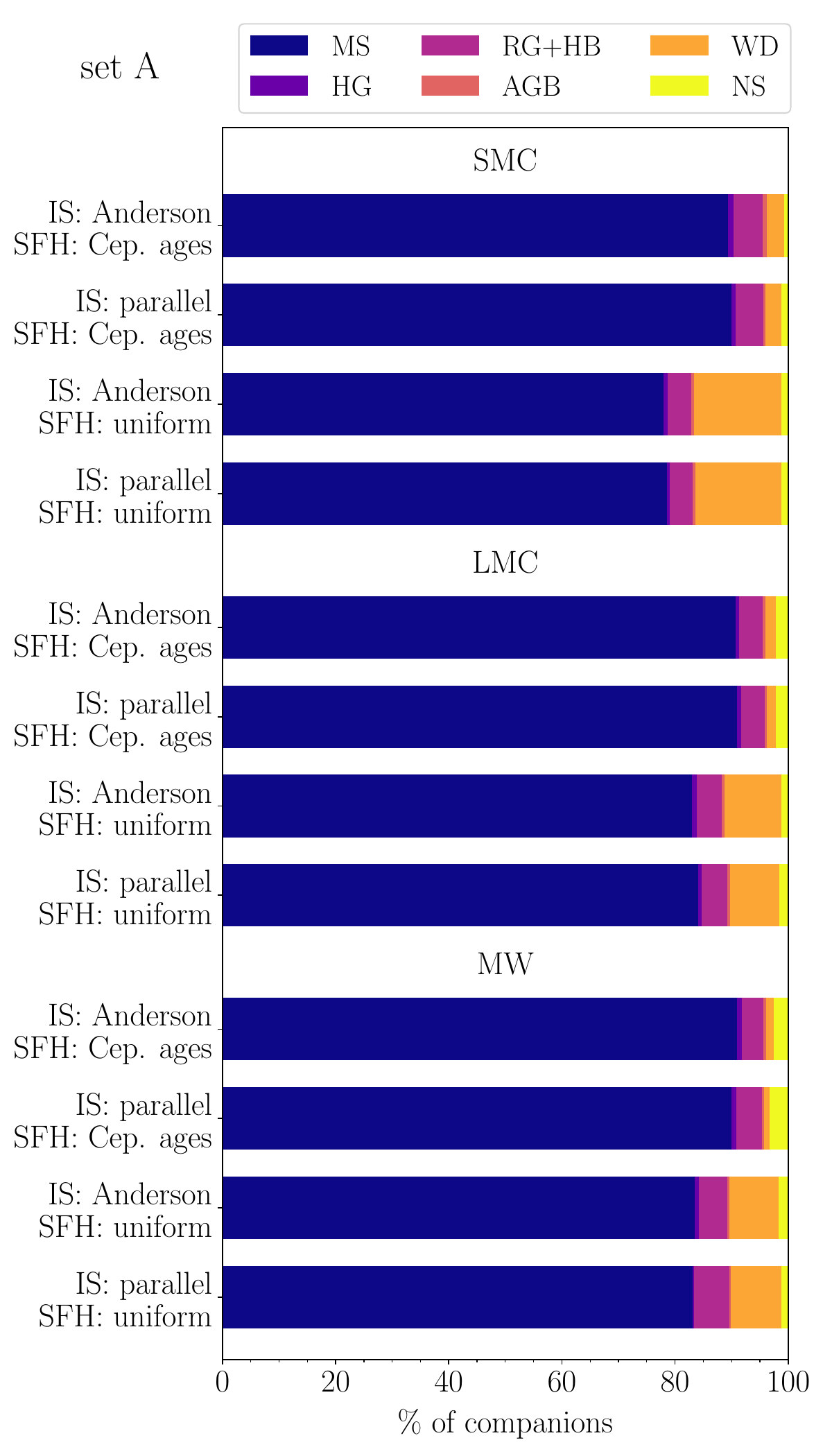}
    \caption{Evolutionary stages of companions in 12 variants of synthetic populations, for set A of the initial parameters. Images for sets B, C, and D can be found in Appendix \ref{apdx:characteristics}.}
    \label{fig:comp_evoltype}
\end{figure}

\begin{figure}
    \centering
    \includegraphics[width=1.0\columnwidth]{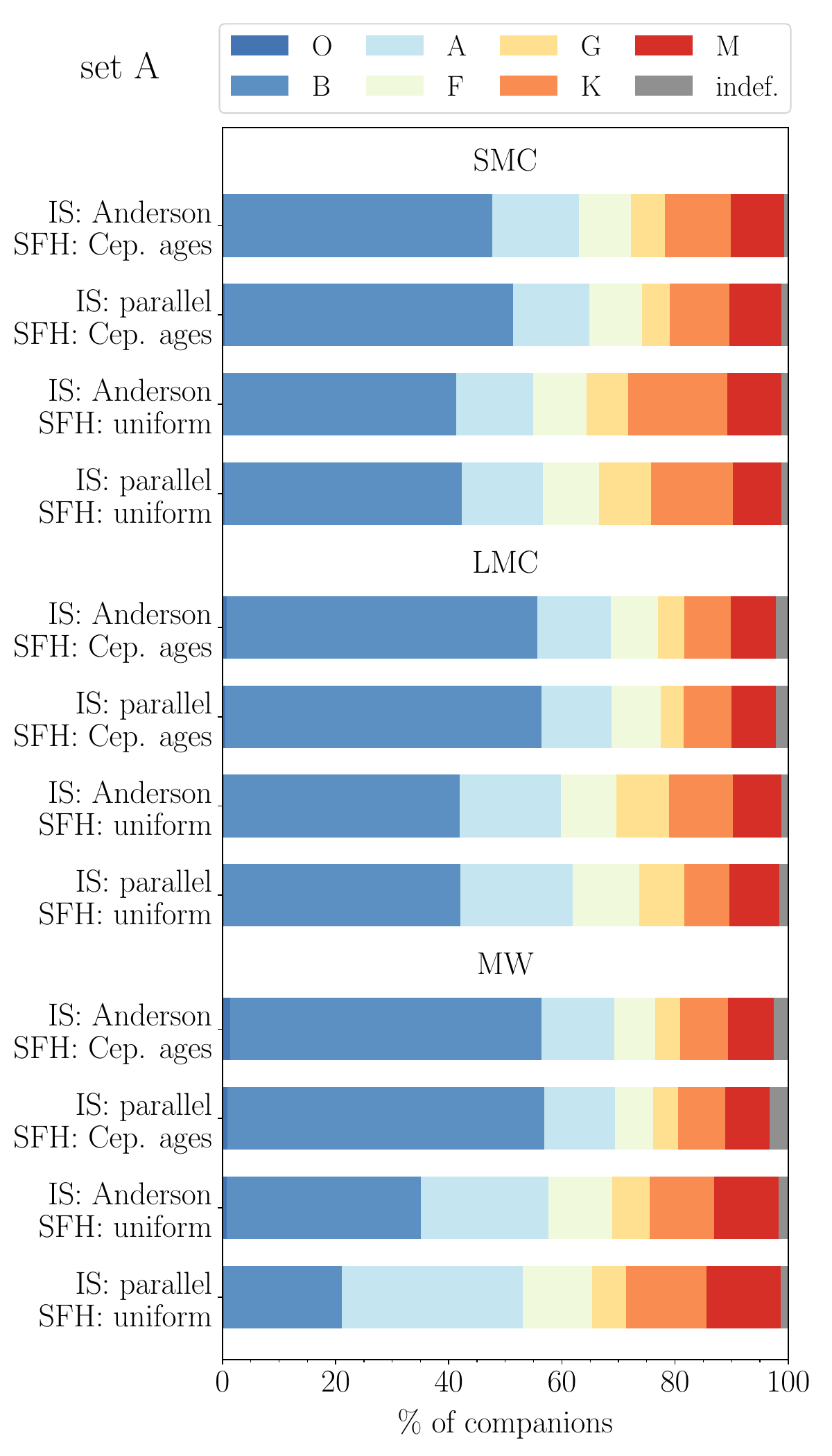}
    \caption{Spectral types of companions in 12 variants of synthetic populations, for set A of the initial parameters. Images for sets B, C, and D can be found in Appendix \ref{apdx:characteristics}. Indefinite spectral types belong to Cepheids' companions that evolved into neutron stars.}
    \label{fig:comp_spectype}
\end{figure}

\begin{figure}
    \centering
    \includegraphics[width=1.0\columnwidth]{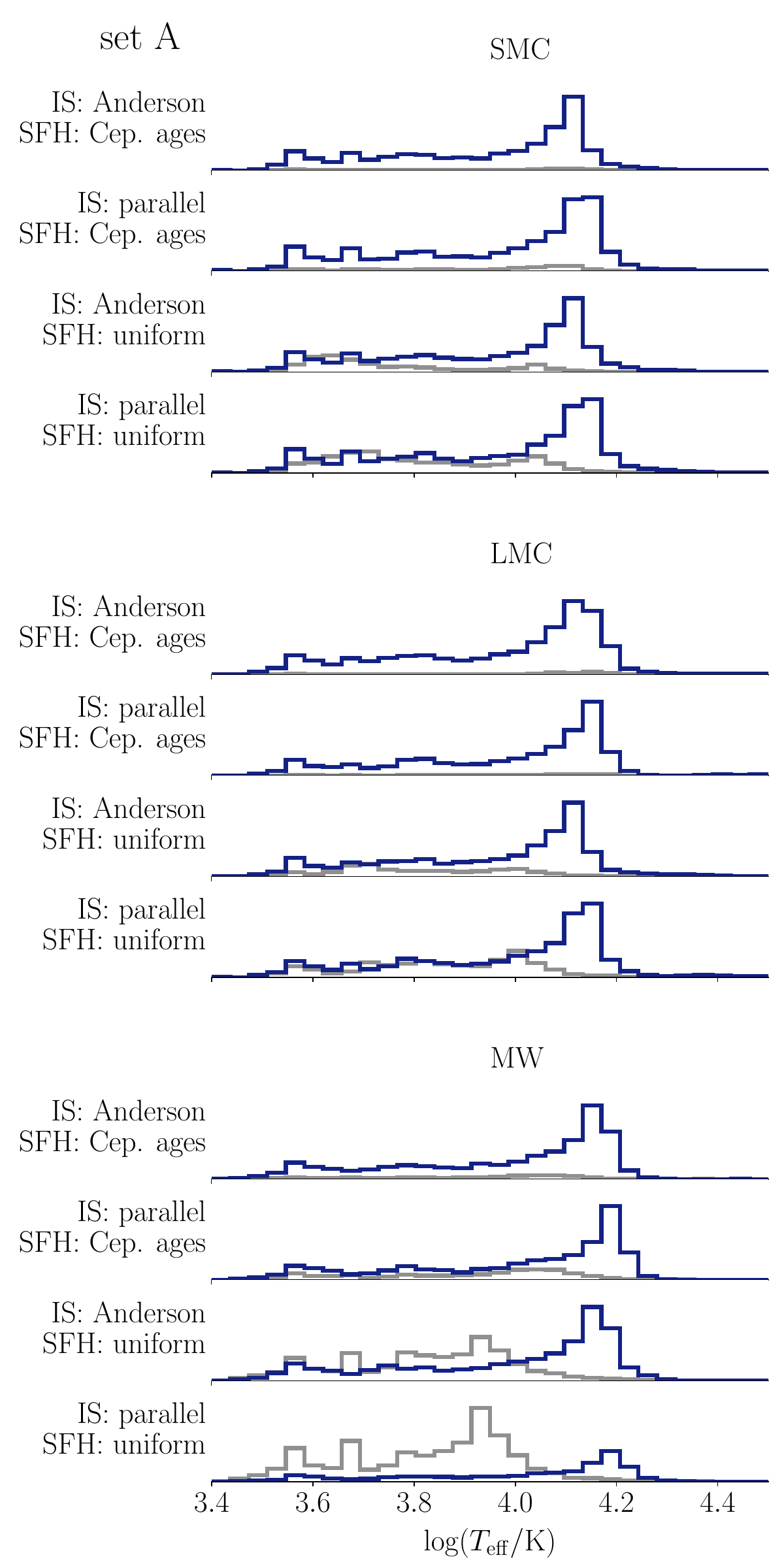}
    \caption{Effective temperatures of companions in 12 variants of synthetic populations, for set A of the initial parameters. IS2+3 Cepheids (navy blue) and IS1 Cepheids (gray) are presented separately. Images for sets B, C, and D can be found in Appendix \ref{apdx:characteristics}.}
    \label{fig:comp_temp}
\end{figure}

Three characteristics of Cepheids’ companions are the following: evolutionary stage, effective temperature, and spectral type. They are presented in Figures \ref{fig:comp_evoltype}, \ref{fig:comp_spectype}, and \ref{fig:comp_temp}, respectively. These figures present the results for a set A of the initial parameters, while the figures for sets B, C, and D are specified in Appendix \ref{apdx:characteristics}.

In most cases ($70-90$\%), companions to Cepheids are MS stars in every set of initial parameters (Figure \ref{fig:comp_evoltype}). The highest percentage of MS companions is in variants with the SFH determined by Cepheids' ages, for all metallicity environments. Evolved companions are as follows: red giants or horizontal branch stars (cumulatively denoted as RG+HB), AGB stars, WDs, and NS are also possible. We report no Cepheid binaries with black holes. Such systems have disrupted their orbits in a supernova event, preceding the creation of a black hole, and as such were excluded from our sample at the stage of data filtering (see Section \ref{sec:syntpop}). Among evolved companions, WDs dominate in scenarios with the uniform SFH for all sets of initial parameters and all metallicity environments, while in scenarios with the SFH based on the observed-age distribution of Cepheids, all types of evolved companions are similarly numerous. Notably, RG+HB companions are more common in the sets A, C, and D ($3-5$\%) but extremely scarce in the set B (0.8\%). 

The presented distributions support the idea that binary Cepheids are indeed common, but hard to observe using photometric methods, since the majority of companions are MS stars and their contribution to the overall luminosity of the system is minuscule. In the case where IS1 Cepheids are the primary components, their companions are MS stars by default, because they are less massive and therefore less evolved. Cepheids as secondary components can have companions at any evolutionary stage, but they constitute a marginal fraction of the sample (see Figure \ref{fig:CepProportions}).

Cepheids' companions have diverse spectral types, and no spectral type is strongly favored over others. Stronger preference for early-type companions (O, B) is visible in sets A and B, but it is only moderate in sets C and D. On the other hand, sets C and D tend to favor K-type companions more than sets A and B. Minuscule differences in spectral types for different metallicities and combinations of IS and SFH variants suggest that the companions' spectral types are solely correlated with initial parameters, among which the most important is the initial mass ratio distribution. Indeed, in sets C and D, the initial mass ratio distribution shows a peak at $q=1$, which favors components of a similar mass and therefore a similar evolutionary stage, increasing the probability of late-type companions. Figure \ref{fig:comp_spectype} shows that $20-40$\% of MW Cepheids have a companion of spectral type B, which agrees with \citet{evans92_uv} who reported that at least 20\% of all binary Cepheids in the MW have a companion of a spectral type earlier than A0V.

Surface temperatures of companion stars, although coarsely encoded by their spectral types, are worthy to be investigated on their own. We found that $\log T_\mathrm{eff}$ ranges from 3.5 to 4.5, but in the case of the SMC and LMC (top and middle panel), it tends to cluster around 3.6, 3.8, and 4.2 for B, C, and D sets, respectively (the equivalent spectral types are K5, F6, and B4, respectively). For the set A, a clear preference for companions having $\log T_\mathrm{eff}$ around 4.2 is visible. Temperatures of companions to MW Cepheids (bottom panel) cluster around $\log T_\mathrm{eff}=4.2$ for sets A and B, and are mostly uniform for sets C and D, with a mild concentration around 3.8. Such characteristics are unique for companions to IS2+3 Cepheids, while companions to IS1 Cepheids tend to present more erratic distributions, rarely coinciding with that of the IS2+3 group.

\subsection{Binaries with two Cepheids}
\label{subsec:2bincep}
In a handful of cases, both components cross the IS simultaneously, and create a Cepheid--Cepheid (C--C) binary. Only two\footnote{Nine candidates await spectroscopic confirmation to exclude a by-chance coincidence of two independent variables: MACHO*05:21:54.8-69:21:50, *04:59:17.5-69:14:18, *05:04:02.3-68:21:32 \citep{alcock95}, OGLE-GD-CEP-0291 \citep{udalski18}, OGLE-SMC-CEP-1526, -2699, -2893, -3115, -3674 \citep{soszynski10}.} such systems have been reported so far: CE Cassiopeiae \citep[CE Cas AB][]{berdnikov90,kervella19b} and OGLE-LMC-CEP-1718 \citep{soszynski08,gieren14,pilecki18}, suggesting that C--C binaries are extremely rare. Our synthetic populations support this conclusion, as presented in Table \ref{tab:cepcepbin}. In almost all cases, C--C binaries constitute less than 1\% of all Cepheid binaries but usually are much more scarce. Values in parentheses show the median duration of the simultaneous Cepheid phase. This phase can occur for pairs of Cepheids on the same IS crossing (e.g. IS2+2, IS3+3) or two different ones (e.g. IS2+3, IS1+3). Variants with IS1 Cepheids are extremely rare and short-lasting; thus the majority of C--C binaries are second- or third-crossers. 

Masses of IS2+2, IS2+3, or IS3+3 pairs are virtually the same; the relative differences in mass are smaller than about 2\%. Relative differences in radii are smaller than 25\%, and relative differences in $\log (L/\lsun)$ are up to 5\%. These values mark upper limits for expected relative differences of masses, radii, and luminosities in C--C binaries in the LMC. Indeed, OGLE-LMC--CEP-1718 has a relative mass difference of 1.2\%, relative radius difference of 19\%, and relative $\log (L/\lsun)$ difference of 4.6\% \citep{pilecki18}. CE Cas, which belongs to the MW open cluster NGC~7790 \citep{berdnikov90}, is a visual C--C binary on an extremely wide orbit; its projected separation of $\log(a/\rsun) \approx 6.176$ corresponds to the orbital period of about 5000 yr \citep{kervella19b}. A large 2.3 arcsecond angular separation of the components enables it to collect photometric and spectroscopic data for each star and investigate their properties individually \citep{berdnikov90,majaess13}. Unfortunately, this large separation precludes from collecting astrometric and/or velocimetric data, leaving this C--C binary impossible to compare with the synthetic results on the level that was possible for OGLE-LMC-CEP-1718.

Careful inspection of percentages and median lifespans of C--C binaries in Table \ref{tab:cepcepbin} shows that in the majority of cases C--C binaries at metallicities characteristic to the SMC live longer and are slightly more abundant than C--C binaries at metallicities characteristic to the LMC and MW. One could therefore expect to observe more C--C binaries in the SMC than in the LMC and MW; however, with just one C--C binary confirmed to date in each galaxy (OGLE-LMC-CEP-1718 and CE Cas), it is impossible to favor/disfavor any particular variant of synthetic populations, based on this prediction alone.

\begin{deluxetable*}{lcccc}
\tablecaption{Percentages of Cepheid-Cepheid Binaries in Environments of Different Metallicities, for Four Different Sets of Initial Parameters, and Four Combinations of IS and SFH Variants\label{tab:cepcepbin}}
\tablehead{
\colhead{Variant} & \colhead{Set A} & \colhead{Set B} & \colhead{Set C} & \colhead{Set D}
}
\startdata
\multicolumn{5}{c}{SMC}\\
\hline
IS: A, SFH: C & 0.99\% (3.25 Myr)	& 0.06\% (1.80 Myr)	& 1.09\% (2.61 Myr) &	0.94\% (2.99 Myr)\\
IS: p, SFH: C & 0.77\% (1.83 Myr)	& 0.07\% (1.32 Myr)	& 1.11\% (2.11 Myr)	 & 0.69\% (1.58 Myr) \\
IS: A, SFH: u	& 0.61\% (2.79 Myr)	& 0.08\% (3.15 Myr)	& 1.00\% (2.68 Myr) &	0.55\% (2.66 Myr) \\
IS: p, SFH: u	& 0.42\% (2.34 Myr)	& 0.00\% (no data)~~~ & 0.54\% (1.75 Myr) &	0.34\% (2.44 Myr)\\
\hline
\multicolumn{5}{c}{LMC}\\
\hline
IS: A, SFH: C	& 0.49\% (0.92 Myr)	& 0.08\% (2.13 Myr)	& 0.44\% (1.73 Myr)	& 0.27\% (2.32 Myr) \\
IS: p, SFH: C	& 0.64\% (1.28 Myr)	& 0.07\% (1.76 Myr)	& 0.54\% (1.51 Myr) &	0.50\% (1.21 Myr)\\
IS: A, SFH: u & 1.05\% (2.67 Myr)	& 0.08\% (3.82 Myr)	& 0.82\% (3.36 Myr) &	0.56\% (2.77 Myr)\\
IS: p, SFH: u	& 0.69\% (2.05 Myr)	& 0.10\% (0.94 Myr)	& 0.68\% (1.83 Myr) & 0.42\% (1.99 Myr)\\
\hline
\multicolumn{5}{c}{MW}\\
\hline
IS: A, SFH: C 	& 0.79\% (1.44 Myr)	& 0.11\% (1.16 Myr)	& 0.30\% (1.02 Myr)	& 0.41\% (1.42 Myr)\\
IS: p, SFH: C & 1.07\% (0.63 Myr) & 0.02\% (0.12 Myr)	& 0.23\% (0.88 Myr)	& 0.29\% (0.63 Myr)\\
IS: A, SFH: u & 0.54\% (1.20 Myr)	& 0.02\% (1.40 Myr)	& 0.19\% (1.12 Myr) & 0.12\% (0.57 Myr)\\
IS: p, SFH: u 	& 0.18\% (0.74 Myr)	& 0.00\% (no data)~~~ & 0.04\% (0.40 Myr) & 0.06\% (0.56 Myr)\\
\enddata
\tablecomments{Values in parentheses show the median duration of the simultaneous Cepheid phase. IS prescriptions: (A)nderson, (p)arallel. SFH prescriptions: based on (C)epheids' ages, (u)niform}
\end{deluxetable*}

\subsection{The estimated binarity fraction of LMC Cepheids}
\label{subsec:binfracestim}
Different detection methods allow for the discovery of binary Cepheids of different physical and orbital characteristics, but also have their own limitations. For example, binary Cepheids discovered due to eclipses in their light curves have companions of similar size, orbital inclination close to $i=90^{\circ}$, and relatively short orbital periods, so that the eclipses can be observed over a finite time span. Within 29 yr of the Optical Gravitational Lensing Experiment project, five LMC eclipsing binaries, consisting of a classical Cepheid and an evolved companion (RG or another Cepheid) have been discovered \citep{soszynski08,pilecki18}. They share similarly high values of inclination, $i\geqslant 83^{\circ}$, and short orbital periods, $\log (P/\mathrm{d})< 4$. By comparing these observational data and our synthetic populations, we make a first estimate of the number of binary Cepheids in the LMC. 

Let us assume that six binary Cepheids (in five systems, because one consists of two Cepheids) set a lower limit for the number of eclipsing binary Cepheids that can be observed given the detection conditions described above. If the inclination angle is distributed uniformly within a range $[0^{\circ}, 90^{\circ}]$, then six detected binary Cepheids with $i\geqslant 83^{\circ}$ constitute 8\% of all 75 binary Cepheids, out of which 92\% are undetectable due to their unfavorable inclination angles $i< 83^{\circ}$. We assume that binary Cepheids with longer orbital periods cannot be detected due to an insufficient time base of observations, regardless of their inclination angles.

Next, we recall that binary Cepheids with RG+HB companions constitute $3-5$\% of the entire population for sets A, C, D and 0.8\% for set B (Section \ref{subsec:compan} and Figures \ref{fig:comp_evoltype} and \ref{apdxfig:evol+spec}). We extract the fraction of these systems with $\log (P/\mathrm{d})< 4$ and we make this value equal to 75 systems derived from the previous step. Now we can calculate the number of binary Cepheids with RG+HB companions over the entire $\log P$ range and compare it with the percentage of RG+HB companions in the whole population of binary Cepheids. This value, divided by the number of LMC Cepheids \citep[4620,][]{soszynski15ccep}, yields the binarity fraction of classical Cepheids in the LMC. Depending on the variant of our synthetic population, we get the lower limit for a binarity fraction that ranges from 55\% to beyond 100\% (Table \ref{tab:binfrac}). In particular, variants from set B produce nonphysically high values (above 700\%), further confirming that set B should be disregarded altogether. Our crude estimation shows that the binarity fraction of classical Cepheids in the LMC should be at least 55\% and likely much higher. 

\begin{deluxetable}{lcccc}
\tablecaption{Binary Fraction Estimates of Classical Cepheids in the LMC, for Different Sets of Initial Parameters, and Four Combinations of IS and SFH Variants\label{tab:binfrac}}
\tablehead{
\colhead{Variant} & \colhead{Set A} & \colhead{Set B} & \colhead{Set C} & \colhead{Set D}
}
\startdata
IS: A, SFH: C & 95\% & 884\% & 84\% & 146\% \\
IS: p, SFH: C & 111\% & 713\% & 73\% & 129\% \\
IS: A, SFH: u & 104\% & 1181\% & 55\% & 103\% \\
IS: p, SFH: u & 176\% & 1012\% & 80\% & 144\%\\
\enddata
\tablecomments{Note that fractions above 100\% are not realistic. See the explanation in Section \ref{subsec:binfracestim} for the derivation of these estimates. IS prescriptions: (A)nderson, (p)arallel; SFH prescriptions: based on (C)epheids' ages, (u)niform}
\end{deluxetable}

\subsection{Mass ratios of MW Cepheid binaries with MS companions}

\begin{deluxetable*}{lcccc}
\tablecaption{Selected Milky Way Classical Cepheids in Binary Systems with Known Mass Ratios and Contrasts in Either the $H$ or $V$ Bands \label{tab:mwcep}
}
\tablewidth{0pt}
\tablehead{
\colhead{Name} & 
\colhead{$\Delta H$ (mag)} & 
\colhead{$\Delta V$\ (mag)} & 
\colhead{$q$} & 
\colhead{References}
}
\startdata
AX Cir & 5.20 $\pm$ 0.20 & 1.68 & 0.93 $\pm$ 0.04\tablenotemark{a} & 2, 4, 6, 7\\
V1334 Cyg & 3.70 $\pm$ 0.10 & 2.18 & 0.94 $\pm$ 0.04\tablenotemark{b} & 2, 5\\
AW Per & 4.78 $\pm$ 0.30 & 2.5 & 0.70 $\pm$ 0.04\tablenotemark{a} & 2, 3, 4, 6\\
U Aql & 5.58 $\pm$ 0.85 & 4.31 & 0.39 $\pm$ 0.01\tablenotemark{a} & 1, 2, 3, 4\\
FF Aql & 5.63 $\pm$ 0.80 & \nodata & 0.3 & 2, 4, 8\\
S Mus & 5.10 $\pm$ 0.14 & \nodata & 0.82 $\pm$ 0.03\tablenotemark{a} & 2, 4, 8\\
DL Cas & \nodata & 4.14 & 0.42 $\pm$ 0.02\tablenotemark{a} & 2, 4\\
RX Cam & \nodata & 3.94 & 0.38 & 2, 4\\
SU Cyg & \nodata & 3.01 & 0.66 $\pm$ 0.03\tablenotemark{a} & 2, 4\\
V350 Sgr & \nodata & 3.97 & 0.46 $\pm$ 0.03\tablenotemark{a} & 1, 2, 4\\
\enddata
\tablecomments{References: (1) \citet{evans92_uv}; (2) \citet{evans95}; (3) \citet{gallenne15}; (4) \citet{evans15}; (5) \citet{gallenne18}; (6) \citet{evans94}; (7) \citet{gallenne14}, (8) \citet{gallenne18interferometry}.}
\tablenotetext{a}{Error assessed as half difference between maximum and minimum values reported in the literature.}
\tablenotetext{b}{Value taken from \citet{gallenne18}, which is the most precise measurement up to date.}
\end{deluxetable*}

\begin{figure*}
    \centering
    \includegraphics[width=0.95\columnwidth]{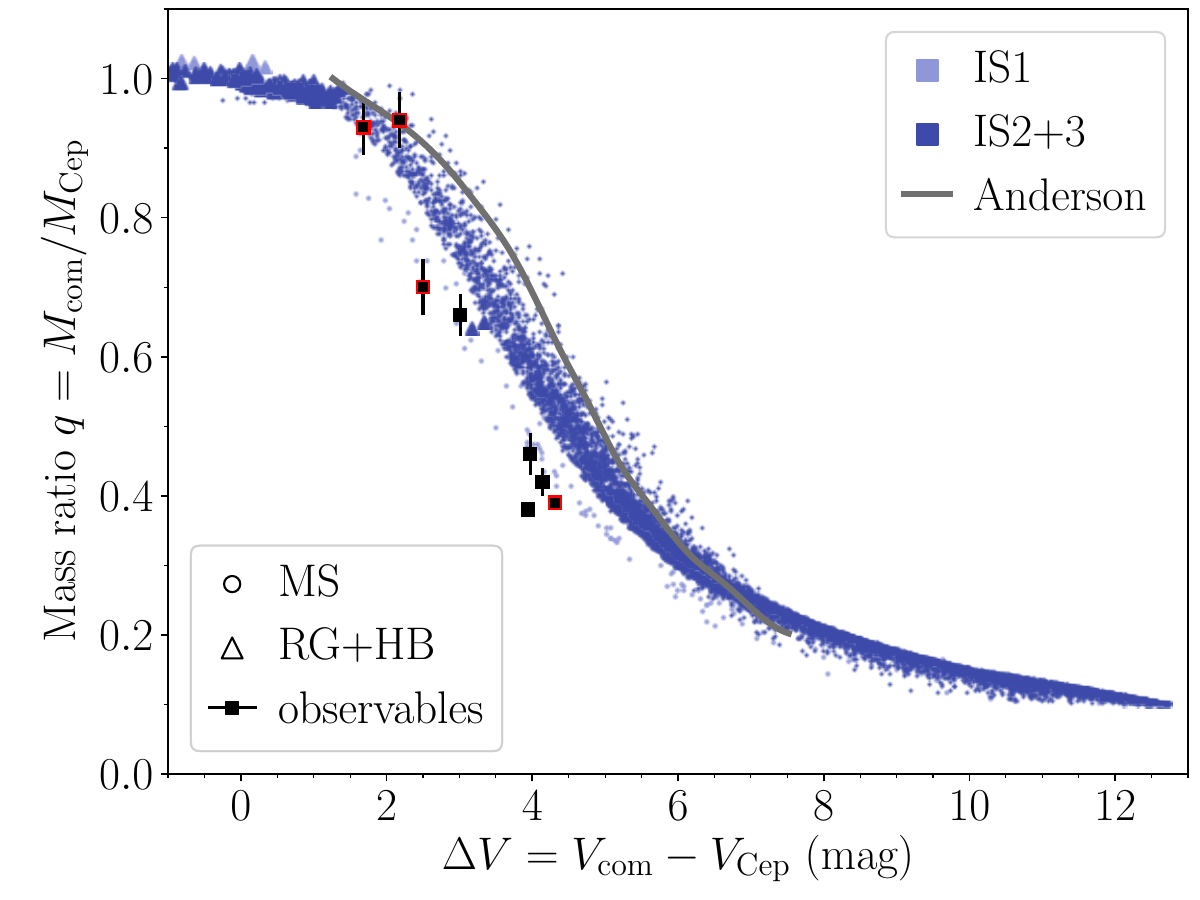}%
    \hspace{1em}
    \includegraphics[width=0.95\columnwidth]{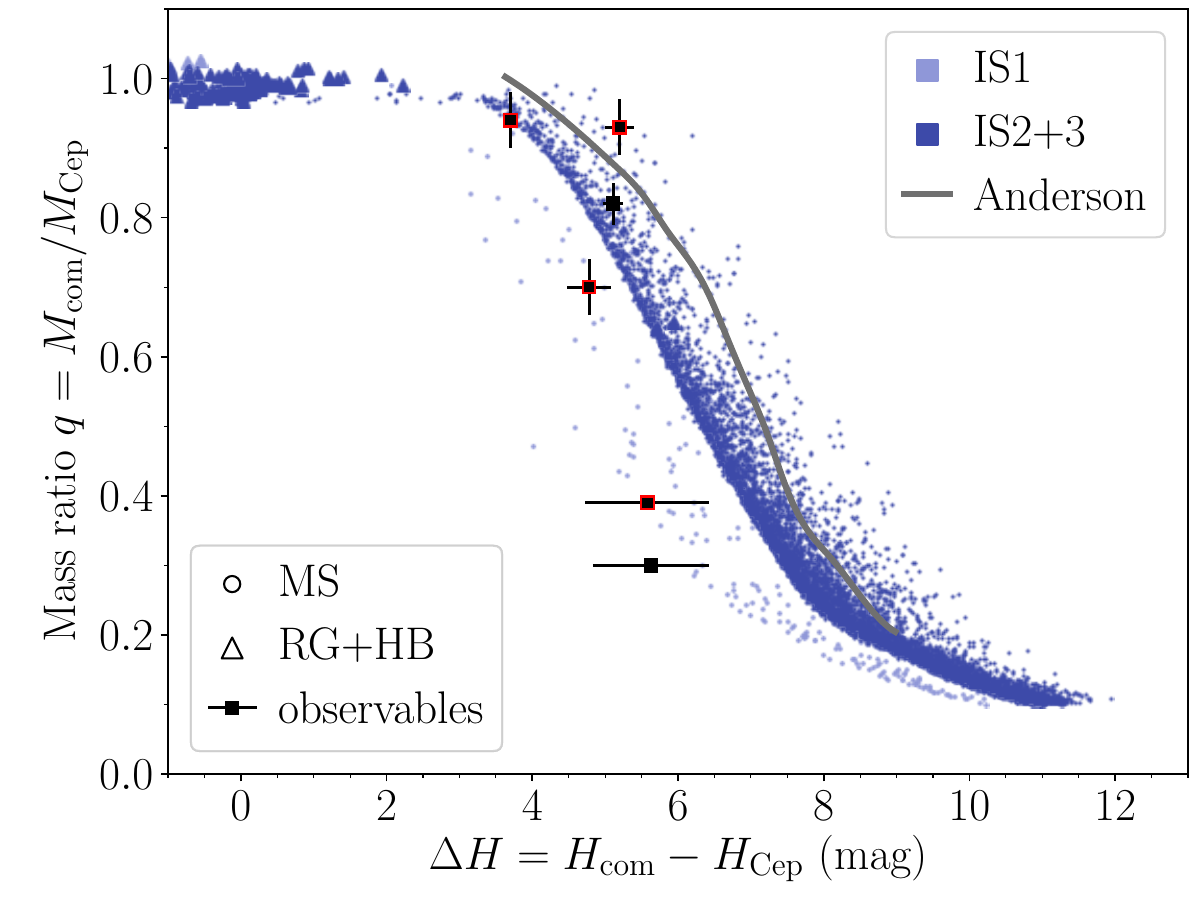}\\ \vspace{1em}
    \includegraphics[width=0.95\columnwidth]{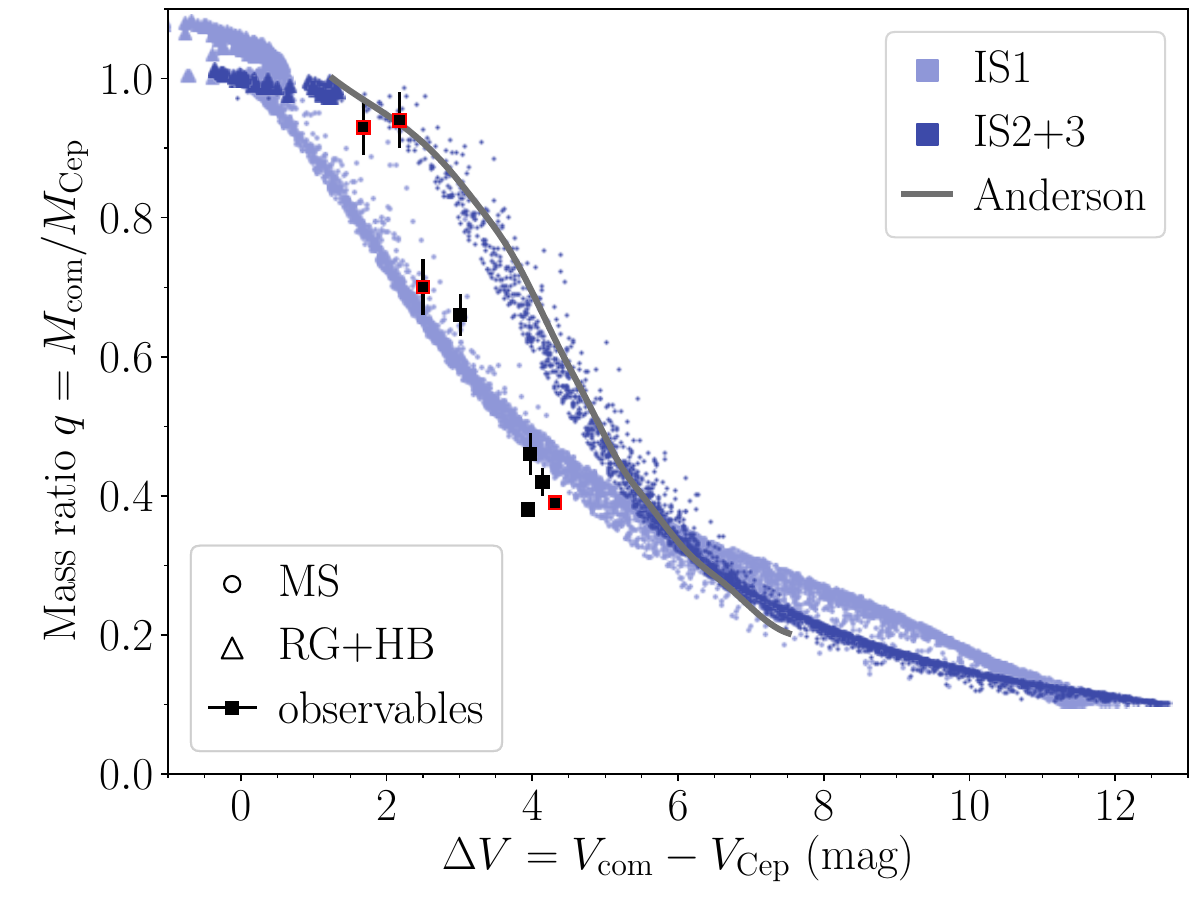}%
    \hspace{1em}
    \includegraphics[width=0.95\columnwidth]{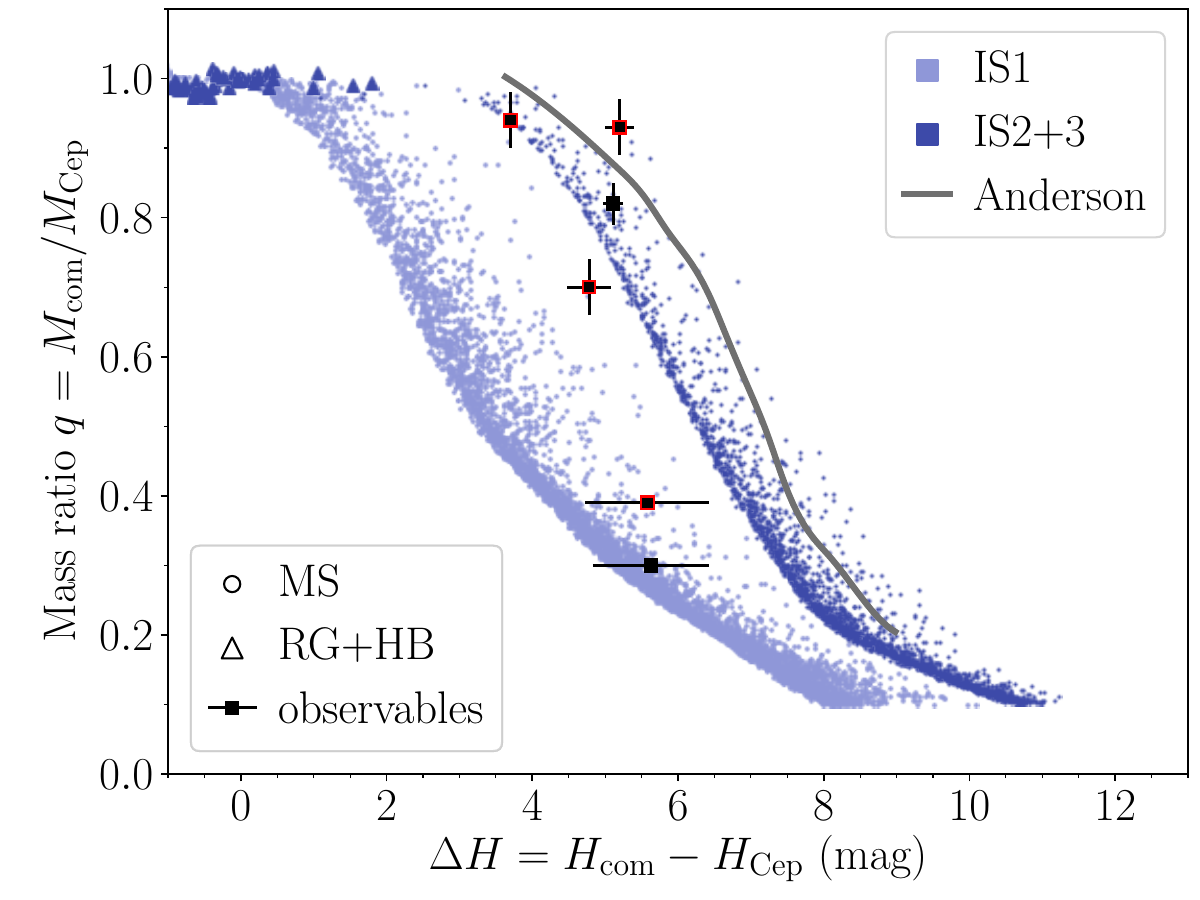}
    \caption{Magnitude difference (contrast) between MW Cepheids and their companions in the $V$ (left) and $H$ (right) bands vs. mass ratio. Empirical data are taken from Table \ref{tab:mwcep}. Systems present on both panels are additionally marked in red. The underlying synthetic populations correspond to the following parameters: set D, Anderson IS, SFH based on Cepheids' ages (top); set A, parallel IS, uniform SFH (bottom). The rest of populations fit between these two most divergent variants. Complementing panels for the $B$, $I$, $J$, and $K$ bands are presented in Figure \ref{apdxfig:contrast_Mq}.}
    \label{fig:contrast_Mq}
\end{figure*}

In the MW, binary Cepheids with MS companions can be detected with interferometric observations; however, such detections are limited to large orbital separations (\,$\geqslant 100$\,milliarcseconds) and relatively bright companions, i.e. the difference in the magnitude between the companion and the Cepheid (the so called \emph{contrast}) must be smaller than 6\,mag in $H$ band \citep{gallenne18,gallenne18interferometry}. In order to determine the physical properties of the binary components (especially masses), interferometric observations have to be supplemented with radial-velocity measurements, which is a time-consuming task, because the orbital period of a binary with a Cepheid is at least one year long \citep{neilson15,pilecki18}. Moreover, in the majority of cases, spectroscopic observations are performed in the visual domain, where hot MS companions are too faint to be detected. For such companions, more challenging UV spectroscopy has to be carried out from space \citep{gallenne18}.

Our synthetic populations offer a unique insight into the relations between physical parameters of binary components and their observed properties. In particular, Figure \ref{fig:contrast_Mq} presents mass ratio as a function of contrast in the $V$ and $H$ bands for the two most divergent variants: set D, Anderson IS, SFH based on Cepheids' ages; and set A, parallel IS, uniform SFH. Complementing figures for $B$, $I$, $J$, and $K$ bands are available in Appendix \ref{apdx:contrast_Mq}. For comparison purposes, we overplot gray curves showing the contrast-mass ratio relation derived and averaged over second and third IS crossings by \citet[][their Figure 2]{anderson18}. We extend our analysis to IS1 crossers, and compare with observational data, summarized in Table \ref{tab:mwcep}. Data points with red borders (V1334~Cyg, AX~Cir, AW~Per, U~Aql) are binary Cepheids with contrasts in both the $V$ and $H$ bands, and are expected to occupy the same region of either IS1 or IS2+3 crossers on all plots in Figure \ref{fig:contrast_Mq}. Indeed, V1334~Cyg and AX Cir lie on the IS2+3 trend, but AW~Per and U~Aql lie below it, in the area of IS1 crossers. While our result does not prove their evolutionary status, it entertains an uncommon idea that IS1 Cepheids might not be as rare as previously thought. Further investigation of the rates of pulsation period change would be required to determine their evolutionary status.

The contrast-mass ratio relation depends on the mass-luminosity relation, which was established by numerous authors based on models of stellar evolution (see Figure \ref{fig:is_ml_lit} and Section \ref{subsec:MLR} for reference). Such models tend to overestimate Cepheid masses by $10-20$\% relative to pulsation models \citep[the so called mass discrepancy problem, ][]{keller08}. The data point of V1334~Cyg, which was derived from the dynamical mass of the Cepheid and its companion \citep{gallenne18}, fits perfectly in the IS2+3 trends in both filters. However, other data points, for which the Cepheid masses were estimated from the evolutionary models via mass-luminosity and period-luminosity relations \citep{evans95}, might be underestimated, meaning that they should be shifted upwards in the plot to fit the IS2+3 trend.

Contrast between a Cepheid and its companion strongly depends on the pulsation phase of the Cepheid and is bigger when the Cepheid is brighter. If the Cepheid is observed at a random phase, an additional statistical error should be added to the error budget of the contrast value. 
$V$-band contrasts in Table \ref{tab:mwcep} were calculated from Cepheid magnitudes averaged over their pulsation cycles. $H$-band contrasts were either calculated from random-phase observation or averaged over a couple of observations, and therefore their errors might be underestimated.

The above relations offer a promising opportunity to estimate the mass ratios using just a single interferometric image, instead of a number of radial-velocity measurements. This method could be applied to very wide binaries, for which spectroscopic observations could take many years to complete.

\subsection{Detection of Cepheid Binaries above the Period-Luminosity Relation}
Following the hypothesis of \citet{pilecki21}, that Cepheids located well above the PLR are binaries, we reproduced their Figure 3 using our synthetic population of LMC Cepheids (set D, IS: Anderson, SFH: Cepheids' ages) with a binarity fraction of 100\%, meaning that every Cepheid has a companion. The total brightness of a binary was calculated as follows:
\begin{equation}
    m_{\rm tot} = -2.5 \log \left( 10^{-m_A/2.5}  + 10^{-m_B/2.5}\right)\,.
\end{equation}

Peak-to-peak amplitude of a binary was calculated as a difference between the minimum and maximum brightness of a binary, $A_{\rm tot} = m_{\rm tot,\,min} - m_{\rm tot,\,max}$. Next, the relative pulsation amplitude with respect to the original pulsation amplitude of a single Cepheid was calculated as $100\% \times A_{\rm tot}/A_{\rm Cep}$. A relative amplitude of 100\% means that the extra light from the companion was negligible, and the Cepheid retains its full amplitude, while a relative amplitude of 0\% means that the Cepheid amplitude was completely diminished by the extra light from the companion. 

\begin{figure}
    \centering
    \includegraphics[width=\columnwidth]{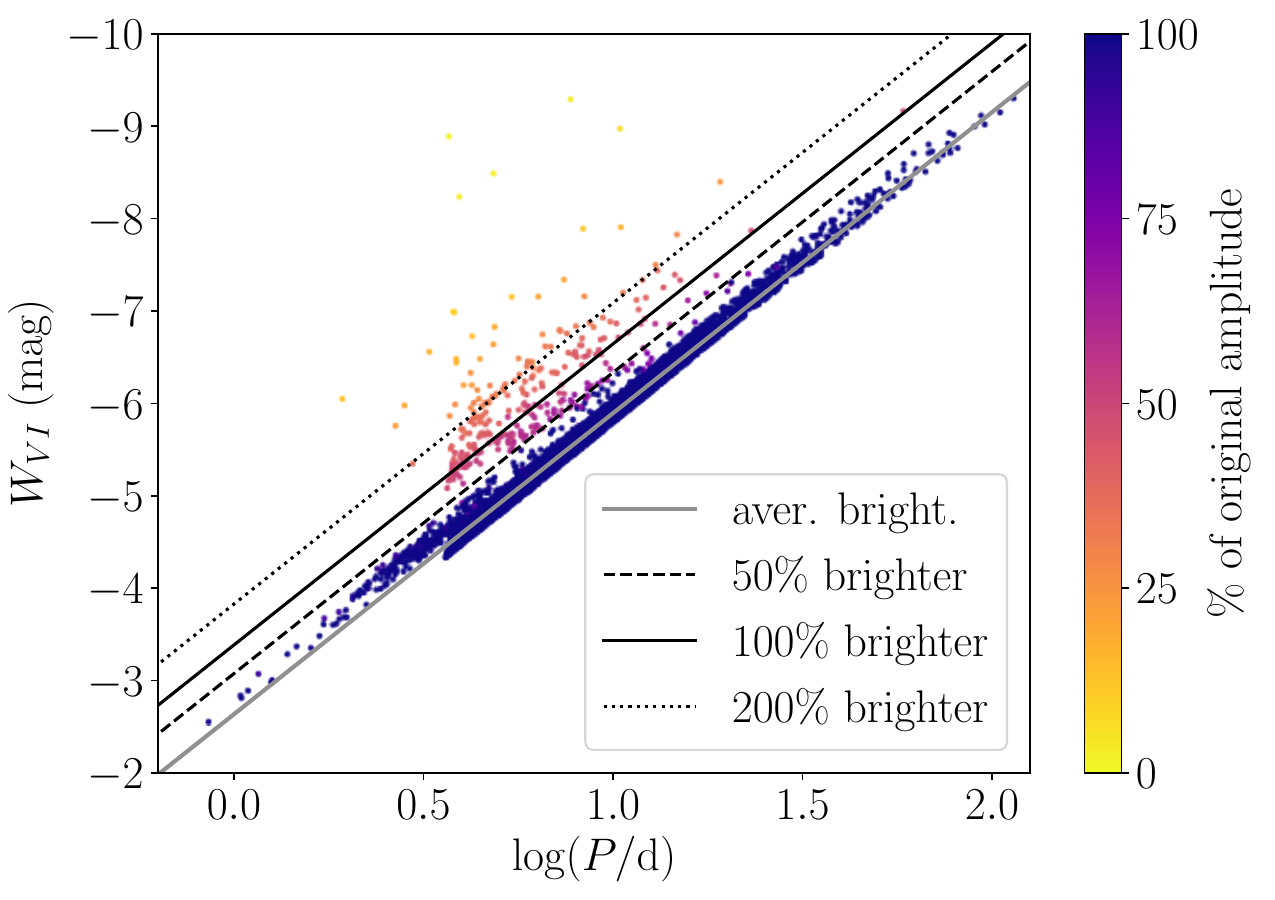}
    \caption{Period-luminosity relation for synthetic Cepheids in the LMC (set D, IS: Anderson, SFH: Cepheids' ages) with binarity fraction 100\%. Black lines mark 50\% (dashed), 100\% (solid), and 200\% (dotted) brighter Cepheids than an average Cepheid of a given period (gray solid line). The gap between navy and magenta data points (i.e. Cepheids with MS and giant companions) reflects the Hertzsprung Gap, where only a few companions are found.}
    \label{fig:outliers}
\end{figure}

Figure \ref{fig:outliers} presents our synthetic population on the period-luminosity plane. Black lines mark thresholds for stars that are 50\%, 100\%, and 200\% brighter than an average Cepheid of a given period (gray line). All outliers (above 50\% threshold) are binary Cepheids with giant or supergiant companions. Colors encode relative amplitudes; particularly interesting are magenta data points, representing binary Cepheids with RG+HB companions, whose light contribution places the Cepheids above the dashed line while preserving about 30--60\% of their pulsation amplitudes. Such stars present an unmatched observational opportunity to find binaries among Cepheids. 

A handful of binary Cepheids located above the dotted 200\% line, marked as yellow or bright orange, have AGB companions. Such systems evolved with Cepheid progenitors as secondary components (less massive at ZAMS). AGB companions dominate the brightness of the system, and also heavily diminish the apparent pulsation amplitude of Cepheids. Consequently, the Cepheid is found well above the PLR, but its pulsation amplitude is so low that it might be mistaken for another type of small-amplitude pulsator or overlooked altogether and treated as a constant star.

MS and WD companions, on the other hand, contribute little or almost no light to their systems (navy blue data points) and lie very close to the main trend and well below the 50\% threshold. However, they outnumber companions of other evolutionary stages and constitute the vast majority of Cepheid binaries. Because of their location on the period-luminosity plane, they are virtually undetectable via photometric methods, yet their cumulative effect on the zero-point and the slope of the PLR is not negligible, and will be thoroughly characterized in Paper II.

\section{Summary}

We presented the synthetic populations of binary Cepheids for three environments of different metallicity: the SMC, LMC, and MW. For each metallicity, we crated 16 different variants of synthetic populations, testing two prescriptions for the shape of the IS, two prescriptions for the SFH, and four sets of initial parameter distributions. Our synthetic populations are free from selection bias, and the percentage of Cepheid binaries is controlled by us via the binarity parameter.

We compared all variants with the literature and concluded that the most realistic synthetic populations are the ones created from the IS prescription of \citet{anderson16}, and the SFH based on the Cepheid ages \citep{bono05}. We dissuade using set B of the initial parameters as it resulted in unrealistically high fractions of binary Cepheids in the LMC, and unrealistically low fractions of binary Cepheids with giant, evolved companions.

Hot MS stars constitute $20-40$\% of all companions, which agrees with the empirical study of \citet{evans92_uv}. Such companions show narrow contrast-mass ratio relations, already suggested by \citet{anderson18}, and replicated by us. Comparison of our theoretical results with empirical values of mass ratios and $V$-, $H$-band contrasts shows satisfactory agreement, and encourages further investigation of the contrast-mass ratio relations as an efficient tool to estimate the mass ratios of binary Cepheids. 

We reported that giant, evolved stars constitute $3-5$\% of all companions, and by comparing observational data with our synthetic populations, we estimated the number of binary Cepheids in the LMC, which is at least 50\% and probably much higher (close to 100\%). We confirmed that Cepheid binaries with giant companions can be easily detected as outliers above the PLR. MS companions lie well below the detection threshold in the period-luminosity plane, but their effect on the PLR is nonnegligible and will be the focus of Paper II.

\acknowledgments
We thank the anonymous referee, whose pertinent comments helped to significantly improve this paper. P.K. thanks P.W. for insightful discussions on the topic of this study and beyond.
The research leading to these results has received funding from the European Research Council (ERC) under the European Union’s
Horizon 2020 research and innovation program (grant agreements No. 695099 and No. 951549). We also acknowledge support from the Polish Ministry of Science and Higher Education grant DIR/WK/2018/09, Polish National Science Centre grant BEETHOVEN 2018/31/G/ST9/03050, and BASAL Centro de Astrofisica y Tecnologias Afines BASAL-CATA grant AFB-170002. K.B. acknowledges support from the Polish National Science Centre grant MAESTRO 2018/30/A/ST9/00050. R.S. acknowledges support from the Polish National Science Centre grant SONATA BIS 2018/30/E/ST9/00598. W.G. gratefully acknowledges support from the ANID BASAL project ACE210002.

%

\vspace{5mm}


\software{
StarTrack \citep{belczynski02,belczynski08}, SciPy \citep{2020SciPy-NMeth}, NumPy \citep{2020NumPy-Array}, Pandas \citep{pandas}, JuPyter \citep{ipython}. All images were proudly made with Matplotlib \citep{matplotlib}.
}




\appendix
\section{Initial distributions}
\label{apdx:initialdistr}
\restartappendixnumbering

We present triangle plots of initial distributions of sets A, B, C, and D. Differences with respect to metallicities are negligible. For sets C and D, we assumed, following \citet{moe17}, that the distributions of initial primary masses and orbital periods are independent, and drew eccentricities and mass ratios from the distributions given in their Table 13. Distributions in sets B--D are based on observational data, and approximated by analytical functions. Such functions are of different types (e.g. power-law, log-normal, log-uniform) and/or have different parameter values (e.g. the exponent of a power-law function) in different ranges of one distribution (e.g. $q$, $e$). As a consequence, such distributions are not smooth, but remain continuous.

\begin{figure*}
\centering
\includegraphics[width=0.49\columnwidth]{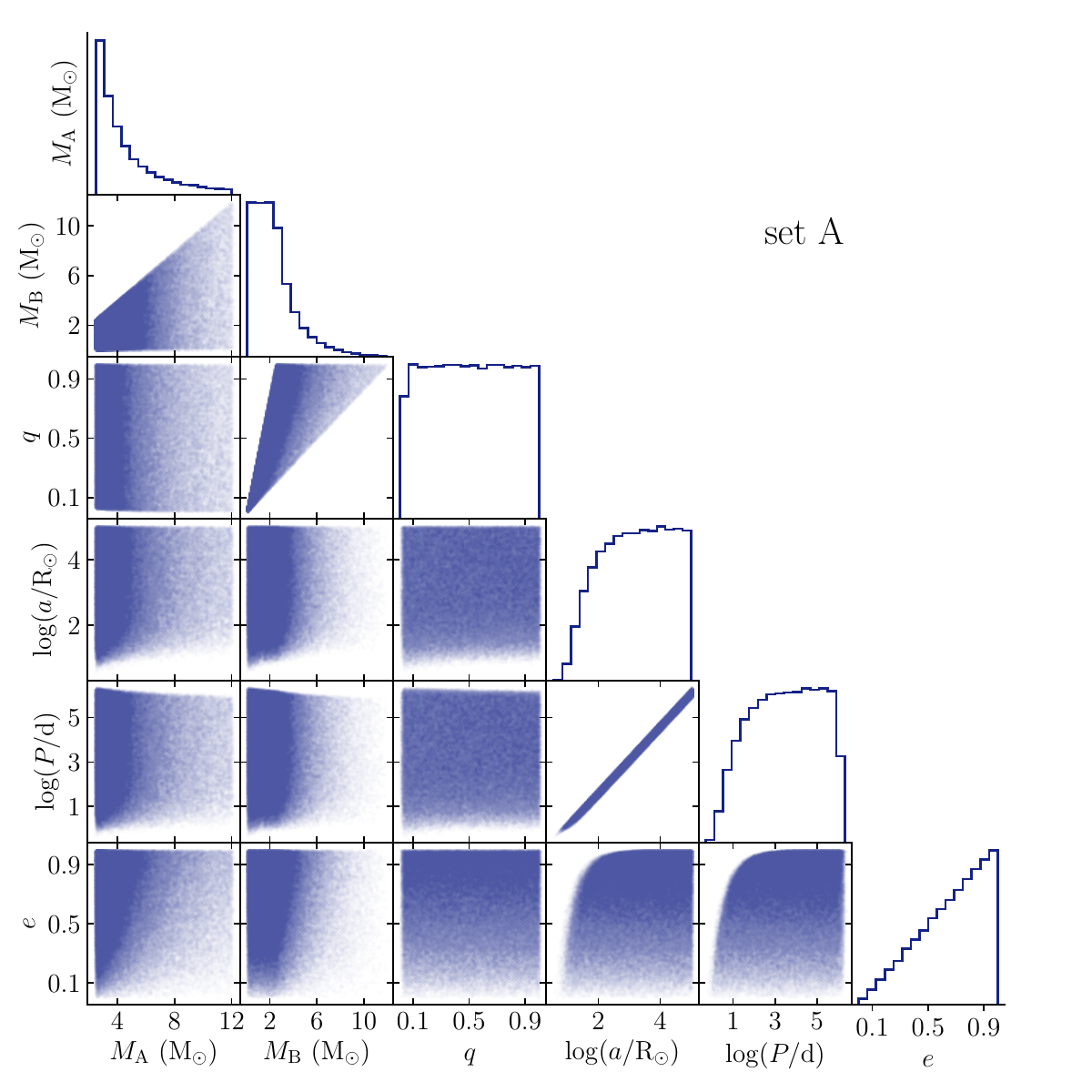}
\includegraphics[width=0.49\columnwidth]{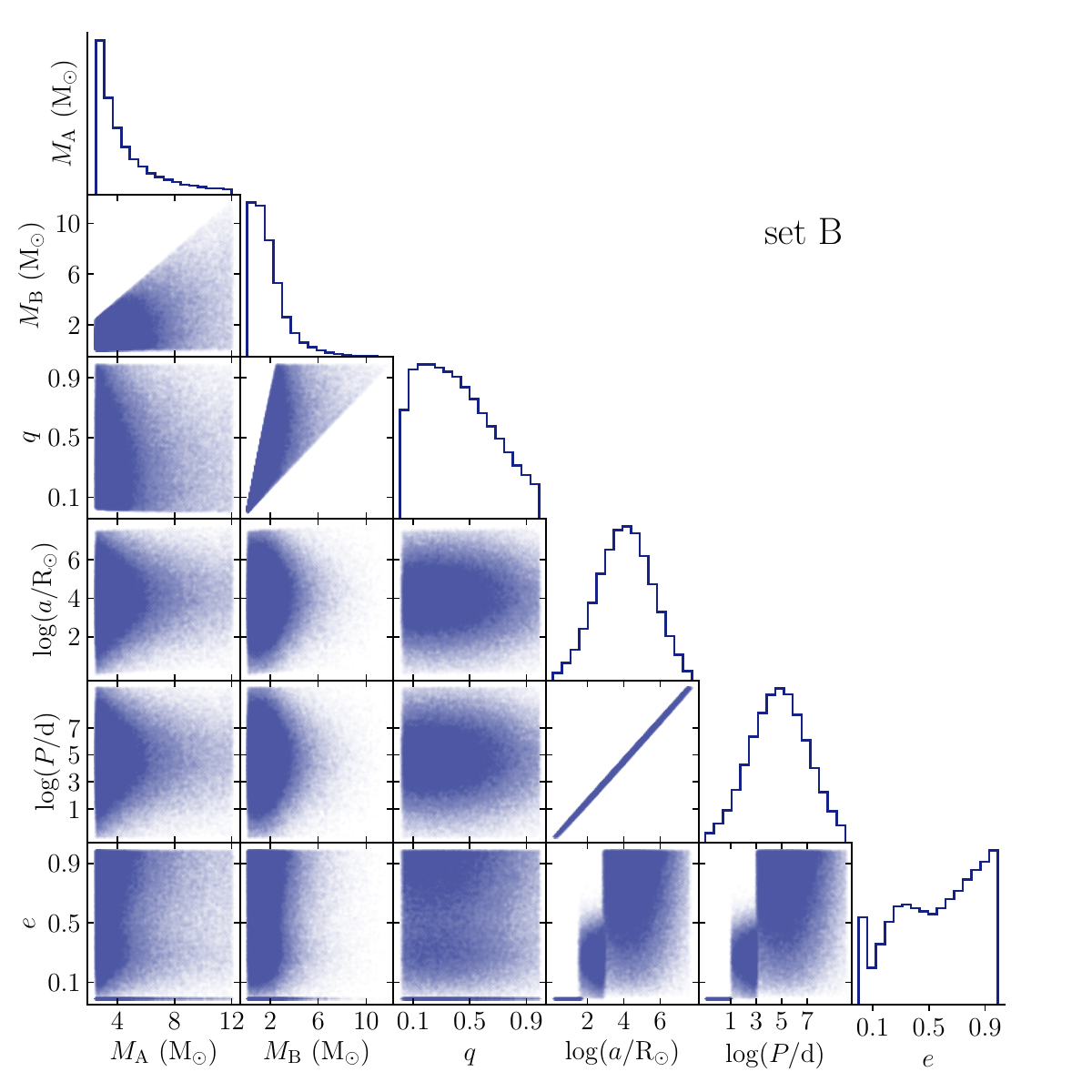}\\
\includegraphics[width=0.49\columnwidth]{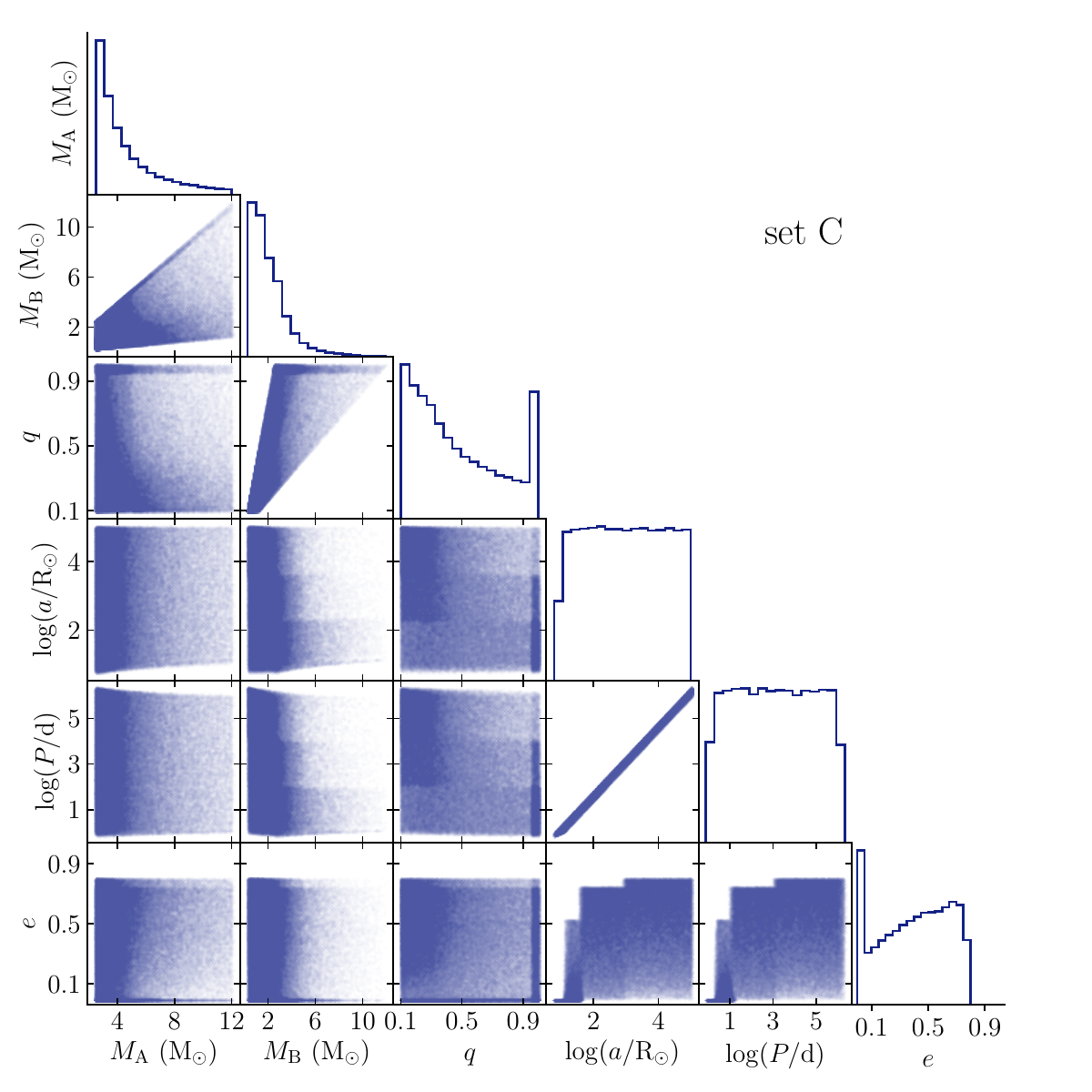}
\includegraphics[width=0.49\columnwidth]{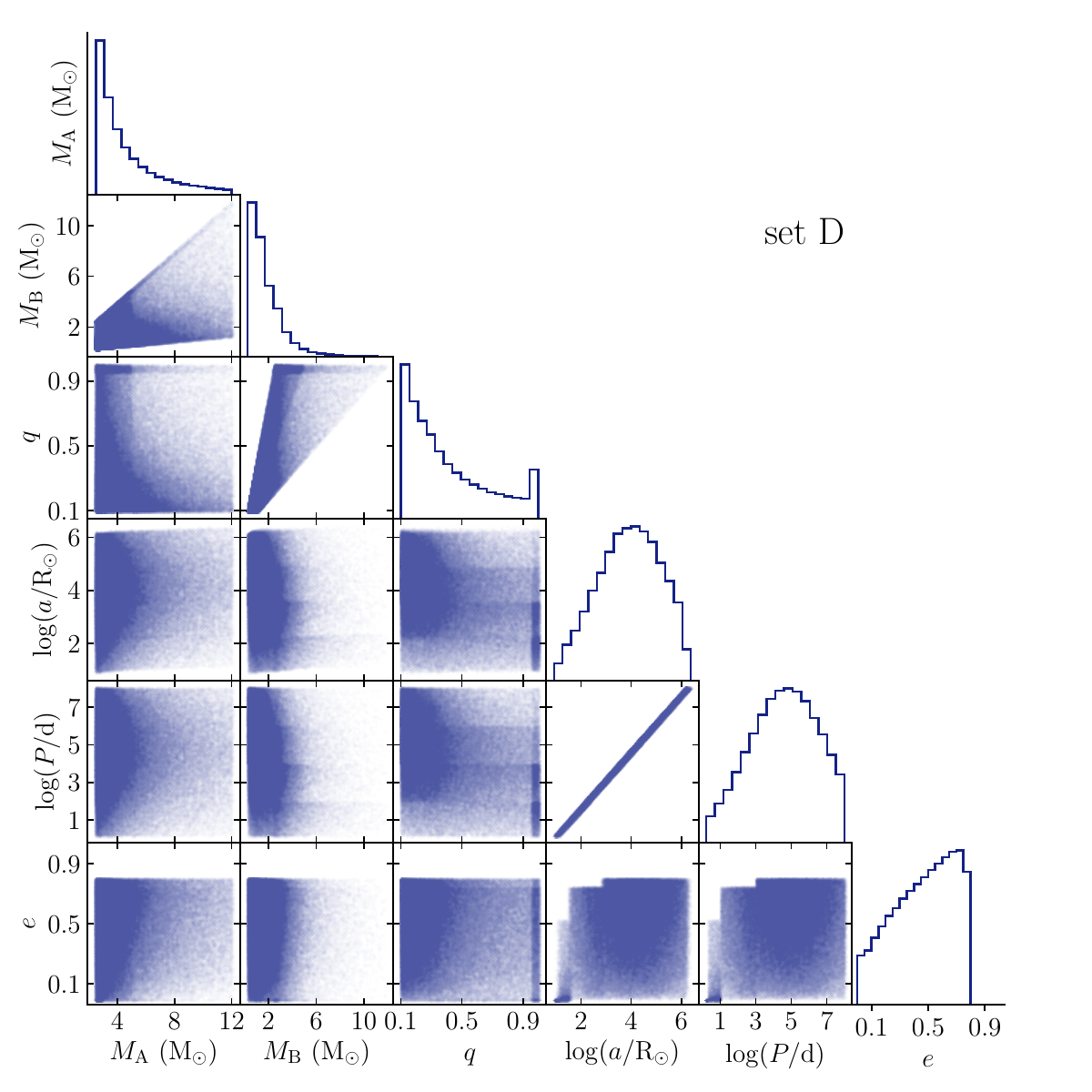}
\caption{Visualization of the four sets of initial distributions described in Table \ref{tab:STmodels}.
\label{fig:setABCD}}
\end{figure*}

\section{Characteristics of binaries and Cepheids' companions}
\label{apdx:characteristics}
\restartappendixnumbering

Figures \ref{fig:porb}, \ref{fig:ecc}, and \ref{fig:q} presented three binary characteristics: orbital period $\log(P/\mathrm{d})$, eccentricity $e$, and mass ratio $q=M_\mathrm{B}/M_\mathrm{A}$, respectively, for set A of the initial parameters. This section complements the presented results with the plots for sets B, C, and D.

\begin{figure}
    \gridline{
    \fig{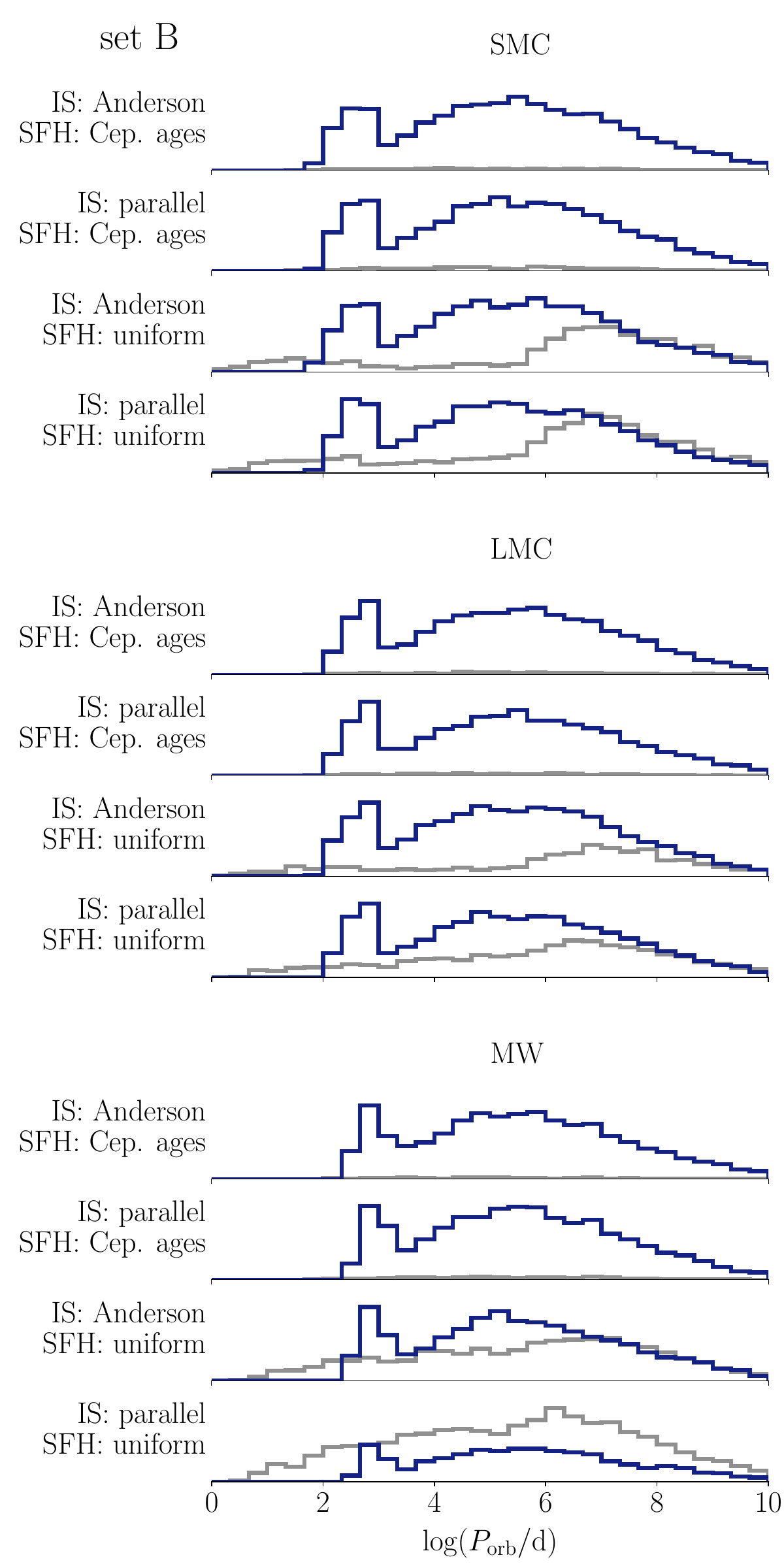}{0.3\textwidth}{}
    \fig{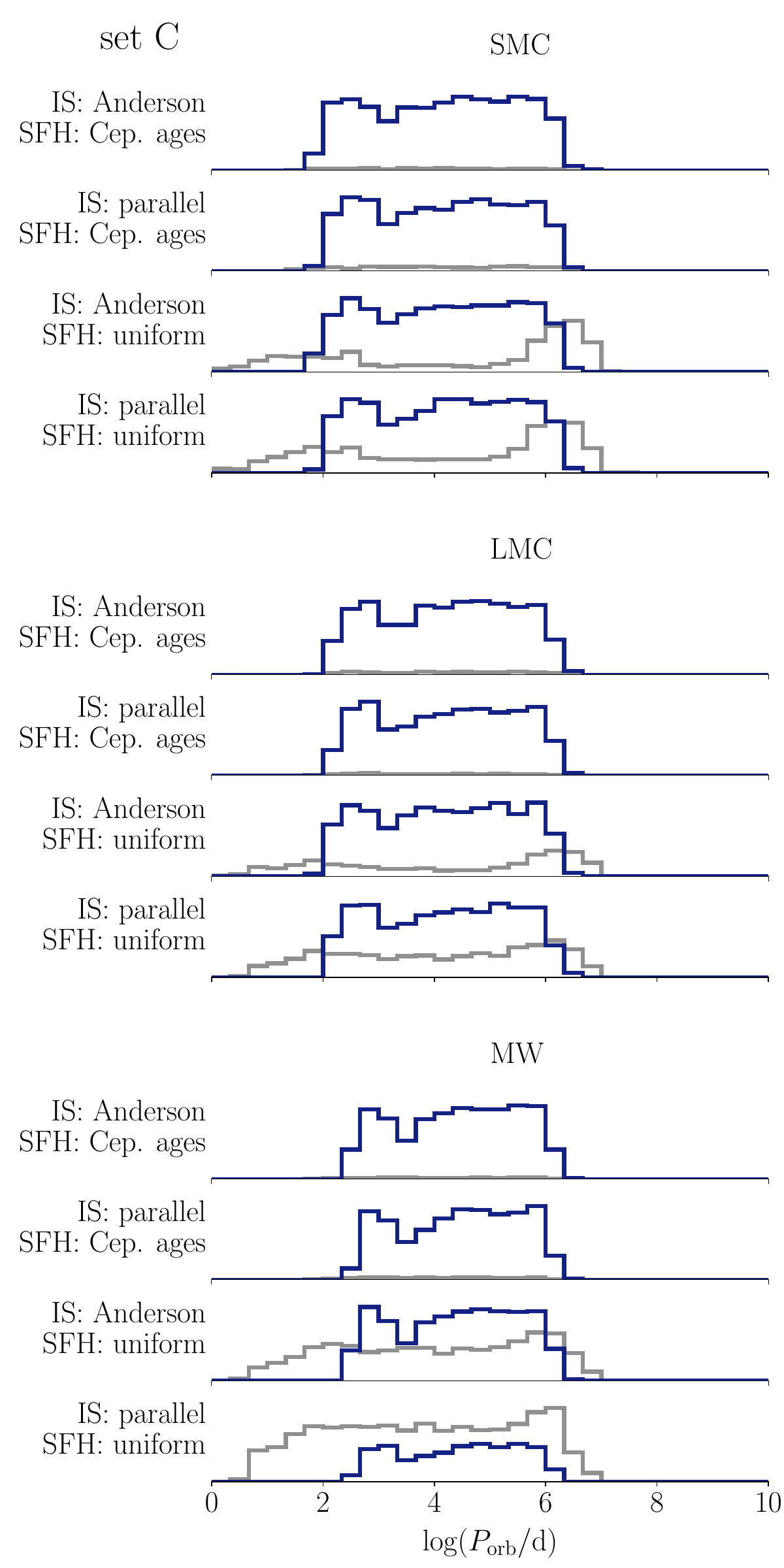}{0.3\textwidth}{}
    \fig{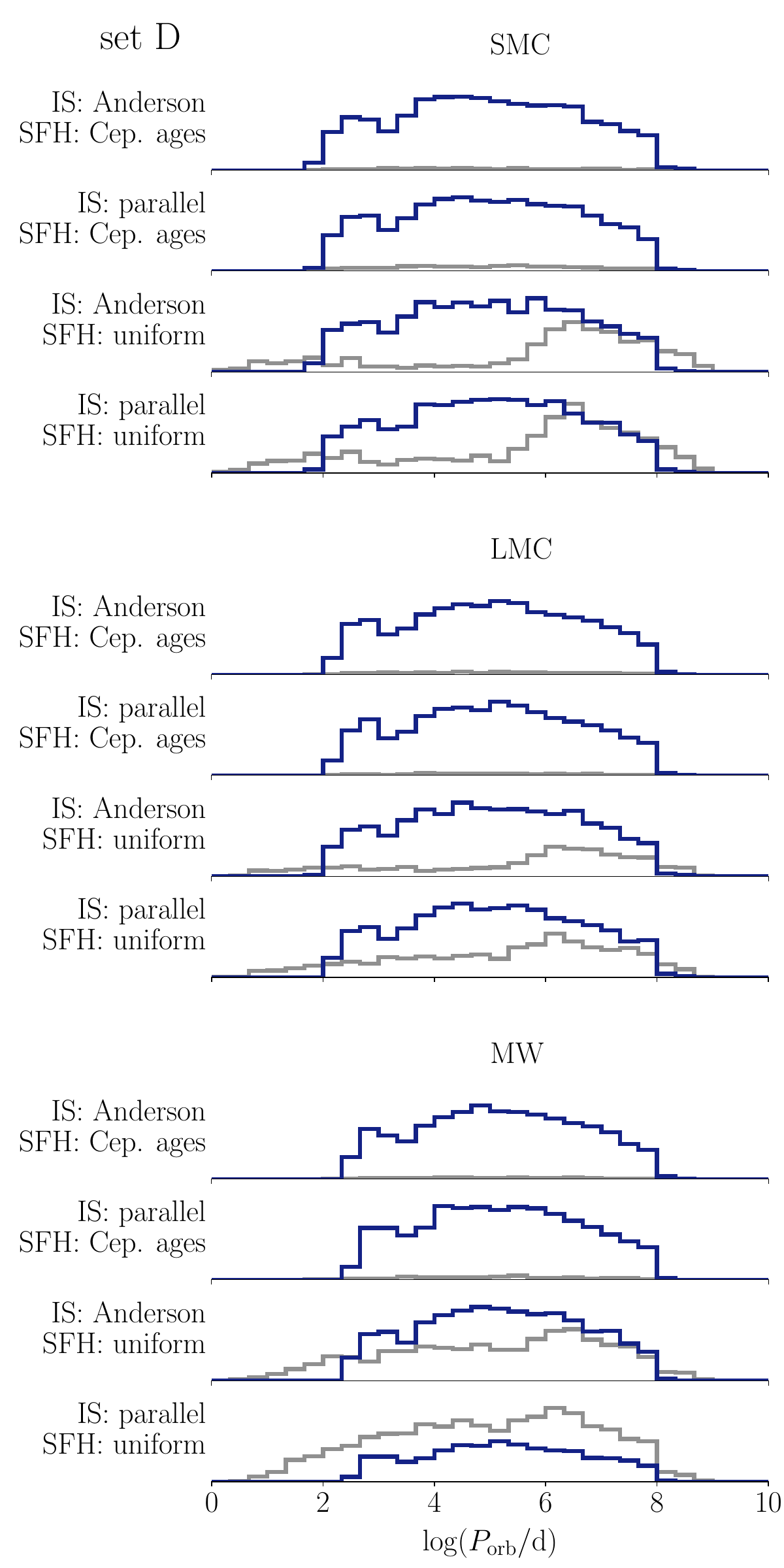}{0.3\textwidth}{}}
    \gridline{
    \fig{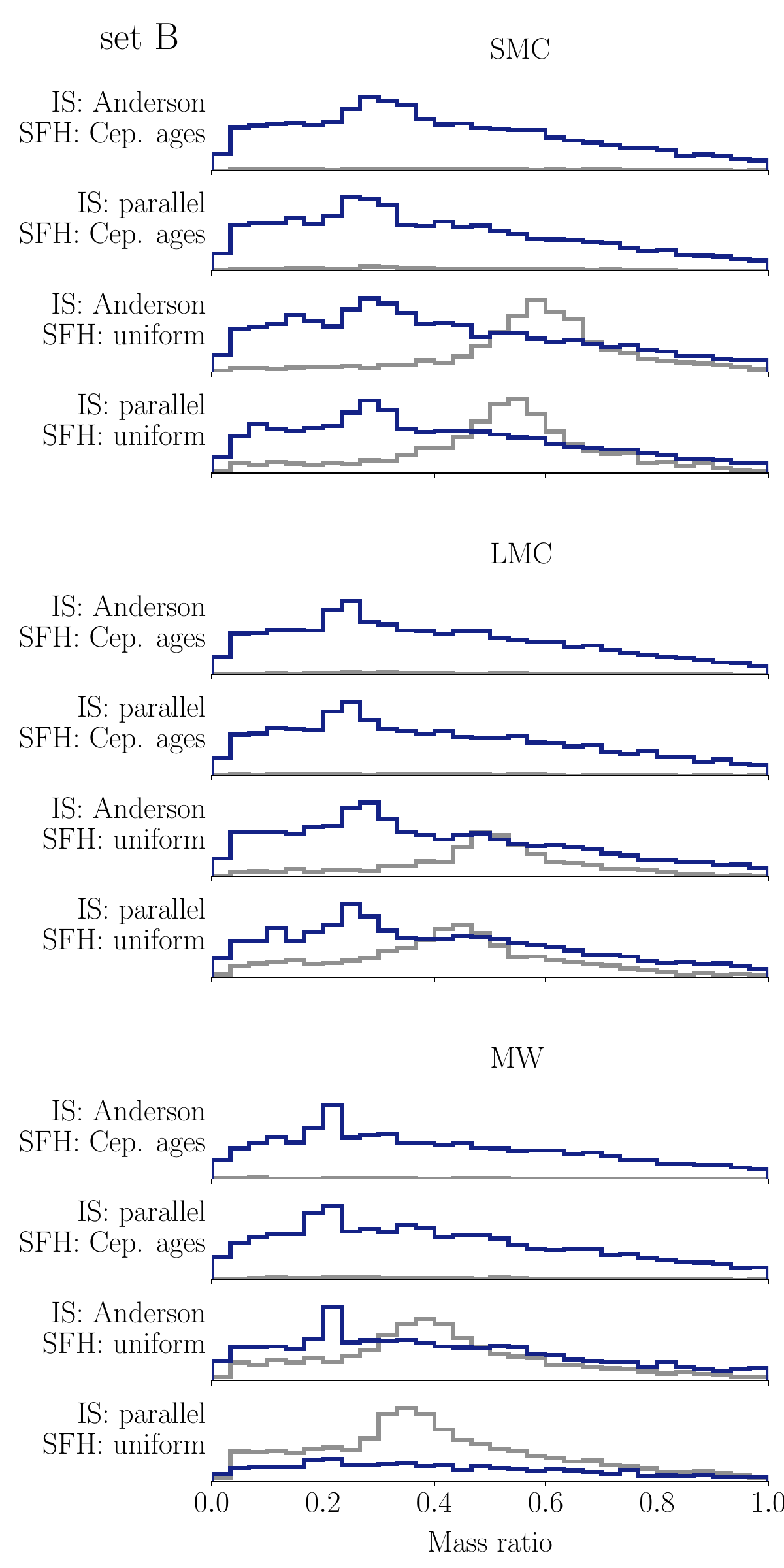}{0.3\textwidth}{}
    \fig{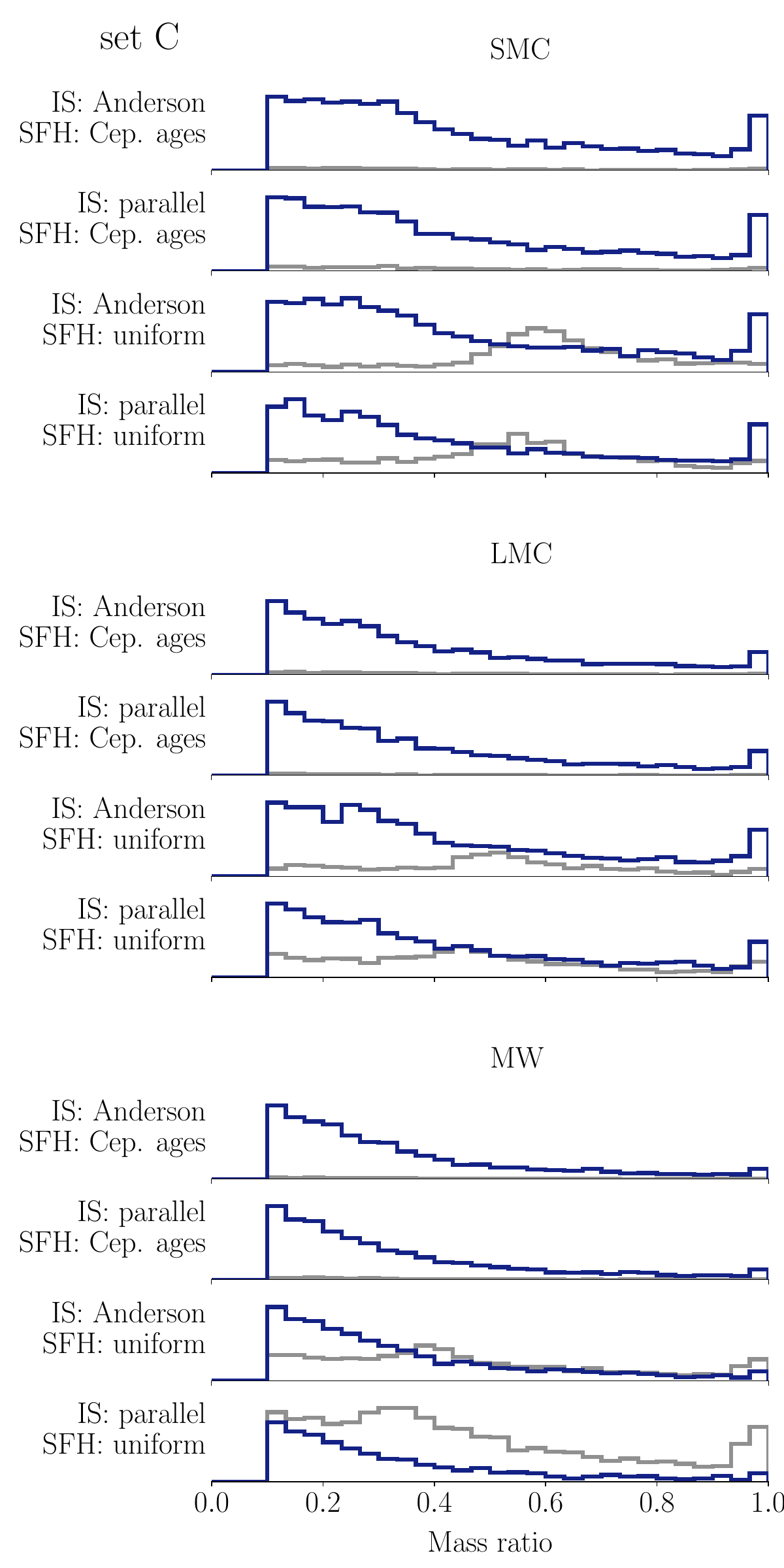}{0.3\textwidth}{}
    \fig{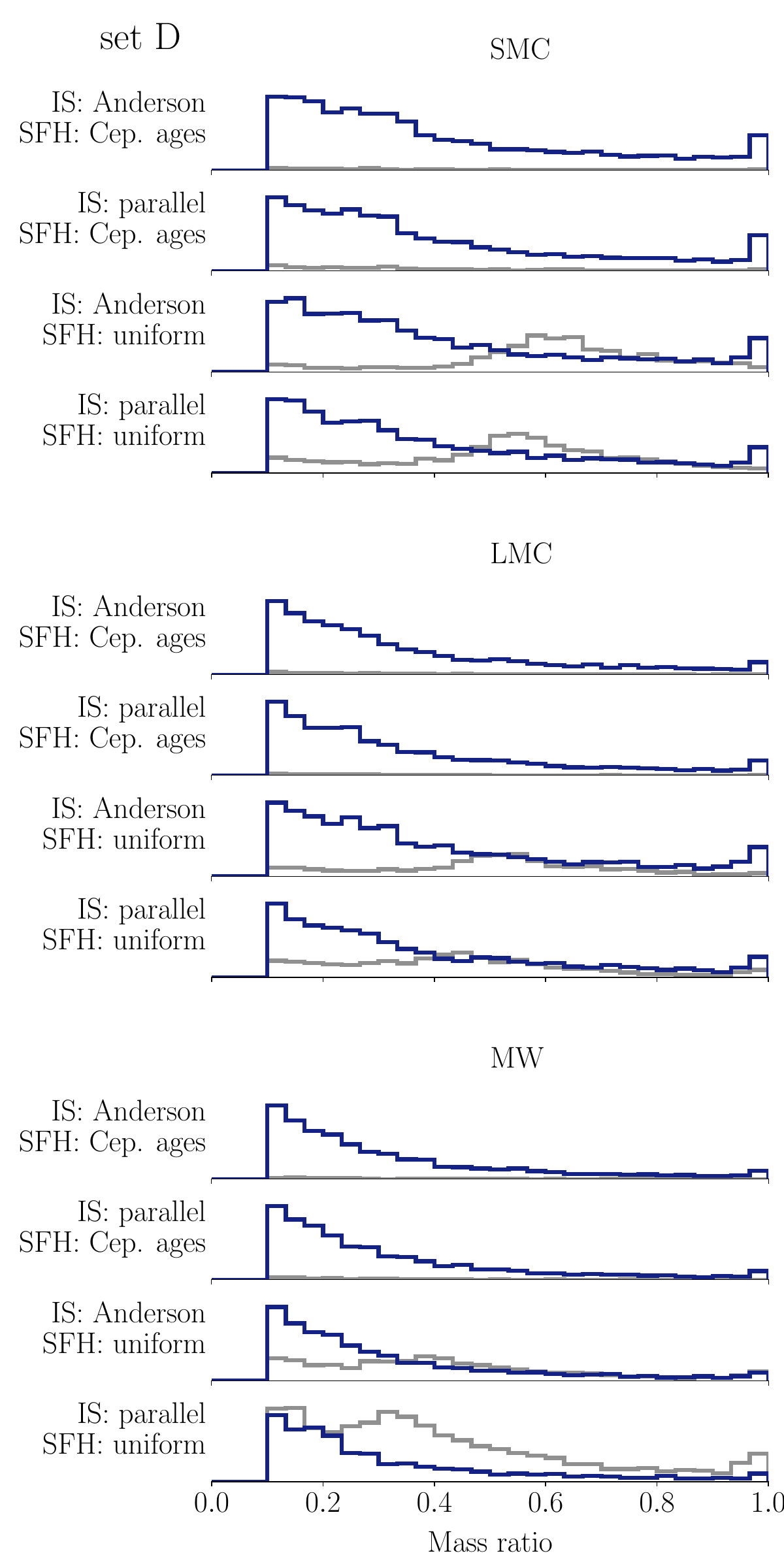}{0.3\textwidth}{}}
    \caption{Distributions of orbital periods (top)  and mass ratios (bottom) in 12 variants of synthetic populations, for sets B, C, D of initial parameters. IS2+3 Cepheids (navy blue) and IS1 Cepheids (gray) are presented separately. The values on all $y$-axes were scaled linearly from 0 to 1, and then omitted for a clearer comparison of the shapes of the distributions.
    \label{apdxfig:porb+q}}
\end{figure}

\begin{figure}
    \gridline{
    \fig{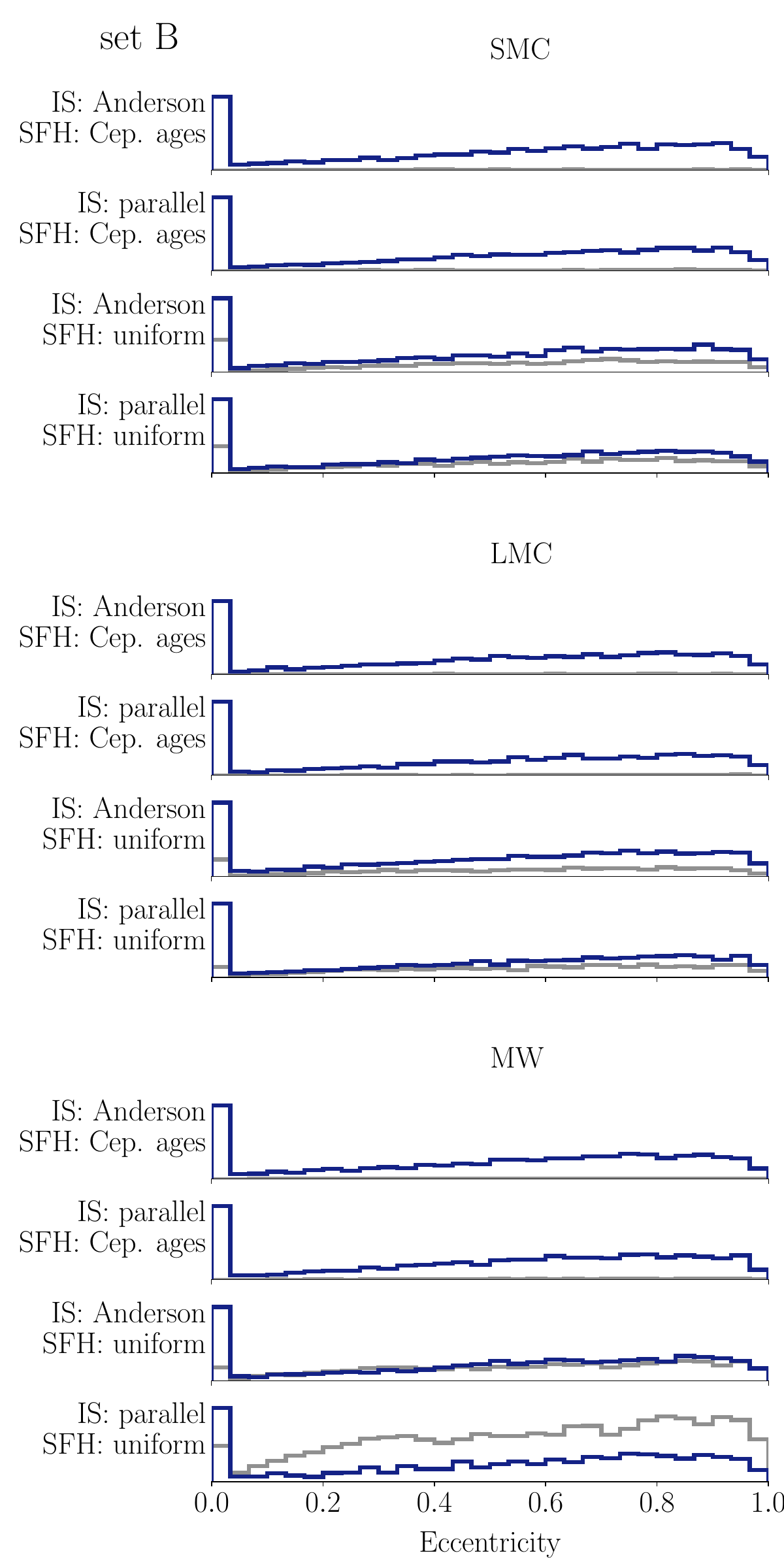}{0.3\textwidth}{}
    \fig{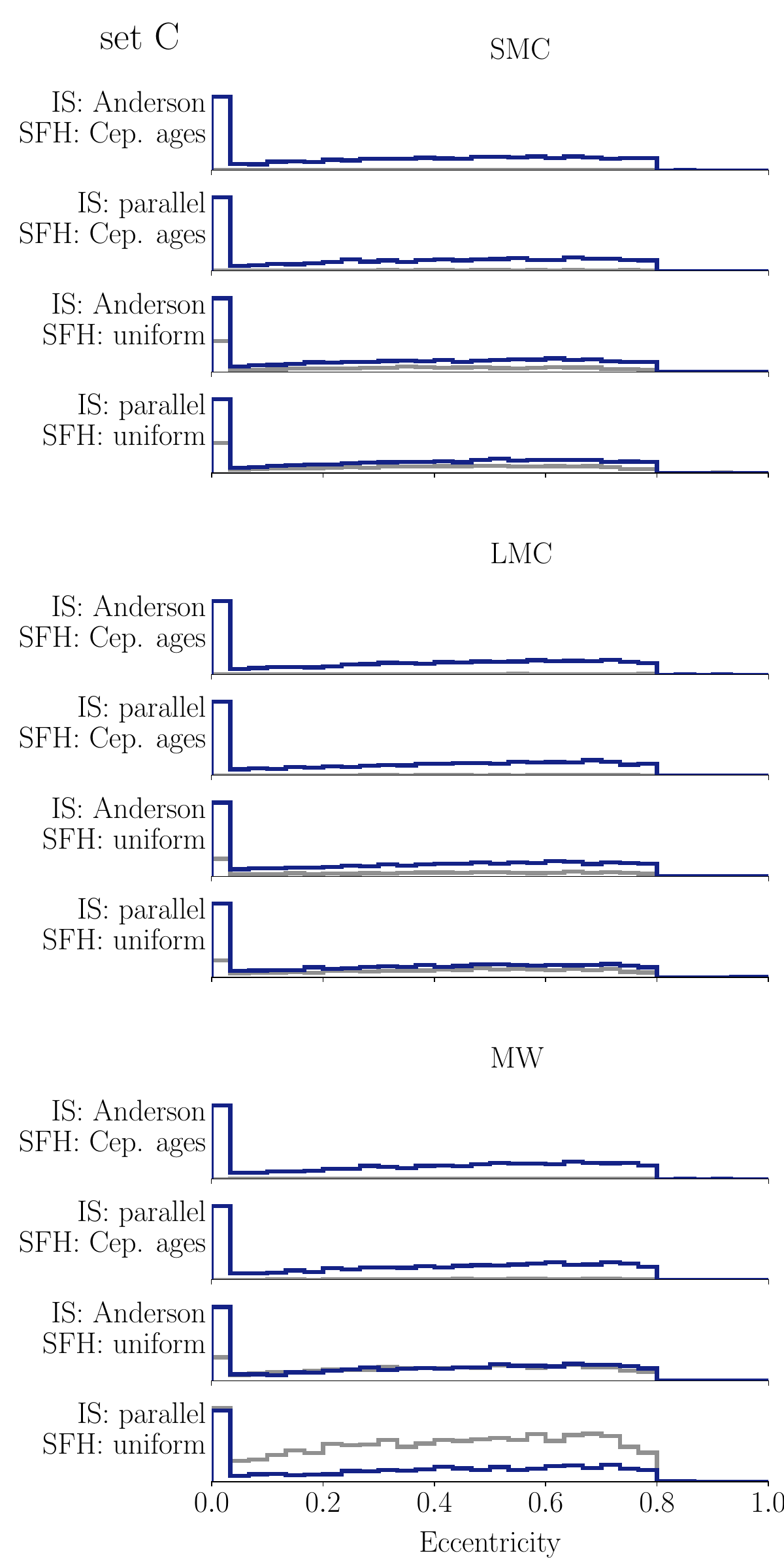}{0.3\textwidth}{}
    \fig{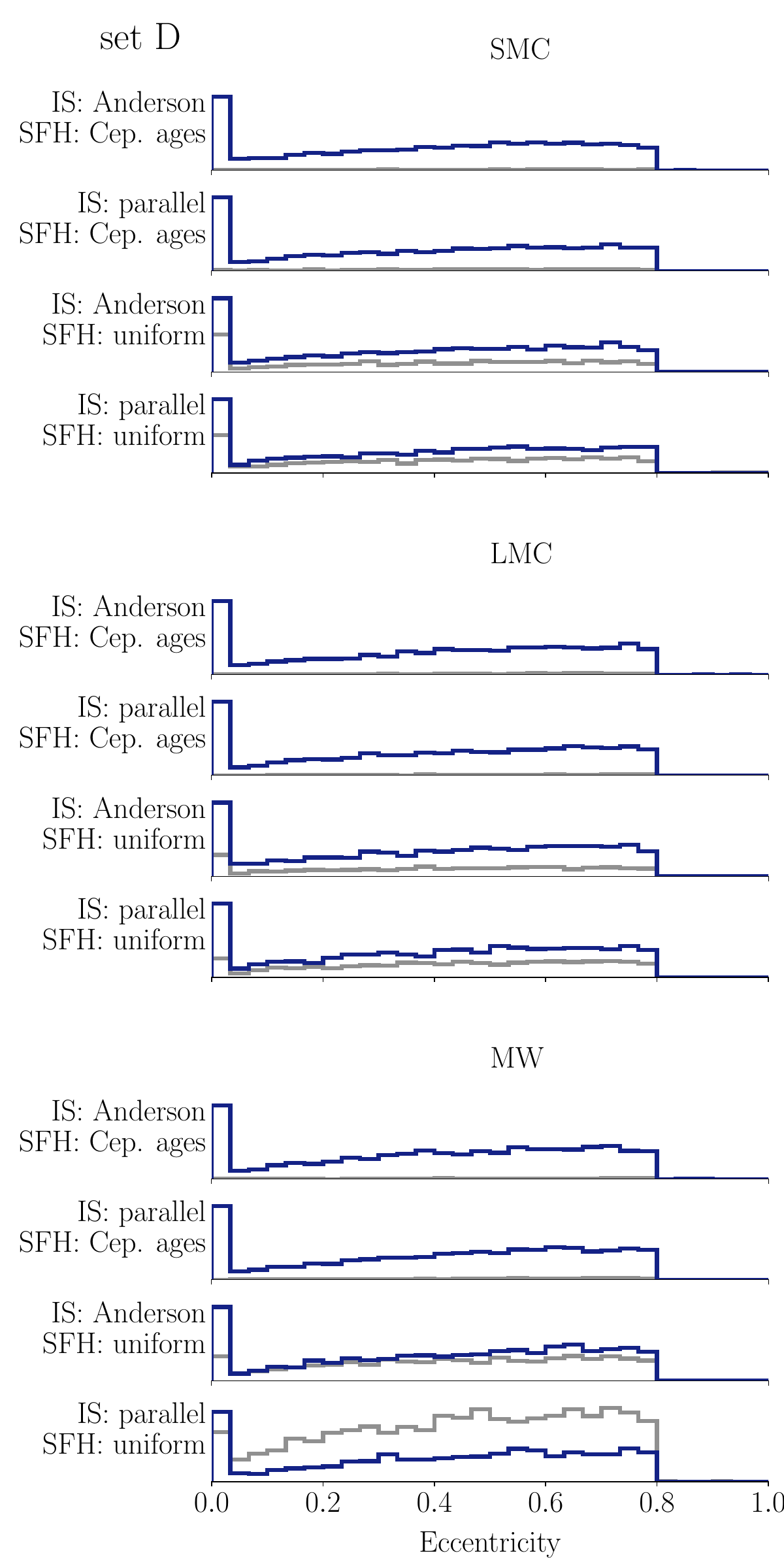}{0.3\textwidth}{}}
    \gridline{
    \fig{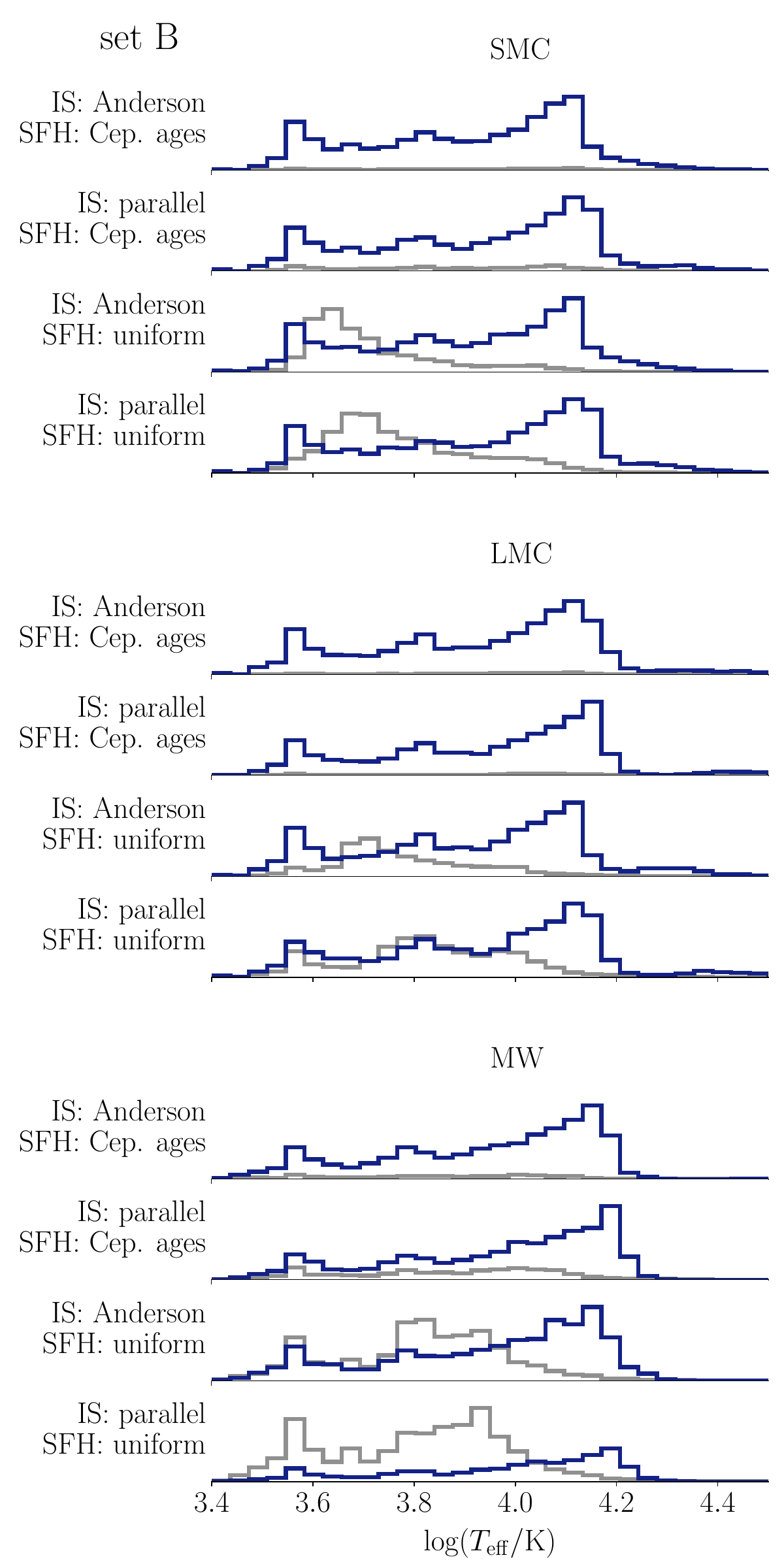}{0.3\textwidth}{}
    \fig{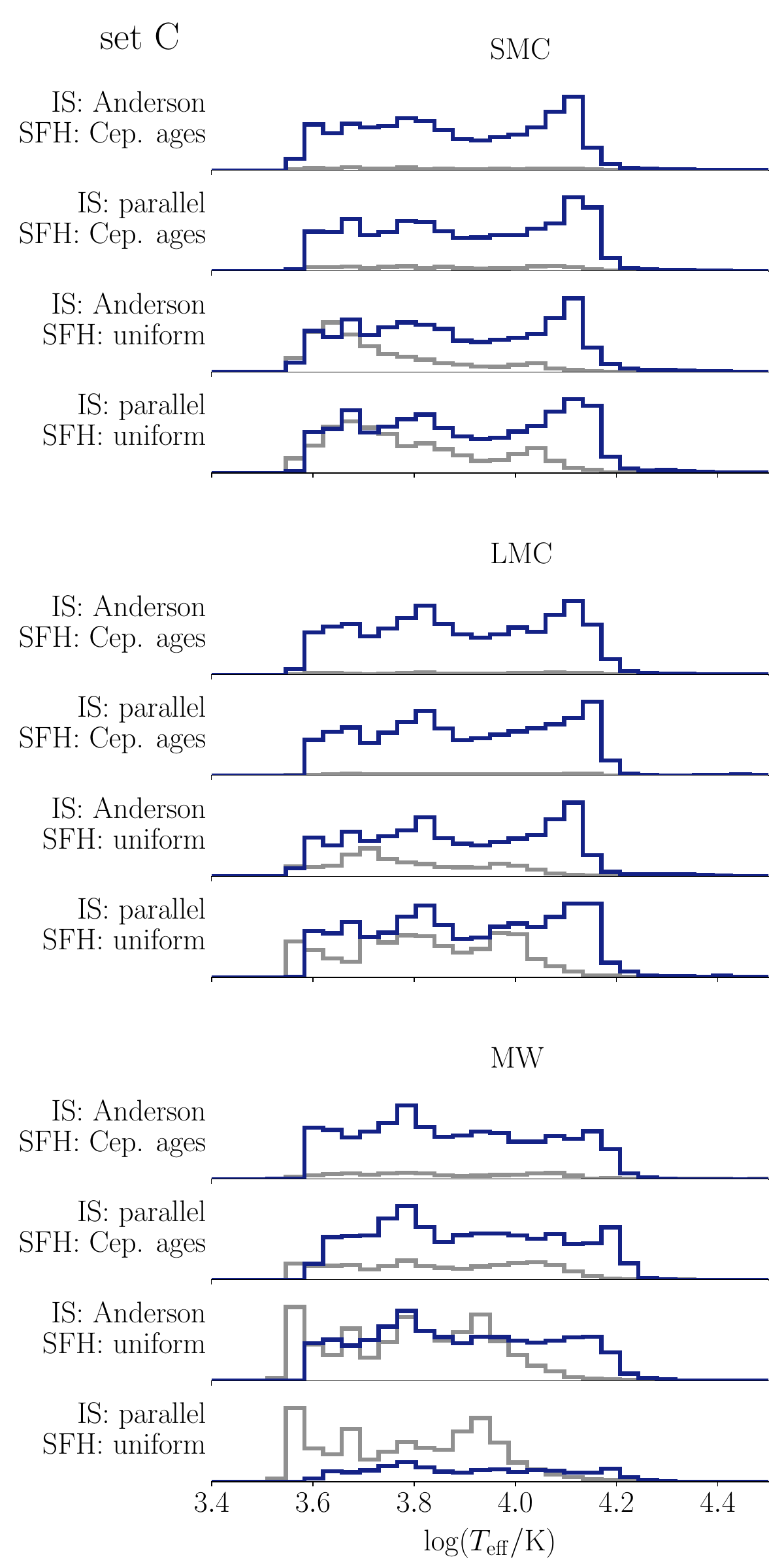}{0.3\textwidth}{}
    \fig{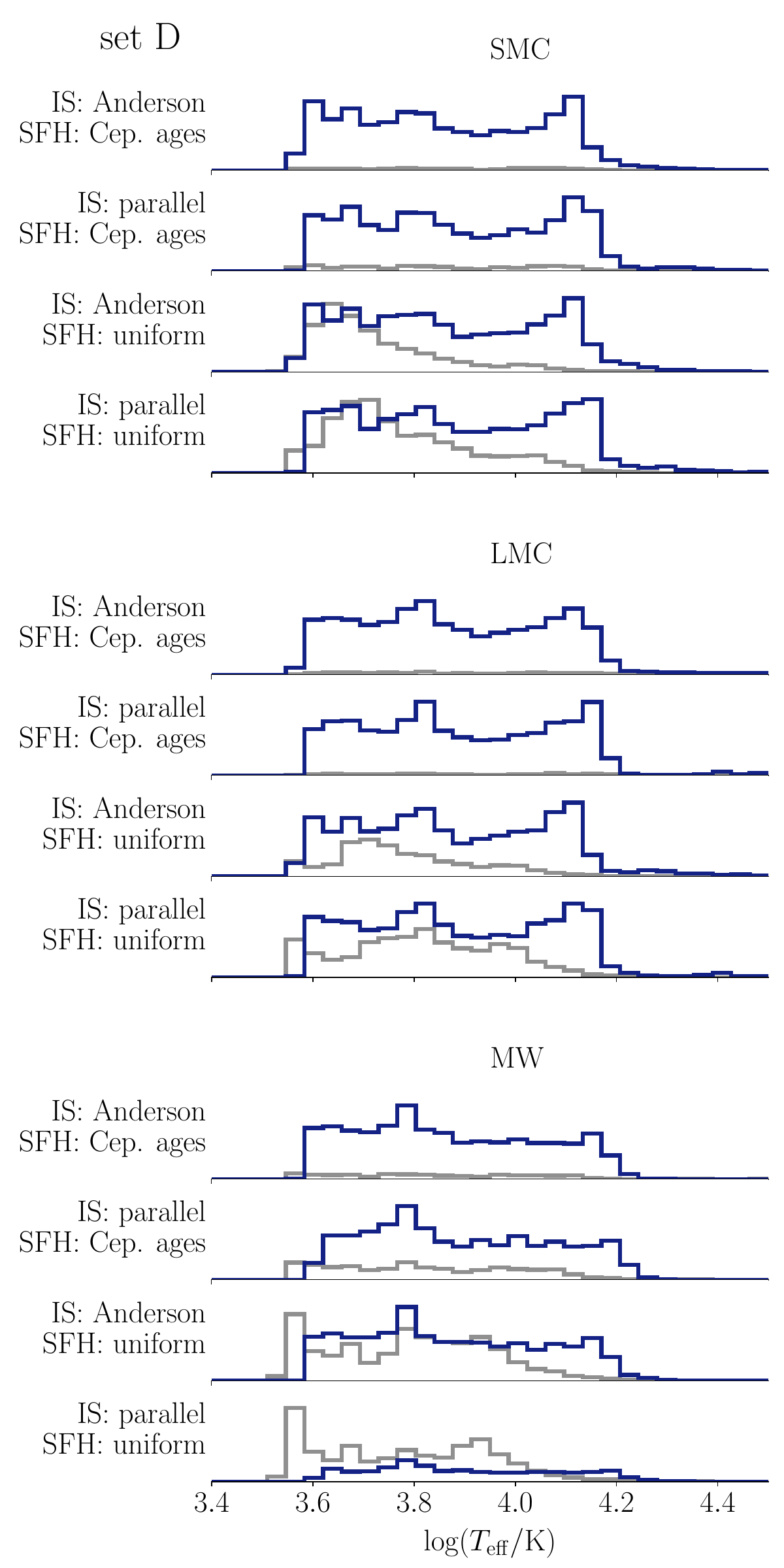}{0.3\textwidth}{}}
    \caption{Distributions of eccentricities (top) and companion effective temperatures (bottom) in 12 variants of synthetic populations, for sets B, C, D of the initial parameters. IS2+3 Cepheids (navy blue) and IS1 Cepheids (gray) are presented separately. The values on all $y$-axes were scaled linearly from 0 to 1, and then omitted for a clearer comparison of the shapes of the distributions.
    \label{apdxfig:ecc+teff}}
\end{figure}

\begin{figure}
    \gridline{
    \fig{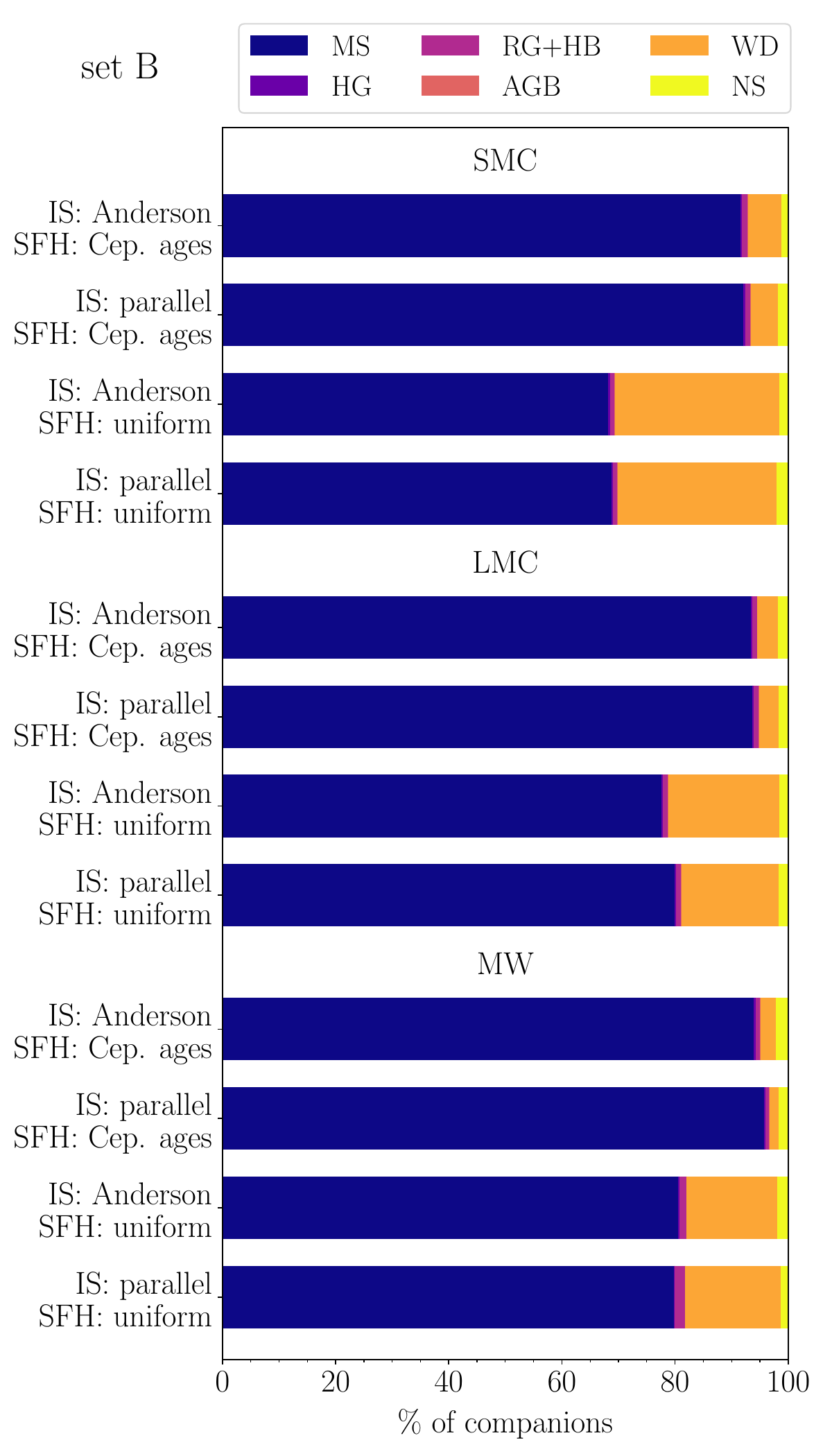}{0.32\textwidth}{}
    \fig{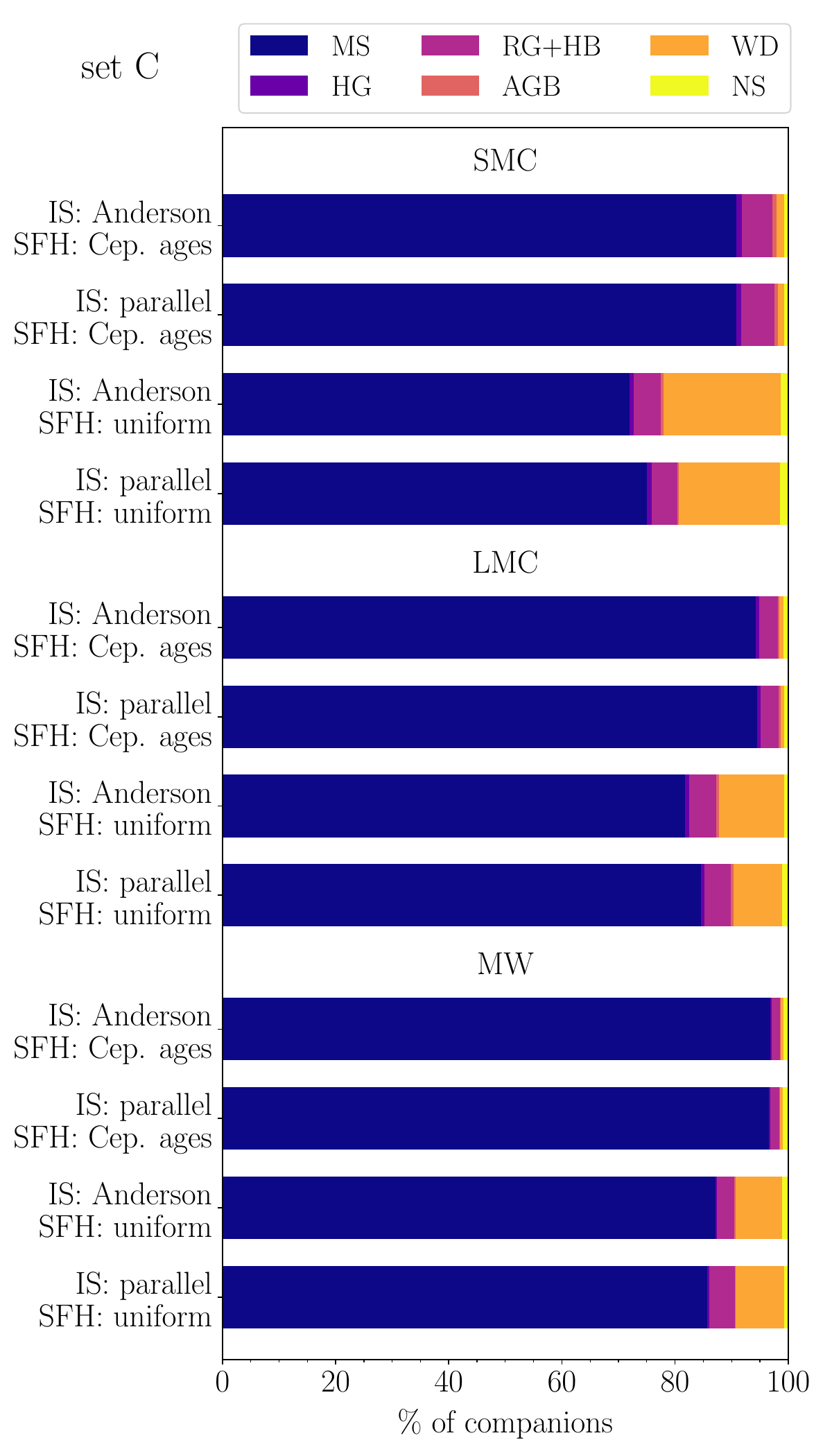}{0.32\textwidth}{}
    \fig{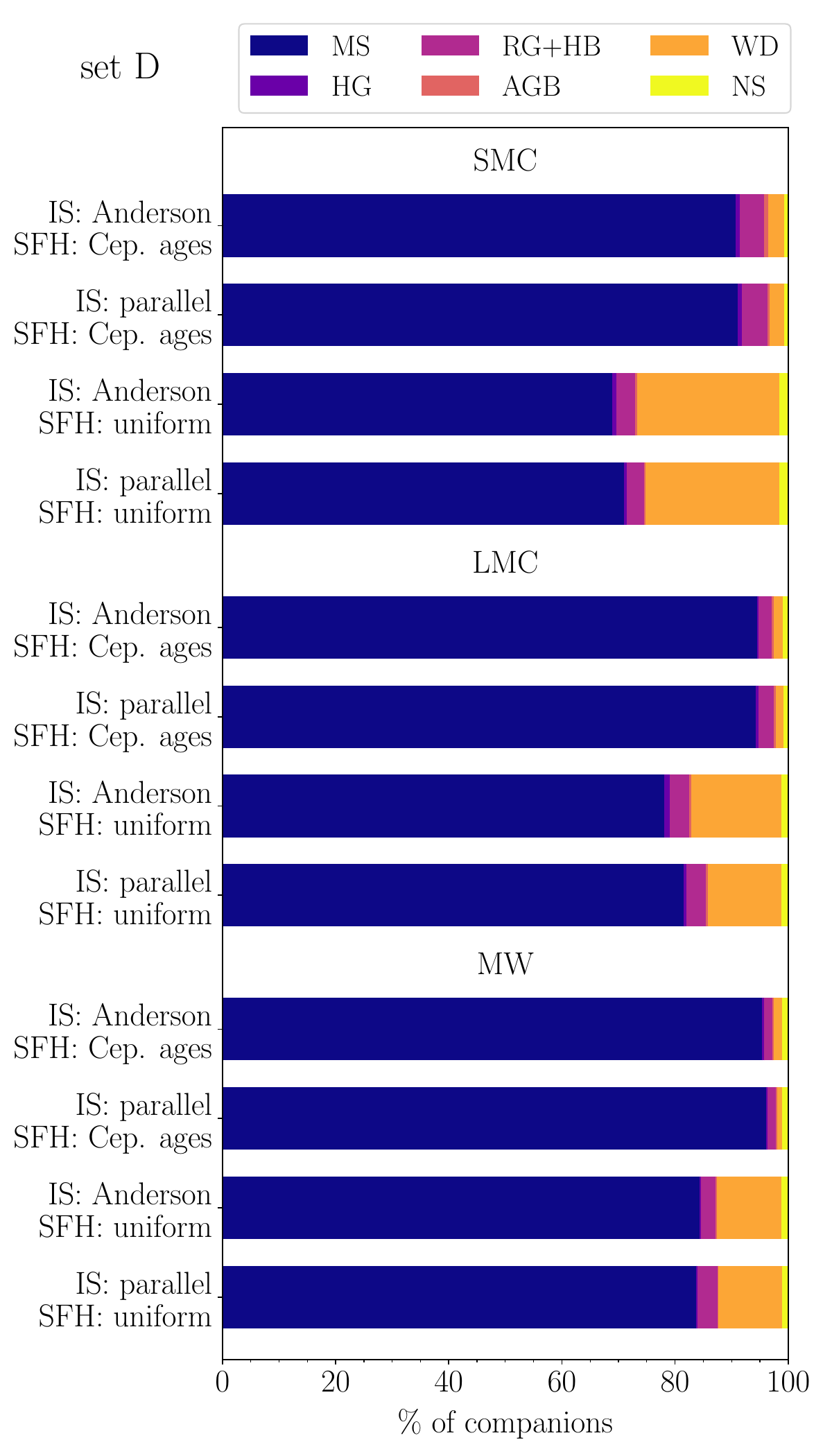}{0.32\textwidth}{}}
    \gridline{
    \fig{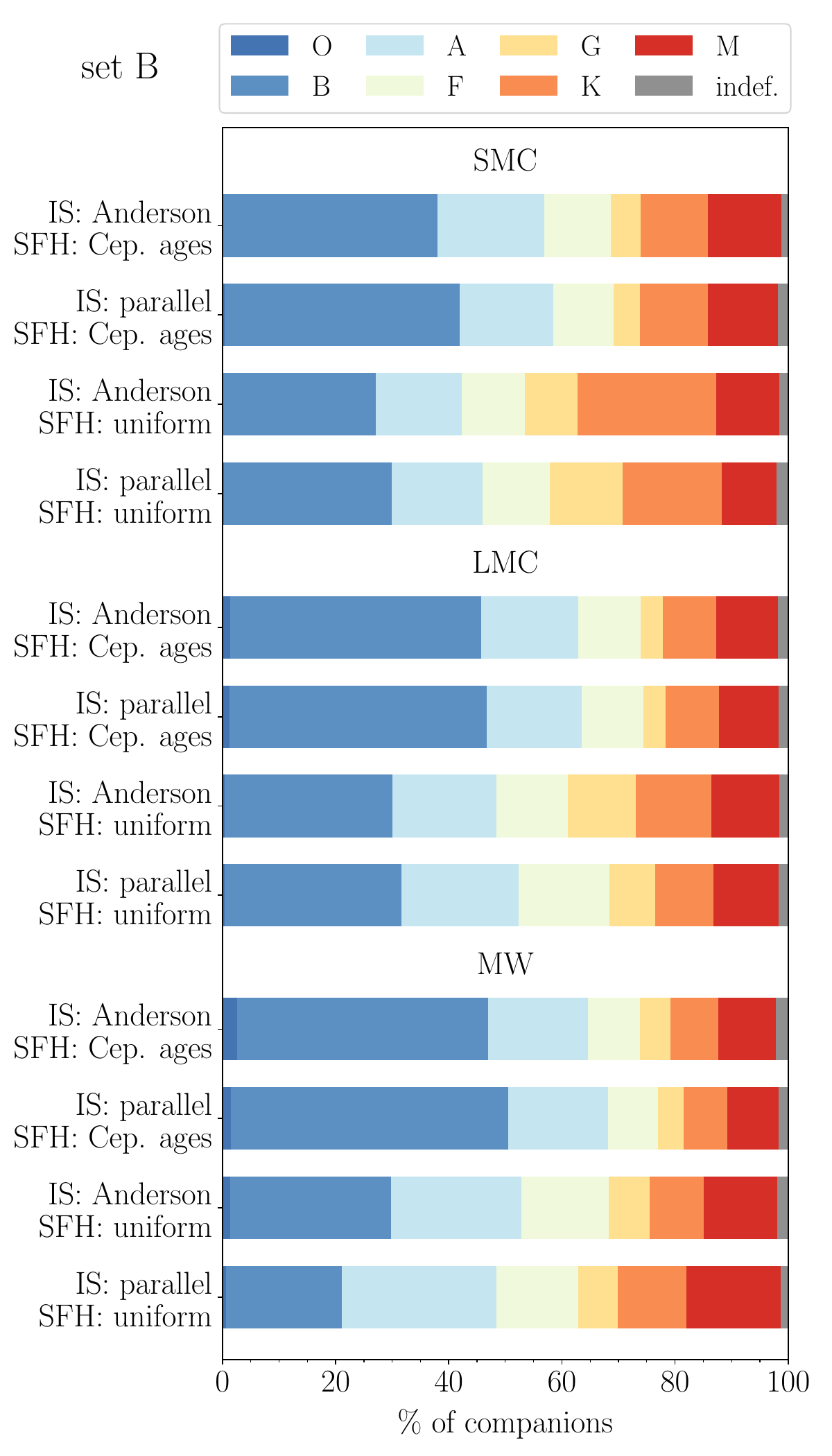}{0.32\textwidth}{}
    \fig{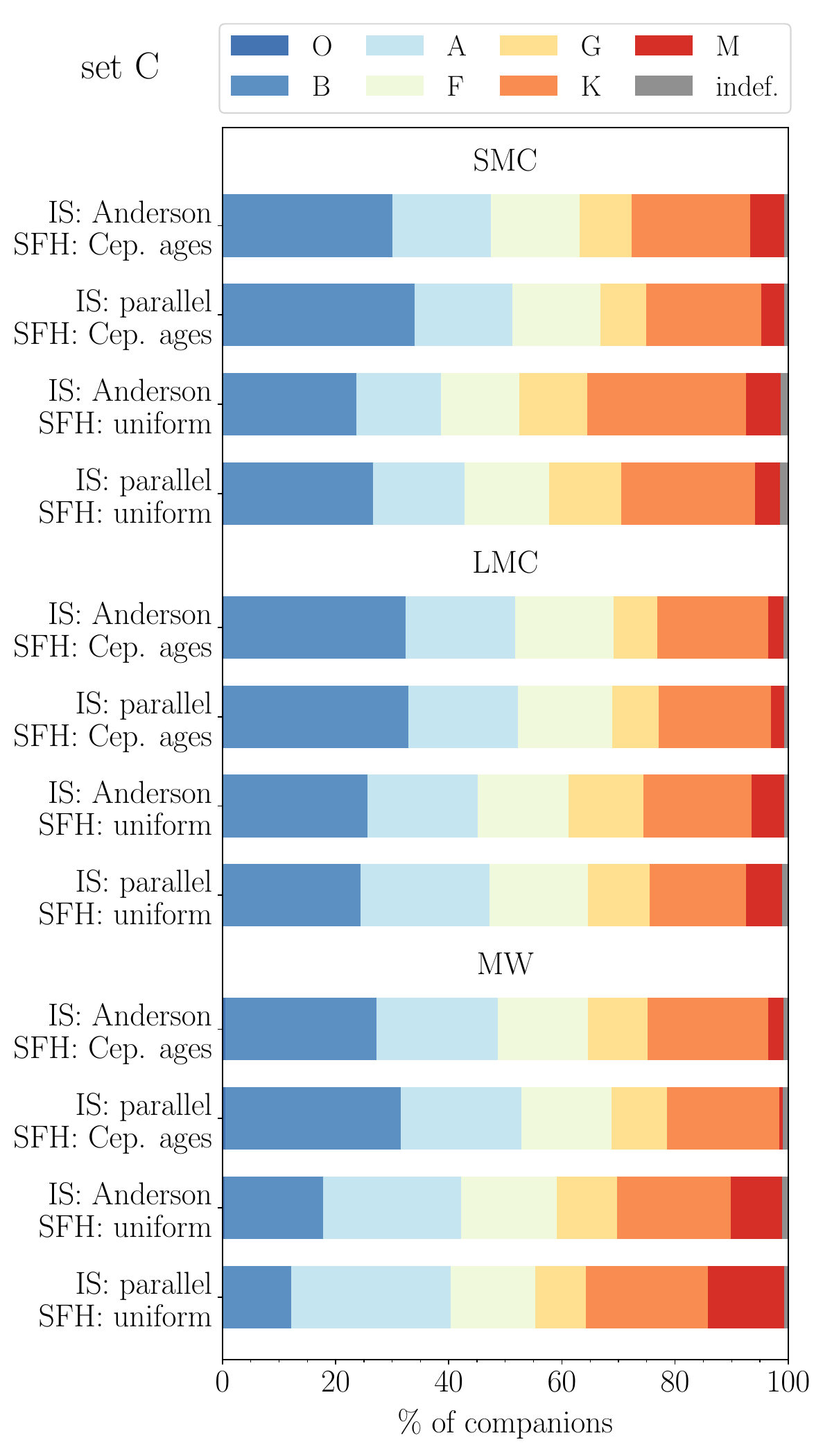}{0.32\textwidth}{}
    \fig{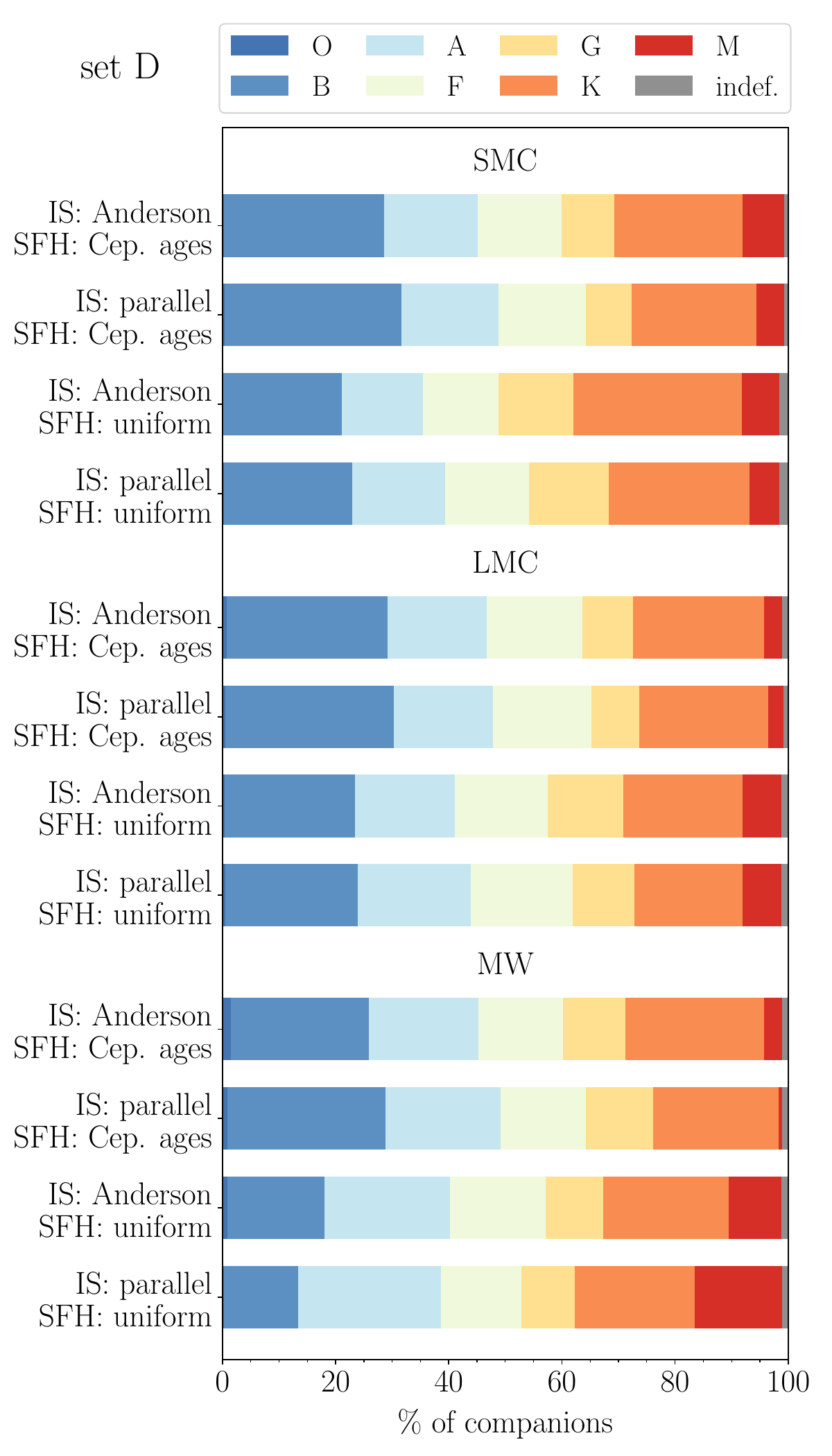}{0.32\textwidth}{}}
    \caption{Proportions of evolutionary (top) and spectral (bottom) types of companions in 12 variants of synthetic populations, for sets B, C, D of initial parameters.
    \label{apdxfig:evol+spec}}
\end{figure}

\section{Multiband contrast-mass ratio relations for MW Cepheids and their companions}
\label{apdx:contrast_Mq}
\begin{figure}
    \centering
    \includegraphics[width=0.4\columnwidth]{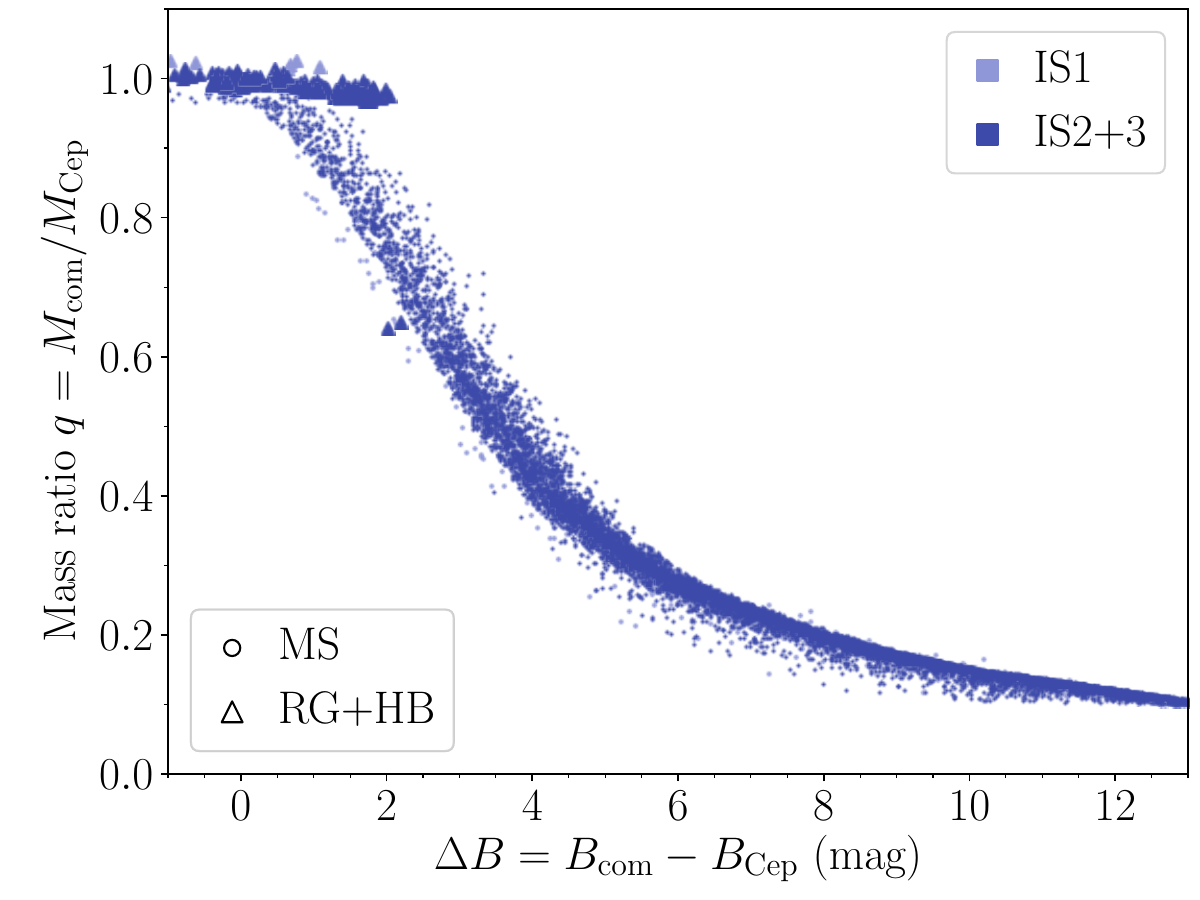}%
    \hspace{1em}
    \includegraphics[width=0.4\columnwidth]{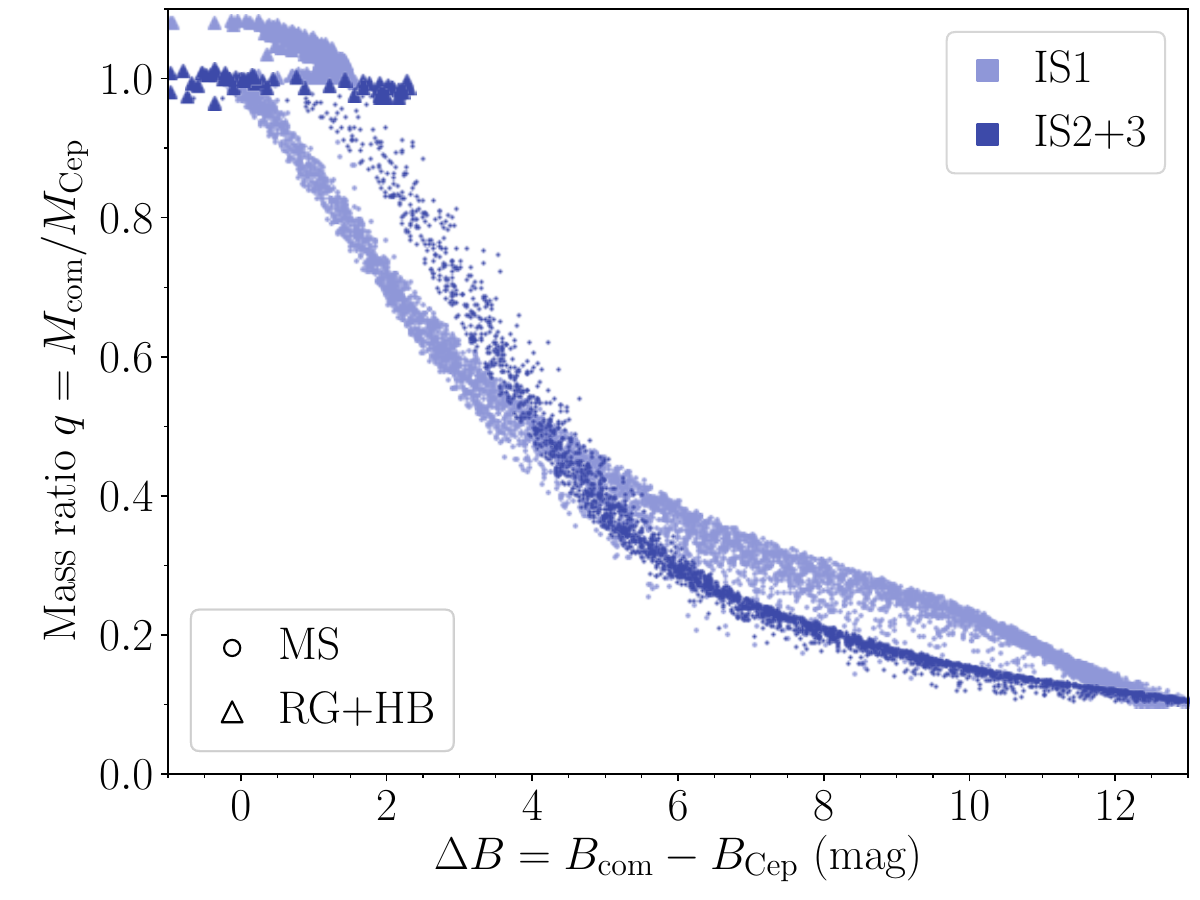}\\ \vspace{1em}
    \includegraphics[width=0.4\columnwidth]{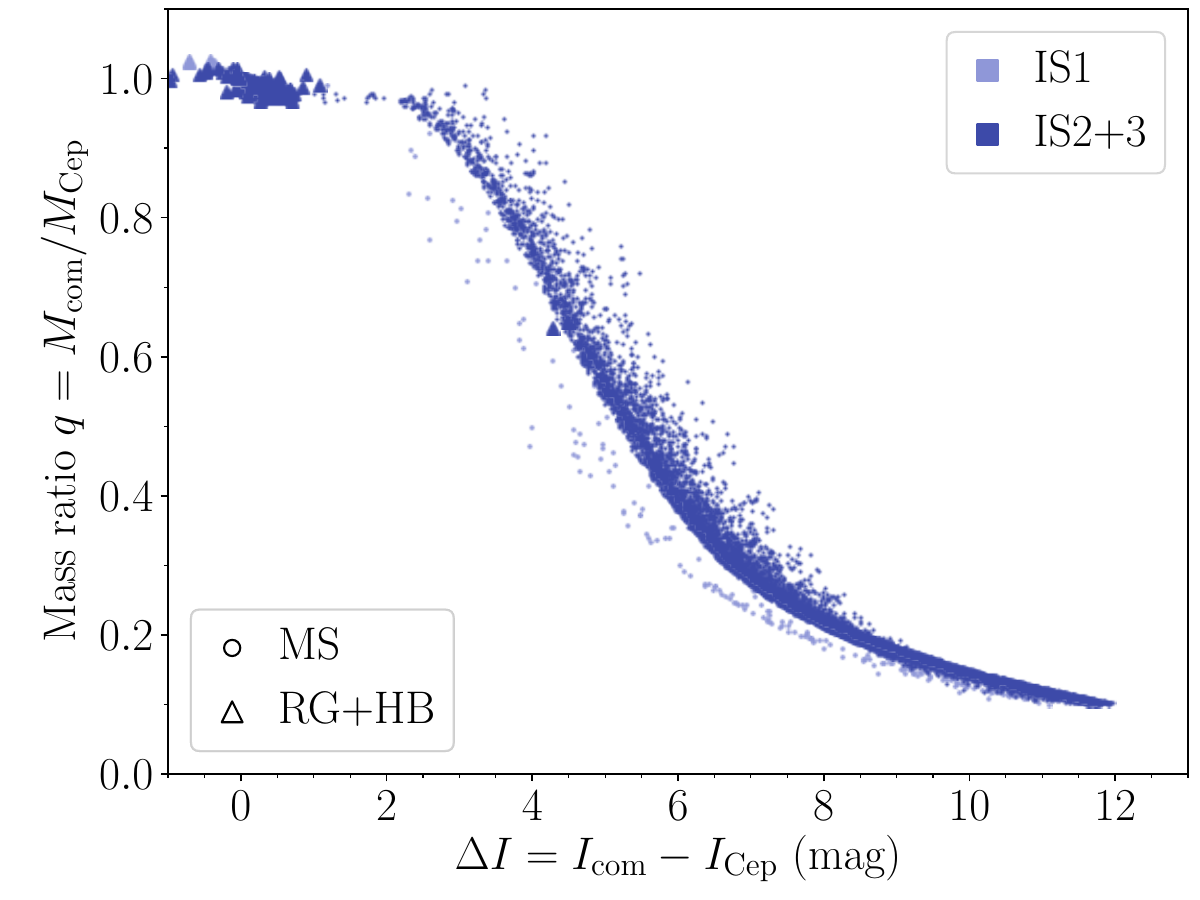}%
    \hspace{1em}
    \includegraphics[width=0.4\columnwidth]{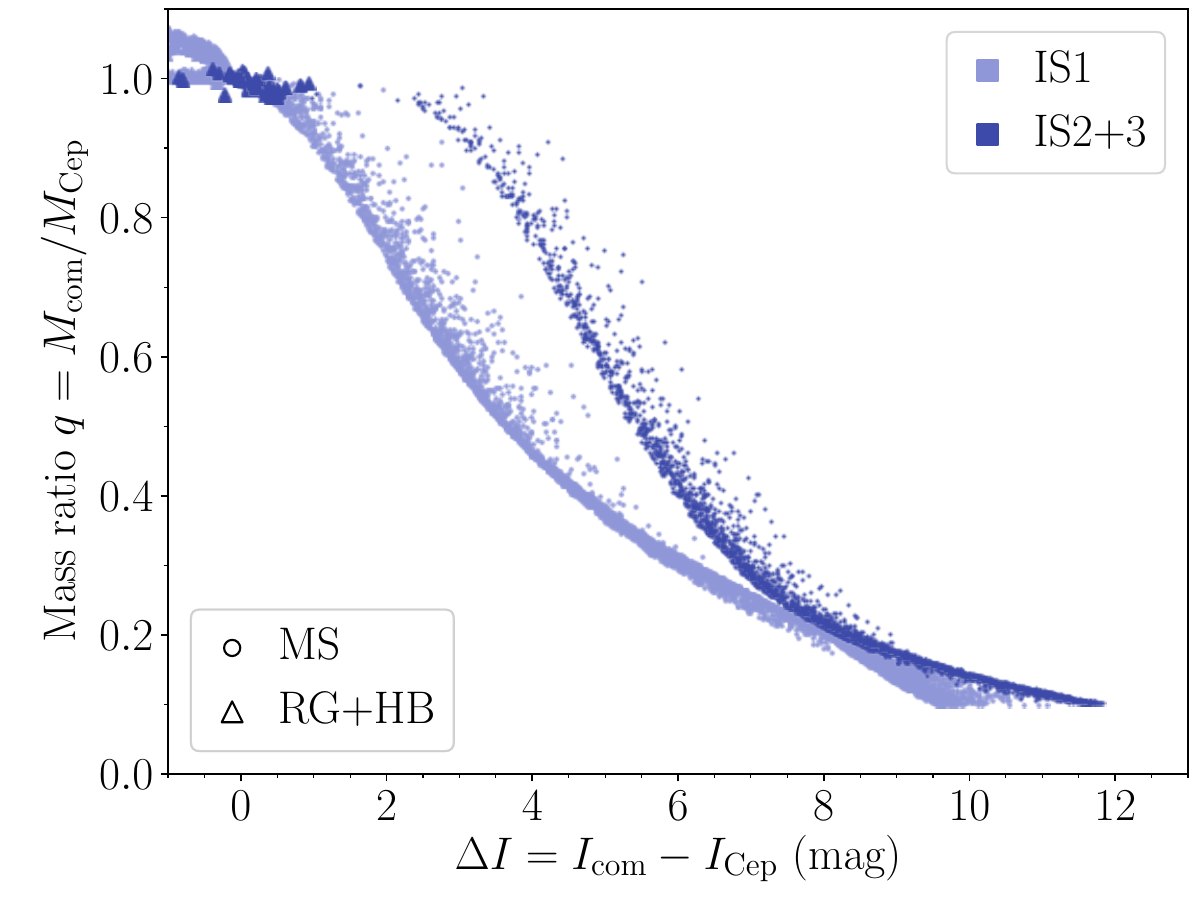}\\ \vspace{1em}
    \includegraphics[width=0.4\columnwidth]{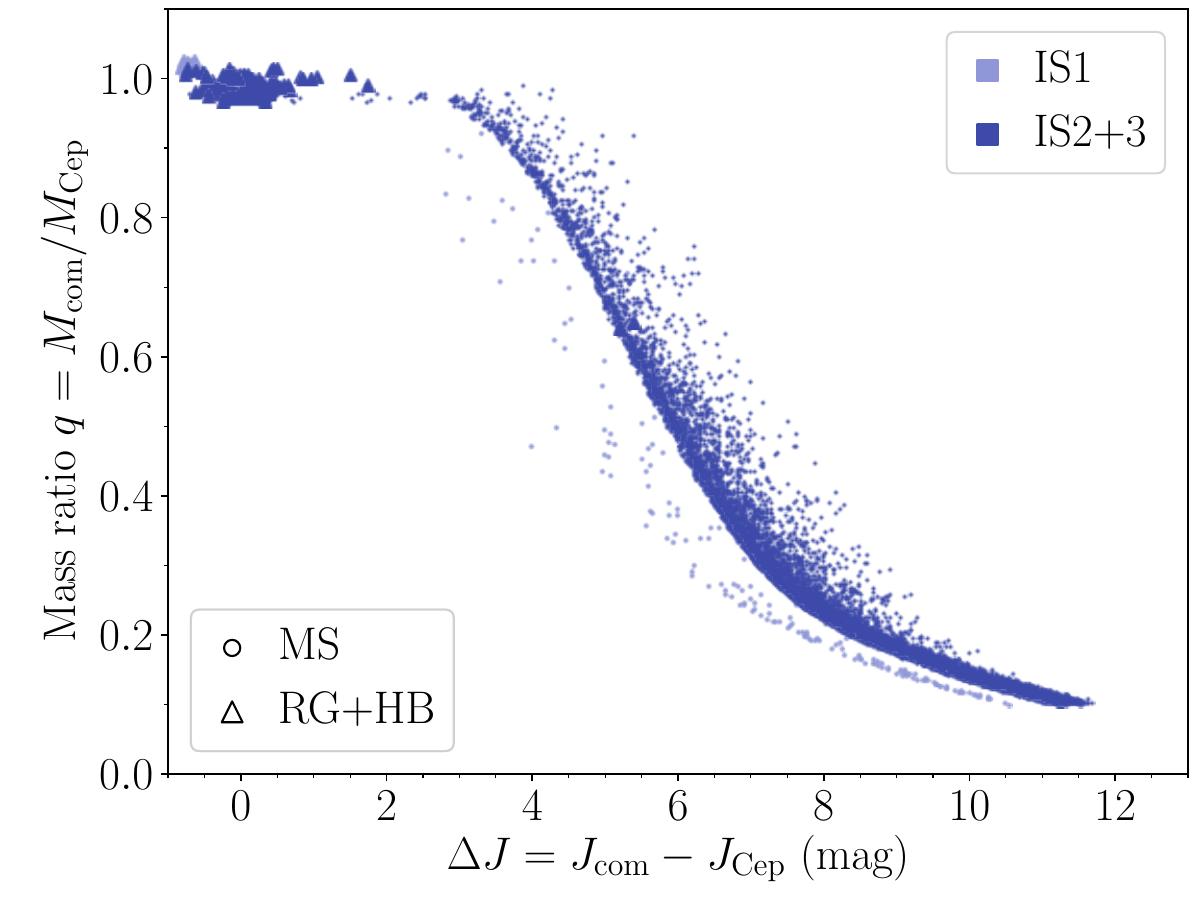}%
    \hspace{1em}
    \includegraphics[width=0.4\columnwidth]{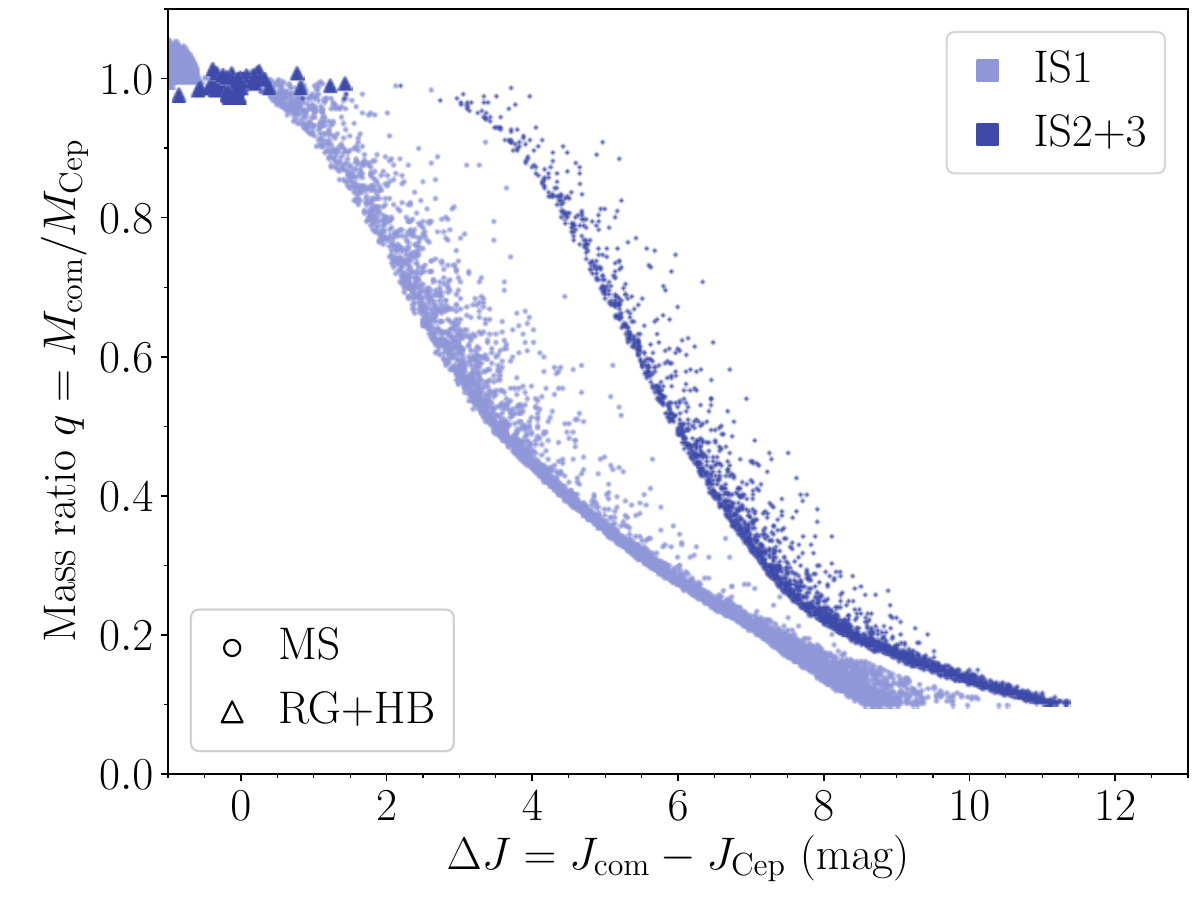}\\ \vspace{1em}
    \includegraphics[width=0.4\columnwidth]{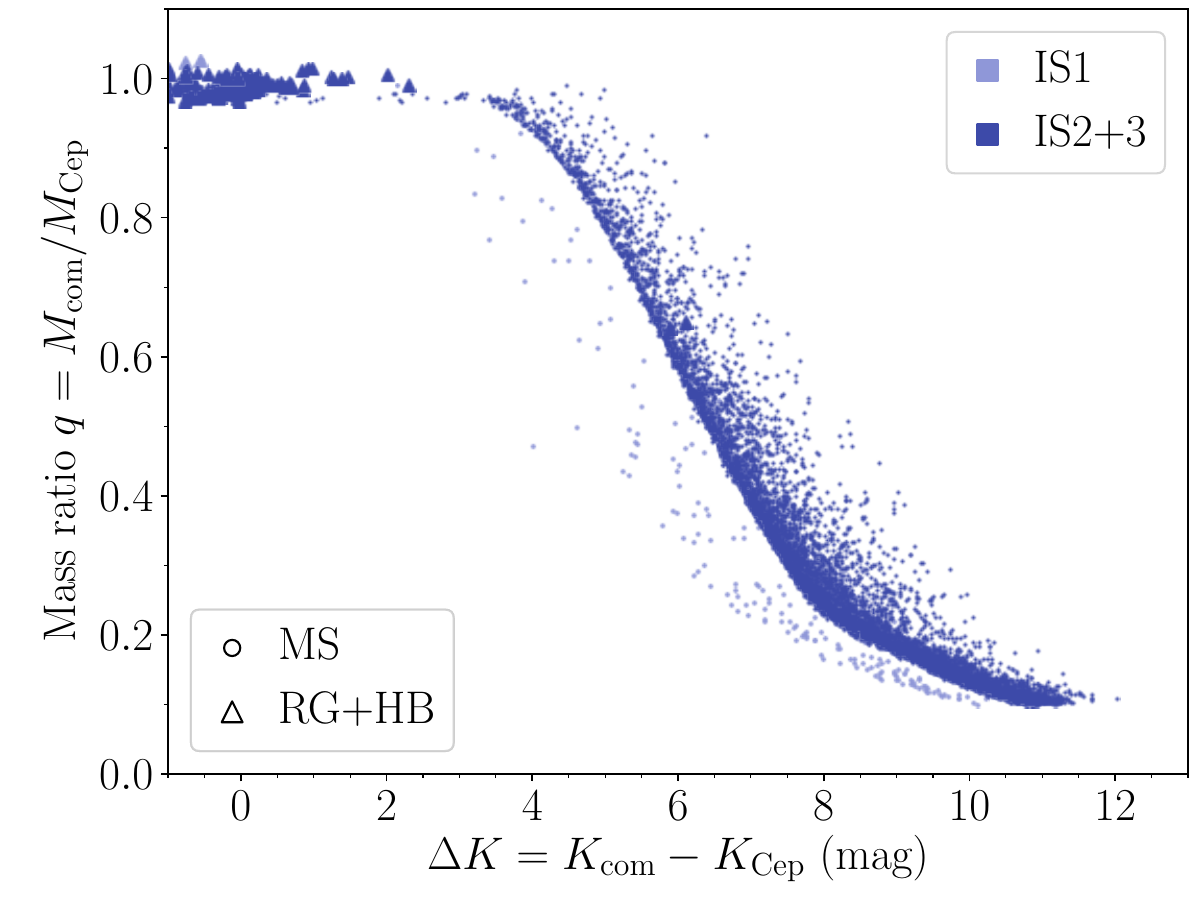}%
    \hspace{1em}
    \includegraphics[width=0.4\columnwidth]{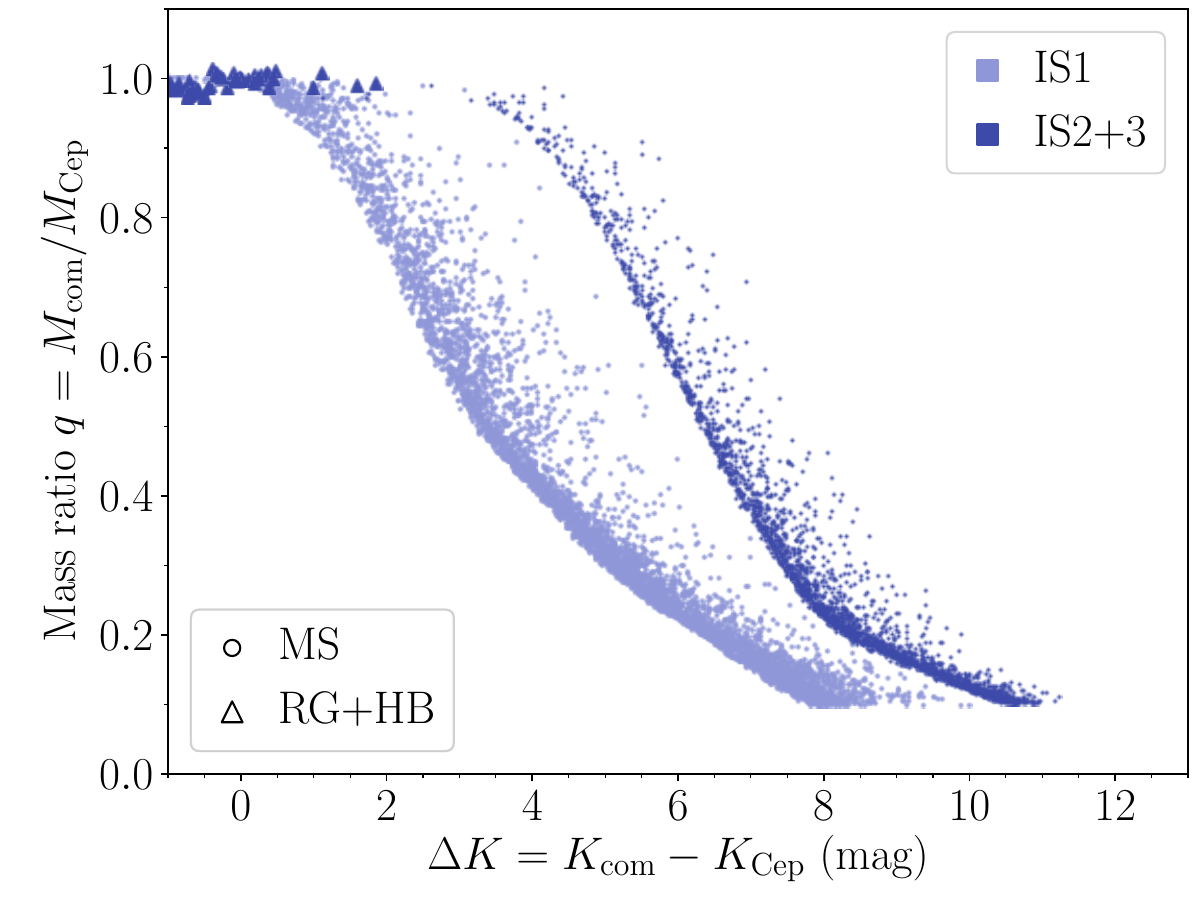}
    \caption{Magnitude difference (contrast) between MW Cepheids and their companions in the $B$, $I$, $J$, and $K$ bands, versus the mass ratio from two populations variants: set D, Anderson IS, SFH based on Cepheids' ages (left); set A, parallel IS, uniform SFH (right). The rest of populations fit between these two most divergent variants. This figure complements Figure \ref{fig:contrast_Mq}, which shows $V$ and $H$-band contrast-mass ratio relations.}
    \label{apdxfig:contrast_Mq}
\end{figure}

\restartappendixnumbering
We present contrast-mass ratio relations for the $B$, $I$, $J$, and $K$ bands, complementing the $V$- and $H$-band relations, presented in Figure \ref{fig:contrast_Mq}. Note that empirical contrasts are available only in $V$ and $H$ band (Table \ref{tab:mwcep}), and therefore Figure \ref{apdxfig:contrast_Mq} presents solely theoretical results.

\clearpage
\bibliography{references}
\bibliographystyle{aasjournal}



\end{document}